\documentclass[aps,twocolumn,nofootinbib]{revtex4-2}
\usepackage{amsmath,amssymb,bm}
\newcommand{\braket}[1]{\langle {#1} \rangle }
\newcommand{\ket}[1]{|{#1} \rangle }
\newcommand{\bra}[1]{\langle {#1}|}
\usepackage{amsmath}
\usepackage{color}
\usepackage{txfonts}

\usepackage[pdftex]{graphicx}
\usepackage{hyperref}
\begin{document}
\title{Nuclear Josephson-like  $\gamma$-emission}
	\author{R. A. Broglia}
	\affiliation{The Niels Bohr Institute, University of Copenhagen, DK-2100 Copenhagen, Blegdamsvej 17, Denmark
		}
        \affiliation{Dipartimento di Fisica, Universit\`a degli Studi di Milano, Via Celoria 16, I-20133 Milano, Italy}
        	\author{F. Barranco}
            \affiliation{Departamento de F\'isica Aplicada III, Escuela Superior de Ingenieros, Universidad de Sevilla, Camino de los Descubrimientos, Sevilla, Spain}
            \author{L. Corradi}
	\affiliation{INFN, Laboratori Nazionali di Legnaro-35020 Legnaro, Italy}
	\author{G. Potel}
	\affiliation{Lawrence Livermore National Laboratory, Livermore, California 94550, USA}
    \author{S. Szilner}
    \affiliation{Ru{d\llap{\raise 1.22ex\hbox
{\vrule height 0.09ex width 0.2em}}\rlap{\raise 1.22ex\hbox
{\vrule height 0.09ex width 0.06em}}}er
Bo\v{s}kovi\'{c} Institute, HR-10$\,$001 Zagreb, Croatia}
	\author{E. Vigezzi}
	\affiliation{INFN Sezione di Milano, Via Celoria 16, I-20133 Milano, Italy}
	\date{\today}	
\begin{abstract}
  Josephson-like junctions, transiently established  in heavy ion collisions between superfluid nuclei, few MeV below the Coulomb barrier, allow for the back and forth transfer of a nuclear Cooper pair of effectively charged nucleons and thus the emission of $\gamma$-rays. The semiclassical description of  single Cooper pair alternating current is shown to contain the gauge phases and gauge rotational frequencies as required by the Josephson (ac) effect, in keeping with the derivation of the transfer (tunneling) Hamiltonian in a gauge invariant representation. The fact that such reaction description is equivalent to a second order DWBA $T$-matrix formulation extensively used in the study of pairing rotational bands with two-particle transfer reactions, together with the nuclear structure result that the Bardeen-Cooper-Schrieffer (BCS) condensation order parameter $\alpha_0=\sum_{\nu>0}U_\nu V_\nu$ (number of Cooper pairs), sum of the coherence factors $U_\nu V_\nu$ (proportional to the two-nucleon transfer spectroscopic amplitudes), is quite stable with respect to  model description, is found to be connected with  the emergence of two strongly convergent parameters (conserved quantities) within the time the abnormal densities of the two superfluid nuclei overlap: a) the correlation length (dc); b) the number of emitted $\gamma$-rays per cycle (ac), and thus the dipole moment of the successively transferred nucleons. Result which  leads to a  nuclear parallel with the direct current (dc) and alternating current (ac) Josephson effects, and  which testifies to the validity of BCS theory of superconductivity down to few Cooper pair condensates, and single Cooper pair alternating currents. The physics at the basis of a quantitative description of Cooper pair tunneling between weakly coupled superconductors or superfluid nuclei at energies below the Coulomb barrier, is that the process is dominated by the successive transfer of the two partner fermions entangled over distances of the order of the coherence length, $\approx10^4$\AA ${}$ in the case of lead, and 13.5 fm in the case of the reaction $^{116}\text{Sn}+^{60}\text{Ni}\to^{114}\text{Sn(gs)}+^{62}\text{Ni(gs)}$ at few MeV below the Coulomb barrier.
\end{abstract}
\maketitle
\tableofcontents
\section{Overview}
At the basis of the understanding of nuclear structure one finds the specific probes of the variety of elementary modes of excitation, namely inelastic scattering and Coulomb excitation (surface vibrations and rotations), one-particle transfer reactions (independent particle motion) and two-particle transfer processes (pairing vibrations and rotations).

Let us consider the transfer process $a+A\to b+B$ described by the Hamiltonian
\begin{align}
  \label{over_eq:1}
  \nonumber H&=T_{aA}+H_a+H_A+V_{aA},\\
  &=T_{bB}+H_b+H_B+V_{bB},
\end{align}
and associated Schr\"odinger equation being
\begin{align}
  \label{over_eq:2}
  i\hbar\frac{\partial \Psi(\xi,\mathbf r,t)}{\partial t}=H\,\Psi(\xi,\mathbf r,t),
\end{align}
where
\begin{align}
  \label{over_eq:3}
  \Psi(\xi,\mathbf r,t)=\Psi_\beta(\xi) \Psi(\mathbf r)\,e^{-i\frac{E}{\hbar}t},
\end{align}
and
\begin{align}
  \label{over_eq:4}
  \Psi_\beta=\Psi^b(\xi_b)\Psi^B(\xi_B),
\end{align}
$\xi$ being the intrinsic variables, while
\begin{align}
  \label{over_eq:5}
  \nonumber H_b\Psi^b&=E_b\Psi^b,\\
  H_B\Psi^B&=E_B\Psi^B.
\end{align}
For one-particle transfer reactions $a(=b+1)+A\to b+B(A+1)$ and
\begin{align}
  \label{over_eq:6}
  \Psi(\xi,t)=\Psi(\xi)\,e^{-i\frac{\epsilon_\nu}{\hbar}t}
\end{align}
where $\epsilon_\nu=\epsilon_\nu^{(0)}+\Sigma=\epsilon_\nu^{(0)}+\Delta E+i\Gamma=\tilde\epsilon_\nu+i\Gamma$, one can write
\begin{align}
  \label{over_eq:7}
  \Psi(\xi,t)=\Psi(\xi)\,e^{-i\frac{\tilde\epsilon_\nu}{\hbar}t}\,e^{-\Gamma t},
\end{align}
$\Sigma$ being the self-energy, $\Delta E$ and $\Gamma$ the corresponding real and imaginary (lifetime) components, resulting from the interweaving of single-particle and collective modes. In other words, Eq. (\ref{over_eq:7}) summarizes the dynamical shell model (see e.g. \cite{Mahaux:85} and refs. therein).

In the case of a two-nucleon transfer process in a heavy ion collision below the Coulomb barrier between  superfluid nuclei (weak link),
\begin{align}
  \label{over_eq:8}
  a(=b+2)+A(\equiv\alpha) \to b+B(=A+2)(\equiv\beta), 
\end{align}
each nucleus being described by a coherent $\ket{BCS}$ state which can be viewed as the intrinsic state of a pairing rotational band (gauge space, Fig. \ref{abs_figII} and App. \ref{rotgauge}, see also App. \ref{AppH}), one can write $E_\beta=2\times \mu_{Bb}=2\times \hbar\dot\phi_{Bb}$ and $E_\alpha=2\times \mu_{Aa}=2\times \hbar\dot\phi_{Aa}$. Here, $\mu$ stands for the Fermi energy (chemical potential) and $\dot\phi$ for the rotational frequency in gauge space. Consequently, the matrix elements of the tunneling (transfer) Hamiltonian between entrance and exit channels will be multiplied by the phase function $\exp\left(i\tfrac{E_\beta-E_\alpha}{\hbar}t\right)$ where,
\begin{align}
  \label{over_eq:9}
  \frac{E_{\beta}-E_{\alpha}}{\hbar}=\frac{Q_{2n}}{\hbar}=2\times\frac{\mu_{Bb}-\mu_{Aa}}{\hbar}=2\times(\dot\phi_{Bb}-\dot\phi_{Aa}).
\end{align}
Translated into a weak link between metallic superconductors $\alpha\to L$ (left) and $\beta\to R$ (right), $2\times\frac{\mu_R-\mu_L}{\hbar}=\frac{2eV}{\hbar}=\omega_J$, $V$ being the direct current bias applied to the barrier, and $\omega_J$ the angular frequency associated with the alternating  current (ac) Josephson effect \cite{Josephson:62,Anderson:64} and associated microwave emitted photons. In the present paper we elaborate on the corresponding nuclear parallel (\cite{Dietrich:70,Dietrich:71b,Hara:71,Kleber:71,Weiss:79,vonOertzen:01}, see also \cite{Potel:21} and App. \ref{AppB}).
\section{Introduction}
In the quest of determining how many Cooper pairs are needed to generate a supercurrent, the Josephson effect occupies a central place. Both because of the role played in it by the relative rotational frequency in gauge space ($\dot\phi_{rel}=\dot\phi_1-\dot\phi_2$), and by the associated energy of the photons emitted by the tunneling Cooper pairs
\begin{align}
  \label{abs_eq:5}
  \hbar\omega_J=2\times\hbar\dot\phi_{rel}=2(\mu_1-\mu_2)=2e\times V,
\end{align}
a phenomenon of universal relevance and validity, given the scientific and technical level of understanding one has concerning  how to operate and record photons, within the framework of the most varied systems.
In particular in connection with open shell nuclei, in which case the BCS condensate is made out of few (2-8) Cooper pairs. Within this context we will have in mind the open shell nuclei $^{60,62}$Ni and $^{116,114}$Sn and the reaction (\ref{abs_eq:6abs}). It is of note that in Eq. (\ref{abs_eq:5}), $\mu_i$ is the Fermi energy of superconductors $i=1,2$ building the weak link. 
The alternating  current (ac) tunneling across the Josephson-like junction (App. \ref{AppB}), transiently established in heavy ion collisions ($\tau_{coll}\approx10^{-21}$ s) between two superfluid nuclei at energies below the Coulomb barrier, in particular in the reaction \cite{Montanari:14,Montanari:16}
\begin{align}
  \label{abs_eq:6abs}
  ^{116}\text{Sn}+^{60}\text{Ni}\to^{114}\text{Sn}+^{62}\text{Ni}\quad(Q_{2n}=1.307\text{ MeV}),
\end{align}
at $E_{cm}=154.26$ MeV and $\theta_{cm}=140^\circ$ for the Sn nucleus, proceeds through the tunneling (of $Q$-value $Q_{2n}$) of a single Cooper pair between individual states. Namely the ground states of the interacting nuclei, each of which is a member of a pairing rotational band (see Apps. \ref{rotgauge}, \ref{AppH}). The fingerprint of such single Cooper ac is predicted to be $\gamma$-rays of nuclear-Josephson-like angular frequency $\omega_J^N=\dot\phi_{rel}=2(\mu_1-\mu_2)/\hbar=Q_{2n}/\hbar$. This is consistent with the fact that $V^N=Q_{2n}/2e_{eff}$ is the nuclear analog to the direct current (dc) bias $V$ (Eq. (\ref{abs_eq:5})), $e_{eff}$ being the effective charge of the transferred nucleons in general, and of neutrons in particular (App. \ref{AppE}). In other words, since the center of mass associated with an intrinsic dipole nuclear oscillation must remain at rest, the oscillating neutron Cooper pair being transferred cause protons to oscillate at the same frequency \cite{Potel:21}. Because of their charge, oscillating protons should emit electromagnetic radiation at this frequency \cite{Magierski:21}.
    \section{Superconductivity}
    In a metal, the passage of an electron through a region of the lattice implies a slight increase of negative charge resulting in a slight attraction of ions and eventually a small excess of positive charge which a second electron will feel. Because the masses of the ion and of the electron fulfill  $M_{ion}\gg m_e$, only electrons far apart from each other will interact through the exchange of lattice phonons. In the case of lead, the lowest peak of the phonon spectral function has an energy  $h\nu\approx4$ meV, and thus a period of $\tau=h/(4\text{ meV})=0.2\times10^{-12}$ s. The Fermi velocity in this metal is $\text{v}_F=1.83\times10^8$cm/s. Thus the electrons which exchange the phonons are correlated over a length $\xi=\text{v}_F\times\tau\approx10^4$ \AA. At such distance the (unscreened) Coulomb repulsion between a pair of electrons is small, $\approx1.4$ meV.

    Assuming that the two electrons which exchange a lattice phonon move in time reversal states ($\mathbf k\uparrow,-\mathbf k\downarrow$)  above a quiescent Fermi sea ($\mathbf k$ linear momentum, $\uparrow$ ($\downarrow$) spin up (down)), Cooper showed that (at zero temperature, $T=0$) the system displays a bound state, no matter how weak the interaction is, provided it is attractive, thus implying the normal state to be unstable \cite{Cooper:56}. The condensation of these ($L=0, S=0$) strongly overlapping, very extended weakly bound, Cooper pairs all carrying the same gauge phase, described by Schrieffer's wavefunction \cite{Schrieffer:64} $\ket{\Psi_{BCS}}=\prod_{\nu>0}\left(U'_\nu+e^{-2i\phi}V'_\nu P_\nu^\dagger\right)\ket{0}$ --where $P^\dagger=a^\dagger_\nu a^\dagger_{\tilde\nu}$ creates, acting on $\ket{0}$, pairs of particles moving in time reversal states-- can be shown to display a pairing gap for both breaking a Cooper pair as well as for making  it move independent of the others, a phenomenon known as Off-Diagonal Long-Range Order (ODLRO) \cite{Penrose:51,Penrose:56,Yang:62,Anderson:96}.

    The transition temperature between the regime in which electrons (e.g. in lead) subject to a potential difference move displaying resistance, and that in which they do not is 7.19K (0.62 meV). One  then expects the pairing gap to be of the order of $\Delta\sim T_c$. In fact,  $\Delta\approx 1.4$ meV, Cooper pairs being bound by $\approx2\Delta$. A binding resulting from the exchange of lattice phonons --weakly overcompensating the  weak (screened) Coulomb repulsion-- as testified by the fact that bad conductors make good superconductors.
    \section{Correlation length}
    \subsection{metals}
    Setting a biasing potential $V$ to a junction it  imparts a momentum $\mathbf s$ to the center of mass of the tunneling Cooper pairs. The quasiparticle energies $E_{\mathbf k}$ are shifted by $\approx\frac{\hbar^2}{m}\, \mathbf k\cdot\mathbf s$ (Fig. \ref{fig:A1}, see also App. \ref{AppA}). The intrinsic degrees of freedom of the tunneling Cooper pairs become unfrozen, and the Cooper pairs broken, when 
\begin{align}\label{eq:1x}
eV=2\Delta=\frac{\hbar^2}{m} k\,  s_{c}.      
\end{align}
In other words, when one applies a momentum $s_c\approx\frac{2\Delta}{\hbar \text{v}}\approx1/\xi$, $\xi$ being the correlation length  of the Cooper pairs. Because only electrons within $\sim kT_c$ of the Fermi energy can play a major role in a phenomenon which sets in at $T_c$, one can assume $\xi\approx\hbar \text{v}_F/(2\Delta)$. It is customary to use for this quantity Pippard's definition (\cite{Pippard:53}, see also \cite{Schrieffer:64}), namely
\begin{align}\label{eq:1}
\xi\approx\frac{\hbar \text{v}_F}{\pi\Delta}.
\end{align}
The momentum $s_c\approx1/\xi$ is known as the depairing, critical momentum (Fig. \ref{fig:1}).
Experimental examples of the above parlance are provided by the Josephson ({\bf S-S})  and Giaever ({\bf S-Q}) currents of carriers of charge $q=2e$ and $q=e$ respectively (\textbf{S}$\equiv$ Superconductor, \textbf{Q}$\equiv$ Quasiparticle). Because the probability for a single electron to tunnel across a Josephson junction is very small, let us say $10^{-10}$ (App. \ref{AppG}), Cooper pair tunneling must proceed through successive tunneling (see also App. \ref{AppC}). This is also consistent with the value of the generalized quantality parameter $q_\xi=(\hbar^2/(2m_e\xi^2))\times\frac{1}{2\Delta}\approx10^{-5}$ \cite{Mottelson:02}. A result which implies that the fermion partners of a Cooper pair are  entangled (anchored) to each other, due essentially to the large correlation length of the pair, and, to some extent, also  because of its  binding ($\approx2\Delta$). Said it differently, the fact that $P_2\approx P_1$ is mainly a result of binding by vanishing quantum kinetic energy of confinement. 

\subsection{Superfluid nuclei}
In the nuclear case it is not  possible to establish a supercurrent and thus a depairing bias across a Josephson-like junction. Nonetheless, and in keeping with the fact that  momentum and coordinates are conjugate variables, one can establish a depairing (critical) width for the junction across which the nuclear Cooper pair has to tunnel. This is in keeping with the fact that successive is the main two-nucleon transfer mechanism, and that the de Broglie reduced wavelength $\lambdabar\approx0.06$ fm associated with the reaction (\ref{abs_eq:6abs}) at the bombarding conditions stated, allows for an accurate determination of the distance of closest approach $D_0$. Making use of the experimental data $\sigma_{1n}$ and $\sigma_{2n}$ concerning the reactions $^{116}$Sn+$^{60}$Ni$\to^{115}$Sn+$^{61}$Ni and $^{116}$Sn+$^{60}$Ni$\to^{114}$Sn+$^{62}$Ni (see Figs. \ref{fig:2} and \ref{fig:7p} and Table \ref{tab:1}), one can introduce the empirical definition of the depairing distance as $(D_0)_{c}$ (see App. \ref{AppF}), as the distance of closest approach associated with the lowest bombarding energy at which the condition 
\begin{align}\label{eq:2}
\left(\frac{4}{\pi}\right)^2\times\left.\frac{\sigma_{2n}}{\sigma_{1n}}\right|_{(D_0)_{c}}\gtrsim0.5,
\end{align}
is satisfied. It corresponds to a center of mass energy $E_{cm}=154.26$ MeV and to the value $(D_0)_c=13.49$ fm. The associated relative velocity is 
\begin{align}
  \label{eq:4}
  \text{v}_{cm}\approx26.49\times10^{21}\text{ fm/s},
\end{align}
leading to a collision time
\begin{align}
  \label{eq:5}
  \tau_{coll}=\frac{13.49\text{ fm}}{26.49\times10^{21}(\text{fm/s})}\approx0.51\times10^{-21}\text{ s}.
\end{align}

From the relation $\text{v}_c=\hbar s_c/m$, and making use of $(\text{v}_F/c)\approx0.27$ and of $\Delta\approx1.27$ MeV, one obtains the value of the depairing (critical) velocity, 
\begin{align}
  \label{eq:3}
  \text{v}_c=\frac{1}{m}\frac{\hbar}{\xi}\approx4.44\times 10^{21}\text{ fm/s}.
\end{align}
One can then estimate the associated width of the junction,
\begin{align}
  \label{eq:6}
  w=\tau_{coll}\,\text{v}_c\approx2.26\text{ fm}.
\end{align}
As seen from Fig. \ref{fig:3}, this value  is consistent  with the distance between the surface of the colliding nuclei at the distance of closest approach $(D_0)_c$, 
\begin{align}
  \label{eq:7}
  \nonumber (D_0)_c-&(R(^{60}\text{Ni})+R(^{116}\text{Sn})+a)\\
  &=13.49\text{ fm}-11.25\text{ fm}=2.24\text{ fm},
\end{align}
where $a\approx0.65$ (diffusivity). This consistency points to the physical process taking place, namely the acquisition of the critical depairing velocity of the Cooper pair while passing across the junction during the collision time.

In fact, to state that at the critical distance of closest approach $(D_0)_c$ ($t=t_2$, see Fig. \ref{fig:4}) the time the second partner of the (transferring) Cooper pair has to traverse the (dynamical, transient) barrier of width (\ref{eq:6}) once the first one has tunneled is  that provided by the collision time (\ref{eq:5}), implies that: \textbf{1)} determining the correlation length in terms of the critical current (momentum $1/\xi$) applied to the cm of the Cooper pair and resulting in the critical violation of time-reversal invariance  $(\mathbf k+1/(2\xi)\,\mathbf {\hat  s},-\mathbf k+1/(2\xi)\,\mathbf {\hat s})$ or, \textbf{2)} in terms of the critical barrier width, again leading to the critical violation of time reversal invariance   $(\mathbf r_1+(\xi/2)\,\mathbf {\hat R}_{cm},-\mathbf r_2+(\xi/2)\,\mathbf {\hat R}_{cm})$, are equivalent. Said it differently, the consistency check provided by (\ref{eq:6}) and (\ref{eq:7}), implies that the single Cooper pair nuclear Josephson-like alternating ``current'' associated with $d\sigma_{2n}/d\Omega|_{\theta_{cm}=140^\circ}=2.58$ mb/sr (Fig. \ref{fig:7p}) or, better, with the corresponding $T$-matrix, displays a value of the cm momentum  of the order of the critical one.

In keeping with the consistency check discussed above --and making use of the fact that momentum and coordinate are conjugate variables-- allows one to expect that Eq. (\ref{eq:7}) is of general validity to provide an empirical expression for the coherence length,  
\begin{align}
  \label{eq:111}
  \frac{(D_0)_c-(R_A+R_a+a)}{\tau_{coll}}=\frac{\hbar}{m\,\xi},
\end{align}
that can be compared to the value of $\xi$  defined in Eq. (\ref{eq:1}) and leading to $\xi\approx13.5$ fm, making use of the values of $\text{v}_F/c$ and $\Delta$ introduced in connection with Eq. (\ref{eq:3}).

Considering the individual nuclei acting as targets (dynamical situations corresponding to a snapshot at times $t=t_1$ and $t=t_3$, Fig. \ref{fig:4}) one can, in principle, only study the Cooper pair under the influence of a strong ``external'' field. Namely the mean field\footnote{$\Delta=G\,\alpha_0$ where   $\alpha_0=\braket{BCS|\sum_{\nu>0}a^\dagger_{\nu}a^\dagger_{\tilde \nu}|BCS}=\sum_{\nu>0}U_\nu V_\nu$ is the number of Cooper pairs; $\Delta=\Delta'e^{-2i\phi}$, $\alpha_0=\alpha_0'e^{-2i\phi}$; see Sect. \ref{Sect10}.} of the nucleus to which it is bound. For example through a high energy $(p,pn)$ reaction \cite{Kubota:20}. It is only at time $t=t_2$ that one has a field (probe) which itself changes particle number in two, namely the tunneling field,
\begin{align}
  \label{eq:8}
  \nonumber F(\mathbf r_1,\mathbf r_2,\mathbf r_{Aa})&\sim\sum_{\nu>0}U'^{A}_\nu V'^{A}_{\nu}\left.\sum_{\nu'>0}U'^a_{\nu'}V'^a_{\nu'}/(E_f+E_F)\right|_{\epsilon_F}\\
  &\sim\frac{\alpha_0'^A\alpha_0'^a}{\alpha_0'^A+\alpha_0'^a}\sim\frac{\Delta'_A\Delta'_a}{\Delta'_A+\Delta_a'},
\end{align}
and thus sensitive to the presence of ODLRO. Not only in terms of general statistical mechanics correlation arguments  with no reference to structure (and reaction) details, but making such ``details'' (basic observables) fully explicit, allowing to foresee, predict, the outcome of concrete experiments. In particular the emission, with determined absolute cross sections, of $\gamma$-rays of (Josephson-like nuclear)  angular frequency $\omega^N_J=Q_{2n}/\hbar$ (App. \ref{AppJx}).
\subsubsection{Consequences of $D_0>(D_0)_c$. Obliterating the coherent state.}\label{Se4.1}
 Critical conditions across a Josephson link --whether in condensed matter or in the heavy ion reaction (\ref{abs_eq:6abs})--   obliterate the entanglement between the fermion partners of the tunneling nuclear Cooper pair between the coherent states involved in the process.  Namely, the superconductors lying at the left (L) and at the right (R) sides of the junction and described by the $\ket{BCS(\text{L})}$ and $\ket{BCS(\text{R})}$ coherent states in condensed matter, and by the coherent states  $\ket{BCS}_{^A\text{X}}$ ($^A$X(gs) $\equiv\,^{116}$Sn,$^{114}$Sn,$^{60}$Ni and $^{62}$Ni)   in connection with the nuclear heavy ion reaction under study. In both cases, one can write the coherent state as
\begin{align}
  \label{eq:9}
  \nonumber  \ket{BCS}&=\prod_{\nu>0}\left(U_\nu+V_\nu P^\dagger_\nu\right)\ket{0}\\
  \nonumber &=\left(\prod_{\nu>0} U_\nu\right)\sum_{n=0,1,2,\dots}\frac{1}{n!}\left(\sum_{\nu>0}c_\nu P^\dagger_\nu\right)^n\\
 &=\left(\prod_{\nu>0}U_\nu\right)\exp\left(\sum_{\nu>0}c_\nu P^\dagger_\nu\right)\ket{0},
\end{align}
 where $c_\nu=V_\nu/U_\nu$.

However, there is an essential difference between the case of two weakly coupled superconductors made out of a number of Cooper pairs of the order of Avogadro's number, and two weakly coupled superfluid nuclei made out of few (3-8) Cooper pairs. In the first case, the S-S$\to$ S-Q phase transition  is sharp, being rather gradual in the second one, strongly blurred by  the important role played by pairing vibrations. Within this context, we refer to studies carried out on pairing fluctuations in rapidly rotating superfluid nuclei   (see for example \cite{Shimizu:89} and refs. therein), in which case rotation plays the role the magnetic field does in superconductors. It is found that the dynamic pairing gap due to pairing vibrations,  amounts to about 40\%-50\% of the BCS ground state pairing gap, keeping essentially unchanged this value even at rotational frequencies a factor of 2 larger than the critical rotational frequency ($\Delta_{BCS}(\omega_c)=0$, equivalent to the critical magnetic field in type I superconductors).

While we are not attempting at translating these results in terms of critical values of the distance of closest approach, one can posit that it is not unlikely that (dynamical) pairing coherence is operative also at the lowest bombarding energies ($E_{cm}<154.26$ MeV), as testified by the fact that the inclusive, two-nucleon transfer absolute differential cross sections are well reproduced in terms of the (gs)$\to$(gs) process (Fig. \ref{fig:7p}). Naturally, with considerably smaller probabilities, as evidenced by the steep decrease displayed by $\sigma_{2n}/\sigma_{1n}$ as a function of $D_0$ passed $(D_0)_c$ (Fig. \ref{fig:2}).

\section{Off-Diagonal Long-Range Order}
In a superconductor there is no electrical resistance because all Cooper pairs are collectively in the same state, the coherent $\ket{BCS}$ state (\ref{eq:9}). Once one  gives a momentum to the center of mass of them and starts a supercurrent, to get one pair or one electron partner away from what all others are doing is very hard (ODLRO). This is at marked variance with ordinary current at room temperature and single electron carriers in which collisions with the impurities,  ions or phonons, can knock one electron or the other out of the regular flow of electrons deteriorating the common momentum.

If one turns a magnetic field on a superconductor\footnote{More correctly, type I superconductors.} there would be a transient electric field which would generate a current  opposing the flux (Lenz law; see also \cite{Meissner:33}). Because the supercurrent once started will keep on going forever (persistent current), the magnetic field is permanently excluded, and this is also what happens starting with a piece of metal at room temperature with a magnetic field going through it, and cooling the system below the critical temperature.
Perfect conductivity and perfect diamagnetism are the hallmark of superconductivity and, in hindsight, a consequence of ODLRO.

The state proposed by Schrieffer (BCS) for a superconductor carrying a supercurrent at $T=0$ is
\begin{align}
  \label{special_eq:15}
  \ket{BCS}=\prod_k\left(U'_k+e^{-2i\phi} V'_k a^\dagger_{\mathbf k+\mathbf s,\uparrow}a^\dagger_{-\mathbf k+\mathbf s,\downarrow}\right)\ket{0}.
\end{align}
In this state all pairs have the same center of mass momentum $2\hbar\mathbf s$. In keeping with the fact that $k\approx k_F$, the pair momentum $2\hbar s$ associated with the finite supercurrent is very small\footnote{In the case of Pb, v$_F=1.83\times10^{8}$ cm/s and v$_F/c\approx0.6\times10^{-2}$ ($c\approx3\times10^{18}$\AA/s). Thus $\xi=\hbar\text{v}_F/\pi\Delta\approx0.25\times10^4$\AA   ($\hbar c\approx 2\times 10^3$ eV\AA) and $1/\xi\approx4\times10^{-4}$\AA$^{-1}$, a quantity which is very small as compared to $k_F=m_e\text{v}_F/\hbar\approx1.58$\AA$^{-1}$ ($m_ec^2\approx0.5$ MeV). In keeping with the fact that $1/\xi$ is the critical momentum leading to depairing and thus breaking of the Cooper pair, one can posit that $2\hbar s\ll k_F$.}. For zero supercurrent $(\mathbf s=0)$ (\ref{special_eq:15}) is the BCS ground state. The parameters $U'_k$ and $V'_k$ are real, being referred to the intrinsic, body-fixed frame (Fig. \ref{abs_figI}; see also Apps. \ref{AppH} and \ref{rotgauge}).

There is a remarkable resemblance between the BCS occupation factors $V^2_k$ as a function of the electron energy and that of the normal metal at $T\gtrsim T_c$ \cite{Tinkham:96,Waldram:96}. Resemblance which is only apparent, because the parts of the wavefunction describing the filled and empty states in BCS have a definite phase coherence while excitations in a Fermi distribution  have no phase coherence.

The BCS  phase factor is arbitrary, but the same for all $k$-states. These phase relations imply the existence of a significant spatial correlation between pairs of electrons, their condensation leading to the superconducting state. In this sense, (Cooper) pairs of electrons described by the many-body wavefunction (\ref{special_eq:15}) behave like (quasi) bosons. There is however an important difference. For real bosons, like e.g. $^4$He, they preexist before undergoing Bose-Einstein condensation into a single momentum state. On the other hand, while one can posit that Cooper pairs are also present in metals at room temperature, they break as soon as they are formed, this being so for temperatures $T>T_c$. In BCS theory, binding and condensation into a single (CM) momentum state, the two sides of ODLRO --namely a pairing gap for single pair dissociation as well as translation-- are not independent events although arguably, neither strictly simultaneous either. Due to the fact that pairs overlap strongly in space, their binding energy becomes cooperative, and the energy gain associated with a new pair joining the condensate depends on how many pairs are already bound.

\subsection{Broken symmetry and macroscopic order}
The connection between violation of particle number and long-range order in the BCS condensate, is associated with the formation of Cooper pairs below the critical temperature $T_c$. This order may be expressed by introducing a correlation function (App. \ref{AppJ}), which involves two Cooper pairs (CP) --assumed for simplicity to be local-- at positions $\mathbf r$ and $\mathbf r'$ respectively, that is,
\begin{align}
  \label{special_eq:16}
  C^{CP,CP}(\mathbf r -\mathbf r')\equiv\braket{\Psi|\hat\psi^\dagger_\uparrow(\mathbf r)\hat\psi^\dagger_\downarrow(\mathbf r)\hat\psi_\downarrow(\mathbf r')\hat\psi_\uparrow(\mathbf r')|\Psi},
\end{align}
where $\ket{\Psi}$ is assumed to be the $\ket{BCS}$ state and
\begin{align}
  \label{special_eq:17}
  \hat\psi_\sigma^\dagger(\mathbf r)=\sum_{\mathbf k} a^\dagger_{\mathbf k\sigma}e^{-i\mathbf k\cdot\mathbf r},
\end{align}
$a^\dagger_{\mathbf {k}\sigma}$ being the creation operator for a particle with momentum $\hbar\mathbf k$ and spin $\sigma$.

In the superconducting state the correlation function (\ref{special_eq:16}) approaches a finite value even in the limit of very large separation of the pairs\footnote{The implicit assumption being made that the condensate extends that far.},
\begin{align}
  \label{special_eq:18}
  \lim_{|\mathbf r -\mathbf r'|\to\infty}C^{CP,CP}(\mathbf r -\mathbf r')\neq0
\end{align}
  expresses the fact that the off-diagonal elements of the two-particle density matrix (\ref{special_eq:16}) (see also (\ref{special_eq:19})) in the position-space representation (i.e. those for $r\neq r'$) show long-range order. It is a consequence that in the pair correlated state the pair non-conserving operator $a^\dagger_{\mathbf k\uparrow}a^\dagger_{-\mathbf k\downarrow}$ has a finite expectation value in the state $\ket{\Psi}=\ket{BCS}$,
\begin{align}
  \label{special_eq:24}
  \braket{\Psi|a^\dagger_{\mathbf k\uparrow}a^\dagger_{-\mathbf k\downarrow}|\Psi}=U_k V_k,
\end{align}
referred to as the pair amplitude \cite{Vollhardt:90}.
\section{Critical Josephson current ($P_2\approx P_1$)}\label{S4}
In order to describe the tunneling of Cooper pairs through a junction (barrier) between two weakly coupled superconductors, one can make use of the Cohen-Falicov-Phillips tunneling Hamiltonian \cite{Cohen:62,Anderson:64b},
\begin{align}
  \label{eq:10}
  H=H_1+H_2+\sum_{kq}T_{kq}\left(a^\dagger_{k\uparrow}a_{q\uparrow}+a^\dagger_{-q\downarrow}a_{-k\downarrow}\right)+hc,
\end{align}
where $H_1$ and $H_2$ are the Hamiltonians which describe the superconductors at each side of the junction, while $T_{kq}$ is the (exponentially) small tunneling matrix element from state $k$ on one side to state $q$ on the other side of the barrier. To second order perturbation theory, the effect of this term can be written as
\begin{align}
  \label{eq:11}
  \Delta E^{(2)}=-2\sum_{kq}|T_{kq}|^2\frac{\left|V_kU_q+V_qU_k\right|^2}{E_k+E_q},
\end{align}
where the inverse quasiparticle transformation relations have been used. Working out the numerator of (\ref{eq:11}) with the help of the BCS relations $2U_\nu V_\nu=\Delta_\nu/E_\nu$ and $|U_\nu|^2-|V_\nu|^2=\epsilon_\nu/E_\nu$ (where $\nu$ stands for either $k$ or $q$) and $(\Delta_k=\Delta_1e^{2i\phi_1},\quad \Delta_q=\Delta_2e^{2i\phi_2})$, together with the replacement of sums by the integrals
\begin{align}
  \label{eq:12}
  \sum_k\to N_1\int d\epsilon_1,\quad \sum_q\to N_2\int d\epsilon_2,
\end{align}
one obtains
\begin{align}
  \label{eq:13}
  \Delta E^{(2)}=-2\pi N_1 N_2\langle|T_{kq}|^2\rangle\cos2\delta\phi\times\left(\pi\frac{\Delta_1\Delta_2}{\Delta_1+\Delta_2}\right).
\end{align}
The quantities $N_1,N_2$ are the density of levels at the Fermi energy for one spin orientation, while the phase difference
\begin{align}
  \label{eq:14}
  \delta\phi=(\phi_1-\phi_2)-\frac{e}{\hbar}\int_1^2\mathbf A\cdot d\pmb \ell,
\end{align}
was obtained replacing $(\phi_1-\phi_2)$ by a gauge invariant phase difference.

The (Cooper pair) supercurrent per unit area can be calculated as\footnote{It is of notice that $\langle|T|^2\rangle=|T^2|_{\text{average}}$, where the average is over the states $k,q$ lying in different sides of the barrier. Furthermore it is assumed that this quantity is an energy squared (eV)$^2$ per unit area $(S)$ ($[\langle|T|^2\rangle]$=(eV$^2$)/S). The units of the vector potential can be obtained from the relation $\pmb\nabla\times\mathbf A=\mathbf B$, $[\mathbf B]=T=\frac{\text{Vs}}{\text{m}^2}$ (V, s, m stand for Volt, second, and meter). Thus $[A]=$(Vs)/m, and $\left[\frac{2e}{\hbar}\int_1^2\mathbf A\cdot d\pmb\ell\right]=\frac{e}{e\text{Vs}}\times\frac{\text{Vs}}{\text{m}}m=1$, in keeping with the fact that $\int_1^2 d\pmb\ell$ is the width of the barrier ($[\ell]=$m). For simplicity, it is assumed that $\mathbf A$ is perpendicular to the barrier.}
\begin{align}
  \label{eq:15}
  \nonumber J=&\frac{1}{w}\frac{\delta\Delta E^{(2)}}{\delta\mathbf A}=-\frac{2\pi}{w}N_1N_2\langle|T|^2\rangle\left(-w\times\frac{2e}{\hbar}\right)\sin2\delta\phi\\
  \nonumber &\times\left(\pi\frac{\Delta_1\Delta_2}{\Delta_1+\Delta_2}\right)\\
  &=\left(\frac{\pi}{e}\frac{\Delta_1\Delta_2}{\Delta_1+\Delta_2}\right)\times\left(\frac{4\pi e^2}{\hbar}N_1 N_2\langle|T|^2\rangle\right)\sin2\delta\phi,
\end{align}
$w$ being the width of the barrier. Assuming, for simplicity, that the two superconductors are equal one can write (see Fig. \ref{fig:6})
\begin{align}
  \label{eq:16}
  J=J_c\sin2\delta\phi,
\end{align}
where
\begin{align}
  \label{eq:17}
  J_c=\left(\frac{\pi}{4}V_{eq}\right)\frac{1}{R_b}=\frac{\pi}{4}\frac{V_{eq}}{R_b}=\frac{\pi}{4}I_N.
\end{align}
The quantity
\begin{align}
  \label{eq:18}
  V_{eq}=\frac{2\Delta}{e},
\end{align}
is the critical dc bias to break the transferred Cooper pair, allowing for $S$-$Q$ tunneling of carriers $q=e$ as in the case of the normal current $(I_N)$ across the barrier of ($N-N$) resistance
\begin{align}
  \label{eq:19}
  R_b=\left(\frac{4\pi e^2}{\hbar}N^2\langle|T|^2\rangle\right)^{-1},
\end{align}
$N$, $S$ and $Q$ standing for Normal, Superfluid and Quasiparticles (Fig. \ref{fig:1}). The units of $V_{eq}$ and $R_b$ are\footnote{$[V_{eq}]=\frac{e\text{V}}{e}=$V (volt); $[R_b]=\frac{\hbar}{4\pi e^2}\frac{1}{N^2\langle|T|^2\rangle}=\frac{e\text{Vs}}{e^2}\frac{e\text{V}^2}{\frac{(e\text{V})^2}{\text{S}}}=\frac{\text{V}}{\left(\frac{e}{\text{s}}\right)}\times \text{s}=\frac{\text{V}}{\text{A}}\times \text{s}=\Omega S$ (Ohm$\times$m$^2$), where V: Volt, A: Ampere, s: second, m: meter.} V (volts) and $\Omega\times S$ (Ohm $\times$ unit surface) respectively.

From the relation (\ref{eq:17}) 
\begin{align}
  \label{eq:20}
  I_c=\frac{\pi}{4}I_N.
\end{align}

\section{Special effects in superconductivity}
Although the tunneling of single electrons from a superconductor through a layer of an insulator into another superconductor known as a weak link has a small probability, as small as $P_1\approx 10^{-10}$ (\cite{Pippard:12}, see also {App. \ref{AppG}), supercurrents of carriers of charge $q=2e$ and mass $2m_e$ (Cooper pairs) were predicted (\cite{Josephson:62}, see also \cite{Anderson:64b}) and observed \cite{Anderson:63} to flow through the junction with probability $P_2\approx P_1$.

In the calculation of the Cooper pair tunneling through a weak link, also known as a Josephson junction, one has to add the phased amplitudes before one takes the modulus squared,
\begin{align}
  \label{special_eq:1}
  P_2=\frac{\left|U'_\nu\sqrt{P_1}+e^{-2i\phi}V'_\nu\sqrt{P_1}\right|^2}{\sqrt{2}}\approx\frac{P_1}{2}\left(1+2U'_\nu V'_\nu \cos 2\phi\right)\approx P_1.
\end{align}
It is like interference in optics, with phase-coherent mixing. We note that in deriving (\ref{special_eq:1}) it is assumed $U'_\nu V'_\nu\approx\frac{1}{2}$ average value of the BCS coherence factor, and $\cos 2\phi\approx1$, in keeping with the fact that this is the value which minimizes the tunneling energy (\ref{eq:13}).

Among the new effects in superconducting tunneling predicted in \cite{Josephson:62}, they include: a) \textit{dc Josephson effect}, a direct supercurrent (of Cooper pairs) flows across the junction in absence of any electric or magnetic field;
b) \textit{ac Josephson effect}, a direct current voltage applied across the junction does not determine the intensity of the supercurrent (Ohm's law) circulating through it, but the frequency of an alternating current $(\nu_J=2e\times V/h)$ and the associated THz frequency photons emitted by it. Effect which has been used in precision determination of the value of $h/e$ (see for example \cite{Rogalla:12}).

Let $\Psi_i=\sqrt{n'_i}\,e^{i\phi_i}$ be the probability amplitude of electron pair on one ($i=1$) and the other ($i=2$) side of the junction, $|\Psi_i|^2=n'_i$ being the pair (abnormal) density $n'_i=(\alpha_0')_i/\mathcal V$, $\mathcal V$ being an appropriate volume element. The time-dependent Schr\"odinger equation applied to the two amplitudes gives
\begin{align}
  \label{special_eq:2}
  i\hbar\frac{\partial \Psi_1}{\partial t}=\hbar T\Psi_2;\quad i\hbar\frac{\partial \Psi_2}{\partial t}=\hbar T\Psi_1,
\end{align}
where $\hbar T$ is the transfer interaction across the insulator, $T$ having the dimensions of a rate (frequency). One can then write,
\begin{align}
  \label{special_eq:3}
   \frac{\partial \Psi_1}{\partial t}&=\frac{1}{2\sqrt{n'_1}}e^{i\phi_1}\frac{\partial n'_1}{\partial t}+i\Psi_1\frac{\partial \phi_1}{\partial t}=-iT\Psi_2,\\
  \frac{\partial \Psi_2}{\partial t}&=\frac{1}{2\sqrt{n'_2}}e^{i\phi_2}\frac{\partial n'_2}{\partial t}+i\Psi_2\frac{\partial \phi_2}{\partial t}=-iT\Psi_1.
\label{special_eq:4}                                      
\end{align}
Multiplying (\ref{special_eq:3}) by $\sqrt{n'_1}e^{-i\phi_1}$ and (\ref{special_eq:4}) by $\sqrt{n'_2}e^{-i\phi_2}$ and equating real and imaginary parts assuming the two superconductors to be equal ($n'_2=n'_1=n'$), one obtains
\begin{align}
  \label{special_eq:5}
  \frac{\partial n'_1}{\partial t}=2Tn'\sin\phi_{rel} \;\text{ (\textbf{a})};\quad \frac{\partial n'_2}{\partial t}=-2Tn'\sin\phi_{rel} \;\text{ (\textbf{b})},
\end{align}
and
\begin{align}
  \label{special_eq:5x}
  \frac{\partial \phi_{rel}}{\partial t}=0,
\end{align}
where
\begin{align}
  \label{special_eq:6}
  \phi_{rel}=\phi_2-\phi_1.
\end{align}
The current flowing 1$\to$2 is naturally equal to minus that flowing 2$\to$1, and can be written as
\begin{align}
  \label{special_eq:7}
  J=J_c\sin2\phi_{rel}.
\end{align}
With no applied bias a direct current flows across the link with a value between $J_c$ and $-J_c$ defined by the relative gauge phase. In other words, the phase difference $\phi_{rel}$ (Eq. (\ref{special_eq:6})) appears as the local velocity potential of a collective flow superimposed on the Cooper pair intrinsic motion, carrying a momentum current $\mathbf i_p\sim n'\pmb\nabla\phi_{rel}$ \cite{Bohr:88}.

Because the junction is an insulator one can apply a dc bias $V$ across it. An electron pair experiences a potential energy difference $-2eV$ in passing through, equivalent to saying that a pair on one side is at potential energy $-eV$ while one pair at the other side is at $eV$. Parallel to (\ref{special_eq:3}) one can write,
\begin{align}
  \label{special_eq:8}
 \frac{\partial \Psi_1}{\partial t}=\frac{1}{2\sqrt{n'_1}}e^{i\phi_1}\frac{\partial n'_1}{\partial t}+i\Psi_1\frac{\partial \phi_1}{\partial t}=-iT\Psi_2+i\frac{eV}{\hbar}\Psi_1  
\end{align}
and similarly for $\partial \Psi_2/\partial t$ in connection with (\ref{special_eq:4}). Multiplying the relations by $\sqrt{n'_1}e^{-i\phi_1}(\sqrt{n'_2}e^{-i\phi_2})$ and again assuming $n'_1=n'_2=n'$, one finds that the real part leads to (\ref{special_eq:5}) (\textbf{a}) as in the case without bias $V$, the imaginary part leading to
\begin{align}
  \label{special_eq:9}
\frac{\partial\phi_1}{\partial t}=-T\cos\phi_{rel}+\frac{eV}{\hbar},  
\end{align}
while $\partial \Psi_2/\partial t$ gives,
\begin{align}
  \label{special_eq:10}
\frac{\partial\phi_2}{\partial t}=-T\cos\phi_{rel}-\frac{eV}{\hbar}.
\end{align}
Thus
\begin{align}
  \label{special_eq:11}
  \frac{\partial\phi_{rel}}{\partial t}=-\frac{eV}{\hbar}.
\end{align}
Integrating this relation on obtains
\begin{align}
  \label{special_eq:12}
  \phi_{rel}(t)=\phi_{rel}(0)-\frac{eV}{\hbar}t,
\end{align}
the superconducting current is then
\begin{align}
  \label{special_eq:13}
  J=J_c\sin\left(2\phi_{rel}(0)-\left(\frac{2eV}{\hbar}t\right)\right)
\end{align}
and oscillates with angular frequency (App. \ref{AppJx})
\begin{align}
  \label{special_eq:14}
  \omega_J=\frac{2eV}{\hbar}.
\end{align}
This is the ac Josephson effect.

\section{Dipole moment and correlation length: emergent properties of ac tunneling}\label{S7}
Within the framework of the Josephson effect, two emergent properties occupy a special place (Fig. \ref{abs_figA}): a) the number $\mathcal N$ of photons of frequency $\omega_J$ emitted per cycle (ac); b) the critical momentum and thus the associated correlation length related to the critical Josephson current $I_c=\frac{\pi}{4}I_N$ (dc, ac).

In order to translate into the nuclear language of the successive transfer process between the superfluid nuclei Sn and Ni of a Cooper pair, one needs to insert into the corresponding $T$ matrix   the radial part of the dipole operator
\begin{align}
  \label{eq:107}
  \mathbf D= e^{(1)}_{eff}\mathbf r_{O1}+e^{(2)}_{eff}\mathbf r_{O2}'.
\end{align}
Because $e^{(1)}_{eff}=e^{(2)}_{eff}=e_{eff}$, and making use of
\begin{align}
  \label{eq:108}
  \mathbf r_{2n}=\mathbf r_{O1}+\mathbf r_{O2}'\quad\text{and}\quad \bar{\mathbf r}_{2n}=\left(\mathbf r_{O1}+\mathbf r_{O2}'\right)/2,
\end{align}
one can write
\begin{align}
  \label{eq:59}
  \mathbf D=2e_{eff}\times \bar{\mathbf r}_{2n},
\end{align}
with spherical components
\begin{align}
  \label{eq:60}
 \mathbf D_{m_\gamma}=e_{eff}\sqrt{\frac{4\pi}{3}}\left(r_{O1}Y^1_{m\gamma}(\hat r_{O1})+r'_{O2}Y^1_{m\gamma}(\hat r'_{O2})\right), 
\end{align}
associated with the transferred Cooper pair, that is,
$\mathbf r_{O1}+\mathbf r_{O2}'$ ($\equiv \mathbf{\hat O}_1$) and $\mathbf r_{2b}'-\mathbf r_{1b}'$ ($\equiv\mathbf {\hat O}_2$) (see inset Fig. \ref{abs_figA}).  One then finds a convergent  average value $\braket{\hat O_i^2}^{1/2}=\left(\left|T(\mathbf k_i, \mathbf k_f,\hat O_i^2)\right|/\left|T(\mathbf k_i, \mathbf k_f)\right|\right)^{1/2}$,  ($\hat O_i^2\equiv \mathbf{\hat O}_i\cdot \mathbf{\hat O}_i$) as a function of the number of partial waves. Consequently, the first quantity provides the radial component of the dipole moment of the transferred nucleons (see also Sect. \ref{AppI2}), the second one the corresponding correlation length, that is, the mean square radius of the transferred Cooper pair.  

\section{Possible extension of the Josephson effect to the single nuclear Cooper pair tunneling}
The ($\gamma$-dipole) Cooper pair transfer amplitude in a heavy ion collision between two superfluid nuclei is dominated by the $T_{m_\gamma}$-matrix describing successive tunneling. The ``reduced'' amplitude associated with the $T$-matrix (labeled $T_{m_\gamma}$), that is the formfactor connecting the relative motion of the heavy ion reaction in the entrance ($d r'_{Cc}\,\chi_i(\mathbf r_{aA},\mathbf k_i)$) with that in the exit ($d r_{Cc}\,\chi_f(\mathbf r_{Bb},\mathbf k_f)$) channel, divided by $\hbar$, can be written as
\begin{align}
  \label{eq:26}
  J_c^N=\frac{\pi}{4}V_{eq}^N\times \frac{1}{R_b^N},
\end{align}
where
\begin{align}
  \label{eq:27}
  V_{eq}^N=\frac{2\Delta_N}{e_{eff}},\quad \Delta_N=\left( \text{v}_p^{bare}+\text{v}_p^{ind}\right)\alpha_0'
\end{align}
and
\begin{align}
  \label{eq:28}
\nonumber   \frac{1}{R_b}&=\frac{4\pi e^2}{\hbar}\frac{1}{\tilde {\text{v}}}
  \int\left[\phi_{j_f}\phi_{j_f}\right]^{0*}_0U^{(b)}(r_{b1})\left[\phi_{j_f}\phi_{j_i}\right]^{K}_M\\
  \nonumber &\times \mathbf D_{m_\gamma}\,d\mathbf r_{b_1}\,d\mathbf r_{A_2}\int G(\mathbf r_{Cc},\mathbf r'_{Cc})\left[\phi_{j_f}\phi_{j_i}\right]^{K*}_M \\
 & \times U(r'_{b_2})\left[\phi_{j_i}\phi_{j_i}\right]^{0}_0\,d\mathbf r'_{b_1}\,d\mathbf r'_{A_2},
 \end{align}
where $\mathbf d^1_{m_\gamma}$ is the dipole moment associated with the transferred Cooper pair (\ref{eq:60})), while 
\begin{align}
  \label{eq:29}
  \tilde {\text{v}}=2\pi^2\left( \text{v}_p^{bare}+\text{v}_p^{ind}\right),
\end{align}
is proportional to the renormalized pairing strength entering the expression of the pairing gap (\ref{eq:27}).

The induced pairing interaction $\text{v}_p^{ind}$ arises from the exchange of collective surface vibrations between pairs of nucleons moving in time reversal states around the Fermi energy. Its contribution to the pairing gap is similar to that of the bare pairing interaction (see  \cite{Barranco:99,Terasaki:02b,Saperstein:12,Avdenkov:12,Lombardo:12} and references therein).

It is of notice that the nuclear quantities $J_1^N$, $V_{eq}^N$ ($\Delta_N$) and $R_b^N$ have the same dimensions as their condensed matter counterparts. This is because we have introduced in the $T$-matrix, the coupling to the electromagnetic field, in the dipole approximation, associated with the transfer of neutrons carrying an effective charge.  It is of notice that this (dipole) coupling to the electromagnetic field seems  natural  when dealing with electrons, while tunneling of neutrons (but also, within the present context, of protons), is not experienced as the motion of (effectively) charged particles. That is a current of a single nucleon Cooper pair.

Within this context, the standard transfer expressions should, in principle, contain the dipole term $\mathbf d^1_{m_\gamma}$  appearing in\footnote{This is in keeping with the fact that alternating currents are associated with $\gamma$ radiation of electromagnetic (dipole) waves.} Eq. (\ref{eq:28}), although in most situations the $\gamma$-emission strength function will have little effect on the pair transfer cross sections. Likely implicitly considered in the optical potentials (imaginary part used in the calculation of the distorted waves).

In keeping with the selfconsistency relation
\begin{align}
  \label{eq:30}
  w=(D_0)_c-(R_B+R_b+a)=\tau_{coll}\text{v}_c,
\end{align}
a quantity equal to $w=2.44$ fm ($w$; ``nuclear junction width'') for the reaction $^{116}$Sn+$^{60}$Ni$\to^{114}$Sn+$^{62}$Ni at $E_{cm}=154.26$ MeV (see Fig. \ref{fig:3}) and $(D_0)_c=13.49$ fm ($r_{Cc}, r'_{Cc}$ fixed, see Eq. (\ref{eq:28}) and Fig. \ref{notes_fig1}), one can view the associated transient Josephson-like junction as a quasi static (adiabatic) situation, in any case concerning the single Cooper pair current discussed above. The similitudes  with the condensed matter situation extending also to the dielectric material of which the junction (barrier) is made of, namely the vacuum.


\section{Structure component of the Josephson current}\label{Sect10}
The  expectation values\footnote{The creation operators used in the calculation of the $\ket{BCS}$ average values are defined as $\Psi^\dagger(\mathbf r)=\sum_\nu\varphi_\nu(\mathbf r) a^\dagger_\nu$ and a similar expression associated with the time reversal one.}  $_1\braket{BCS|\Psi^\dagger(\mathbf r_1)\Psi^\dagger(\mathbf r_2)|BCS}_a=\left(\alpha_0'(\mathbf r_1,\mathbf r_2)\right)_1$ and $_2\braket{BCS|\Psi^\dagger(\mathbf r_1)\Psi^\dagger(\mathbf r_2)|BCS}_2=\left(\alpha_0'(\mathbf r_1,\mathbf r_2)\right)_2$ (Fig. \ref{fig:7} (C)), are related with the tunneling interaction energy (see Eq. (\ref{eq:13})) through the inverse of the summed radius of curvature, namely,
\begin{align}
  \label{eq:32}
  \left(\frac{1}{(\alpha_0')_1}+\frac{1}{(\alpha_0')_2}\right)^{-1}=\left(\frac{(\alpha_0')_1(\alpha_0')_2}{(\alpha_0')_1+(\alpha_0')_2}\right),
\end{align}
or, better, through the reduced gap
\begin{align}
  \label{eq:33}
  \Delta E^{(2)}\sim \text{v}_p\frac{(\alpha_0')_1(\alpha_0')_2}{(\alpha_0')_1+(\alpha_0')_2}\sim\frac{\Delta_1\Delta_2}{\Delta_1+\Delta_2}.
\end{align}
Here, the pairing strength $\text{v}_p$ is a function of the coordinates which can, in principle be zero inside the junction (Eq. (\ref{eq:27}) and (\ref{eq:29} in the nuclear case). A possibility which does not jeopardize Cooper pair tunneling, as seen from Fig. \ref{fig:7} (D) associated with a reaction between superfluid nuclei in a heavy ion reaction at the critical  distance of closest approach $(D_0)_c$. Within this context see also App. \ref{AppD}.

\section{Dependence and independence on time of the Josephson effects}
A persistent supercurrent that flows in a thick ring of radius $R$, made out of a metallic superconductor maintains a flux through the ring of some integral number of fluxoids ($\Phi_0=2\pi\hbar/(2e)\approx2.07\times10^{-15}$ T$\times$m$^2$, where T stands for Tesla and m for meters). A fluxoid cannot leak out of the ring and hereby reduce the persistent current, unless  a minimum volume of the superconducting ring ($R\xi^2$) is, due to a thermal fluctuation, momentarily in the normal state. Estimates for this to happen leads to times longer than the age of the universe.

Let us consider two equal superconductors at $T<T_c$ separated by a thick barrier --through which no electrons can tunnel-- made out of two oxide layers, one of which has thickness of 1--3 nm, the whole system being made into a single connected circuit (conductor joining the ends of the superconductors). Removing from the link the thick oxide layer maintaining at the same time contact with the thin one in such a way that one is now in presence of a weak link, fluctuations --thermal or quantal-- would allow $\sin2\phi_{rel}$ to be different from zero, and thus the start of a direct Cooper pair supercurrent across the link between the two superconductors which have the same Fermi energy.  Such a current, once started, is expected to persist indefinitely. Thus the direct current \textit{(dc) Josephson effect}.

On the other hand, if the junction is  biased by a dc potential $V$, Cooper pairs will tunnel between two superconductors with different values of the Fermi energy. A system  equivalent to a weakly coupled, two rotor   system in gauge space ($\Psi_1=\sqrt{n'_1}e^{-i\phi_1}$, $\Psi_2=\sqrt{n'_2}e^{-i\phi_2}$). Thus, the alternating current \textit{(ac) Josephson effect}, of angular frequency $\omega_J=2\times(\dot\phi_1-\dot\phi_2)=2\frac{\mu_1-\mu_2}{\hbar}=\frac{2eV}{\hbar}$.


\section{Distorted wave diffraction of successive transfer formfactors: nuclear Cooper pair observable}
At the basis of quantum mechanics one finds the aim to get rid of quantities which could not be observed, and build upon concepts referring to entities which, at least in principle, are measurable\footnote{\cite{Heisenberg:25,Born:25a,Born:25b}}.
Within this scenario, the electronic cloud or that of nucleons of the core of atoms in molecules, also of macromolecules like proteins, did not \emph{a priori} met the general quantum criterium, until X-rays, combined with de Broglie\footnote{\cite{Broglie:25}.} wave theory became operative\footnote{The extension of this parlance to NMR, SSRL, etc. occupy these days central stage around strategies to develop coronavirus therapies and vaccines \cite{Kramer:20}.}.

Neither did the order parameter of a superconductor. Namely the finite expectation value of pair of electron fields and its gauge angle\footnote{$\alpha_0'=\sum_{\nu>0}U'_\nu V'_\nu$, number of Cooper pairs, $U'_\nu$ and $V'_\nu$ being the BCS (\cite{Bardeen:57a,Bardeen:57b}) occupation numbers referred to the body-fixed frame attached to the two-dimensional rotor, whose $z'$ axis makes an angle $\phi$ with that of the laboratory system $z$ Fig. \ref{abs_figI}.} $\alpha_0=\alpha_0'\exp(-2i\phi)$, $\alpha_0'$ being the number of Cooper pairs, closely related to the abnormal density $(\alpha_0'/\mathcal V)$ where $\mathcal V$ is an appropriate volume. Said it differently, a superconductor has a rather perfect internal gauge phase order, and one does not as a rule have to deal with probes which have such a property. Exception made another superconductor. This is the reason why the Josephson effect associated with the tunneling of Cooper pairs across a junction (barrier) between two weakly coupled superconductors, is the specific tool which can pin down the order parameter in gauge space\footnote{\cite{Josephson:62,Anderson:64b}.}.

While Josephson-like junctions transiently established in heavy ion collisions $B(=A+2)+b\to A+a(=b+2)$ between superfluid nuclei --through which Cooper pair tunneling proceeds mainly in terms of successive transfer of entangled nucleons-- is deprived from the macroscopic aspects of a supercurrent, it displays many of the special effects associated with spontaneous symmetry breaking in gauge space (BCS condensation). The process of addition or removal of a Cooper pair from a superconductor (as in a Josephson junction) or of a nucleon pair from a superfluid nucleus in a heavy ion collision, constitutes a rotational mode in gauge space, in which particle number plays the role of angular momentum \cite{Bohr:76}. Pairing rotational bands represents the collective mode in gauge space, and are associated with restoration of particle number conservation.

If the tails of the normal densities extend across the junction, that of the abnormal density will similarly do so. Provided that the distance of closest approach $D_0\lesssim (D_0)_c$, phase correlated partners of a nuclear Cooper pair will be transferred mainly successively, equally well phase pairing correlated as if the two nucleons ($2n$) were transferred simultaneously. As a consequence, the absolute cross section $\sigma_{2n}$ will be under these circumstances, of the same order of the absolute cross section $\sigma_{1n}$.

The detailed diffraction pattern and strength (in units of mb/sr) of the absolute gs$\to$gs differential cross section $d\sigma_{2n}/d\Omega$ brought to the detector by the distorted waves $\chi_{Bb}^{(+)}$, $\chi_{Aa}^{(-)}$ depend in an essential fashion  --aside that on the real part of the optical potential $U_{Bb}$ ($E_{cm}<E_B$)--  on the non-local, correlated, successive transfer formfactor $F(\mathbf r_1,\mathbf r_2,\mathbf r_{Bb})$. Quantity which reflects the BCS correlation factors $U_\nu V_\nu$, known in nuclear physics as two-nucleon spectroscopic transfer amplitudes. It is of notice that to be able to gain such insight from the experimental data, one should be able to calculate absolute, two-nucleon transfer differential cross sections, within a 10\% error limit or better.

The theoretical framework of second order DWBA successive two-nucleon transfer reactions, as well as the variety of its computer program embodiments, are well established\footnote{\cite{Bayman:82,Potel:13,Gotz:75} and \cite{Thompson:88} regarding its coupled channel version.}. Also regarding its semiclassical approximation version\footnote{G. Pollarolo, see e.g. \cite{Montanari:14}. See also \cite{Potel:13} and \cite{Thompson:88}.}. One can thus view it as a universal software allowing to relate nuclear structure   with the observed absolute differential cross section $d\sigma_{2n}/d\Omega$.

Because the reaction part of the transfer amplitude ($\chi_{Bb}^{(+)}$, $\chi_{Aa}^{(-)}$, $U_{Bb}$, $U_B$, $U_b$, $G_{fF}$) is univoque and universal, one can posit that the corresponding asymptotic waves provide specific probe information of the successive transfer formfactors. Combining this information with electron scattering input concerning normal density $\rho$, one can posit that $F_{Bb}/(\rho_B\rho_b)$, or its equivalent representations,  are the specific gs$\to$gs nuclear Cooper pair tunneling observables.



\section{Gamma emission in the ($2n$) transfer reaction $^{116}$Sn +$^{60}$Ni$\to$$^{114}$Sn(gs) +$^{62}$Ni(gs) ($E_{cm}$=154.26 MeV, $\theta_{cm}=140^\circ$ for $^{114}$Sn(gs), $D_0=13.49$ fm)}
In heavy ion collisions below the Coulomb barrier with typical values of the ratio $D_0/\lambdabar\approx10^2$ between the distance of closest approach and the reduced de Broglie wavelength, Cooper pair tunneling can be described in terms of the semiclassical second order transfer amplitude (see \cite{Broglia:04a} p. 306 Eq. (23)),
\begin{align}
  \label{abs_eq:1}
  \nonumber (a_\beta)_{succ}&\approx\left(\frac{1}{i\hbar}\right)^2\\
                              \nonumber &\times\sum_{Ff\neq Bb}
  \int_{-\infty}^{\infty}dt\,\braket{\Psi_{Bb}|V_{Ff}-\braket{V_{Ff}}|\Psi_{Ff}}_{\mathbf R_{Bb,Ff}(t)}\,e^{i\frac{\left(E_{Bb}-E_{Ff}\right)}{\hbar}t}\\
  &\times \int_{-\infty}^{t}dt'\,\braket{\Psi_{Ff}|V_{Aa}-\braket{V_{Aa}}|\Psi_{Aa}}_{\mathbf R_{Aa,Ff}(t')}\,e^{i\frac{\left(E_{Ff}-E_{Aa}\right)}{\hbar}t'}.
\end{align}
in keeping with the fact that successive transfer
\begin{align}
  \label{abs_eq:2}
  a(=b+2)+A\to f(=b+1)+F(=A+1)\to b+B(=A+2)
\end{align}
is the overwhelming dominant process. Assuming that $\Psi_{Aa}=\Psi_A(\xi_A)\Psi_a(\xi_a)$ and $\Psi_{Bb}=\Psi_B(\xi_B)\Psi_b(\xi_b)$
are the BCS wavefunctions describing the intrinsic states of the pairing rotational bands of systems $a,A$ and $b,B$, the two quasiparticle excitation energies $E_{Ff}$ can be assumed to be large as compared with the reaction $Q$-value. Furthermore, one expects that the variation of the matrix elements (formfactors) along the trajectories to be smooth. In fact, a decaying exponential function for reactions below the Coulomb barrier (Eq. (\ref{abs_eq:6abs}), for the kinematic conditions stated; see Fig. \ref{fig:form}). In such a case one can make use of the approximation,
\begin{align}
  \label{abs_eq:3}
   \sum_{Ff\neq Bb}\exp\left(i\frac{E_{Ff}}{\hbar}(t'-t)\right)\approx \frac{1}{\Delta E}\int d E \exp\left(i\frac{E}{\hbar}(t'-t)\right)\approx\frac{\hbar}{\Delta E}\delta(t'-t),
\end{align}
$1/\Delta E$ being the density of two quasiparticle levels $Ff$. As a result, the argument of the remaining exponential contains the relative Josephson phase (see App. \ref{AppJx})
\begin{align}
  \label{abs_eq:4}
  \frac{E_{Bb}-E_{Aa}}{\hbar}=\frac{Q_{2n}}{\hbar}=2\frac{\mu_{Bb}-\mu_{Aa}}{\hbar}=2(\dot\phi_{Bb}-\dot\phi_{Aa})=\omega_J^N.
\end{align}
Furthermore (see \cite{Broglia:04a}, Ch. V, Sect. 11), in keeping with the fact that the $N$-member of a pairing rotational band  can be written as
\begin{align}
  \ket{N}=\frac{1}{\sqrt{2\pi}}e^{i N\phi}\ket{BCS},
\end{align}
the spectroscopic amplitude (coherence factors) connecting successive members of a pairing rotational band (Fig. \ref{abs_figII}) are
\begin{align}
  \label{abs_eq:6}
  \nonumber  B_j=&\int d\phi\braket{N_B|T|N_A}=\frac{1}{2\pi}\int d\phi e^{i\phi(N_B-(N_A+2))}\braket{BCS_B|T^\dagger|BCS_A}\\
  &=\delta(N_B,N_A+2)\sqrt{\frac{(2j+1)}{2}}\,U'_j V'_j,
\end{align}
$T'^\dagger=[a'^\dagger_ja'^\dagger_j]_0/\sqrt{2}$ being the two-nucleon transfer operator, while  $U'_j V'_j$ are the coherence factors in the intrinsic system (see Fig. \ref{abs_figI}), i.e. real quantities.

Summing up,  expression (\ref{abs_eq:1}) describing the reaction mechanism of the successive transfer of a nuclear Cooper pair, has been worked out making use of classical physics to describe the relative motion, but fully quantum mechanically in a particle conserving, gauge invariant formalism concerning the Cooper pair transfer process. Consequently, it contains  the (relative) gauge phases and frequencies required to describe the (ac) Josephson effect as well as, within a coarse grain approximation the (dc) Josephson effect, by assuming $Q_{2n}\approx 0.1$ MeV (see Fig. \ref{abs_fig:19}). In other words, (\ref{abs_eq:1}) contains the relative gauge phases and gauge frequencies initiating and modulating the Cooper pair tunneling process, those associated with the isolated BCS systems being  integrated out,  leading to the proper real intrinsic coherence factors.

Because the second order DWBA $T$-matrix
\begin{widetext}
\begin{align}
  \label{abs_eq:7}
  \nonumber T(\mathbf k_f,&\mathbf k_i)=2\sum_{KM}\sum_{j_i,j_f}B^{(A)*}_{j_f}B^{(b)}_{j_i}\int \chi_f^*(\mathbf r_{Bb},\mathbf k_f)\left[\phi^{(A)}_{j_f}(\mathbf r_{A_2})\phi^{(A)}_{j_f}(\mathbf r_{A_1})\right]^{0*}_0 U^{(A)}(r_{b1})\left[\phi^{(A)}_{j_f}(\mathbf r_{A_2})\phi^{(b)}_{j_i}(\mathbf r_{b_1})\right]^{K}_M \, d\mathbf r_{Cc}\, d\mathbf r_{b_1}\, d\mathbf r_{A_2}\\
  &\times\int G(\mathbf r_{Cc},\mathbf r'_{Cc})\left[\phi^{(A)}_{j_f}(\mathbf r'_{A_2})\phi^{(b)}_{j_i}(\mathbf r'_{b_1})\right]^{K*}_M U^{(A)}(r'_{b2})\left[\phi^{(b)}_{j_i}(\mathbf r'_{b_2})\phi^{(b)}_{j_i}(\mathbf r'_{b_1})\right]^{0}_0\chi_i(\mathbf r'_{Aa},\mathbf k_i)\, d\mathbf r'_{Cc}\, d\mathbf r'_{b_1}\, d\mathbf r'_{A_2},
\end{align}
\end{widetext}
gives essentially the same results as (\ref{abs_eq:1}) making use of the $B$-coefficients (\ref{abs_eq:6}), one can conclude it contains the relative Josephson phase (\ref{abs_eq:4}). This is in keeping with the fact that in (\ref{abs_eq:7}) the time representation of (\ref{abs_eq:1}) is changed into an energy one (energy denominator of the Green's function). Being time and energy conjugate variables one can change from one representation to the other through a Fourier transform. However, in the present case it is not through a mathematical identity, but through a physical approximation ($D_0/\lambdabar\approx10^2$ that the correspondence $(\mathbf r_\beta+\mathbf r_\alpha)/2\to (\mathbf R_\beta(t)+\mathbf R_\alpha(t))/2$)  is made. In the above  relation $\mathbf r_\alpha,\mathbf r_\beta$ are the quantal relative center of mass coordinates in entrance and exit channels, $\mathbf R_\alpha(t),\mathbf R_\beta(t)$ being the corresponding classical quantities,  functions of time.

In Eq. (\ref{abs_eq:7}), the distorted waves $\chi_i(\mathbf r'_{Aa};\mathbf k_i)$ and  $\chi_f^*(\mathbf r_{Bb};\mathbf k_f)$ describe the motion (trajectory) of the colliding nuclei in the initial and final channels, respectively. The Green function $G(\mathbf r_{Cc},\mathbf r'_{Cc})$ accounts for the propagation of the system from the  trajectory point $\mathbf r'_{Cc}$  in which the first neutron $n_1$ is transferred, to the position $\mathbf r_{Cc}$ where the transfer of the second neutron $n_2$ takes place. While the reaction aspect of the process is expressed in terms of the distorted waves, the structure content can be extracted by defining the non-local $(2n)$ transfer formfactor (transition density)
\begin{align}
  \label{eq:113}
  \nonumber F^{(nl)}(&\mathbf r_1,\mathbf r_2,\mathbf r_{Cc};\mathbf r'_1,\mathbf r'_2,\mathbf r'_{Cc})=\sum_{KM}\sum_{j_i,j_f}B^{(A)*}_{j_f}B^{(b)}_{j_i}U^{(A)}(r_{b1})\\
  \nonumber&\times \left[\phi^{(A)}_{j_f}(\mathbf r_{A_2})\phi^{(A)}_{j_f}(\mathbf r_{A_1})\right]^{0*}_0 \left[\phi^{(A)}_{j_f}(\mathbf r_{A_2})\phi^{(b)}_{j_i}(\mathbf r_{b_1})\right]^{K}_M G(\mathbf r_{Cc},\mathbf r'_{Cc})\\
         &\times  U^{(A)}(r'_{b2})\left[\phi^{(A)}_{j_f}(\mathbf r'_{A_2})\phi^{(b)}_{j_i}(\mathbf r'_{b_1})\right]^{K*}_M \left[\phi^{(b)}_{j_i}(\mathbf r'_{b_2})\phi^{(b)}_{j_i}(\mathbf r'_{b_1})\right]^{0}_0,  
\end{align}
where the vectors $\mathbf r_1$ and $\mathbf r_2$ indicate the position of neutrons 1 and 2, according to the coordinate system shown in Fig. \ref{notes_fig1}.
The fact that the reaction and structure aspects of the $(2n)$ transfer process do not factorize in Eq. (\ref{abs_eq:7}) is a mathematical expression of the fact that both aspects are inextricably linked. The formfactor $F(\mathbf r_1,\mathbf r_2,\mathbf r_{Cc})$ is obtained from (\ref{eq:113})   by setting $\mathbf r'_{Cc}=\mathbf r_{Cc}$, which amounts to making the substitution
\begin{align}
  \label{eq:114}
  G(\mathbf r_{Cc},\mathbf r'_{Cc})\to \delta(\mathbf r_{Cc}-\mathbf r'_{Cc})
\end{align}
in Eq. (\ref{abs_eq:7}), and corresponds to a situation in which both neutrons $n_1$ and $n_2$ are transferred at the same point of the collisional trajectory,  obtaining
\begin{align}
  \label{eq:112}
  \nonumber F&(\mathbf r_1,\mathbf r_2,\mathbf r_{Cc})=\sum_{j_i,j_f}B_{j_i}B_{j_f}\left[\phi_{j_f}(\mathbf r_{A_1})\phi_{j_f}(\mathbf r_{A_2})\right]^{0*}_0  \\
  \nonumber           &\times \left[\phi_{j_f}(\mathbf r_{A_2})\phi_{j_f}(\mathbf r_{b_1})\right]^{K}_M \,U^{(A)}(r_{b1}) U^{(A)}(r_{c2})\\
  &\times \left[\phi_{j_f}(\mathbf r_{A_2})\phi_{j_i}(\mathbf r_{b_1})\right]^{K*}_M \left[\phi_{j_i}(\mathbf r_{b_2})\phi_{j_i}(\mathbf r_{b_1})\right]^{0}_0.
\end{align}
Making use of this relation one can introduce  a local formfactor according to
\begin{align}
  \label{eq:112}
  F^{(l)}(\mathbf r_{Cc})=\int F(\mathbf r_1,\mathbf r_2,\mathbf r_{Cc}) \,d\mathbf r_1\,d\mathbf r_2,
\end{align}
which, for the case of the reaction (\ref{abs_eq:6abs}) is displayed in Fig. \ref{fig:form}.

Returning back to the Josephson phase function $\exp\left(i\omega_J^N t\right)$ and in keeping with the fact that the effective charge of the neutrons associated with the reaction  (\ref{abs_eq:6abs})  is $e_{eff}=-e(50+28)/(116+60)$, each time a Cooper pair ($q=2e_{eff}=-e\times0.89$) tunnels through the nuclear junction it is expected to emit $\gamma$-rays of energy $\hbar\omega_J^N=Q_{2n}$.
Because the transient and finite size character of the (biased, $V^N$) nuclear Josephson-like junction, the coherent monochromatic photons (\ref{abs_eq:5}) become, in the nuclear case (\ref{abs_eq:4}), spread over an energy range of few MeV, combined effect of $\tau_{coll}$ and of recoil.

\subsection{Differential cross section}\label{Sect.14.A}
Let us consider the  two neutron transfer process in which circularly polarized photons of energy $E_\gamma$ are emitted in the direction $\hat k_\gamma$. The associated $T$-matrix can be written  as \cite{Brink:68},
\begin{align}
      \label{notes_eq:300}
\mathcal T^q(\mathbf k_\gamma,\mathbf k_f)=\sum_{m_\gamma} \mathcal D_{m_\gamma q}^1(R_\gamma)\,T_{m_\gamma}(\mathbf k_f,\mathbf k_i),        
    \end{align}
    where $q=\pm1$ labels the transverse  photon polarization (clockwise, anticlockwise), $\mathcal D_{m_\gamma q}^1(R_\gamma)$ are the Wigner matrices describing the rotation from the quantization axis $\mathbf{\hat  k}_i(\equiv \mathbf{\hat z})$ to the direction $\mathbf{\hat k}_\gamma$,   $T_{m_\gamma}(\mathbf k_f,\mathbf k_i)$ being
    \begin{widetext}
\begin{align}\label {notes_eq:301}
\nonumber T_{m_\gamma}(\mathbf k_f,\mathbf k_i)&=\sum_{j_i,j_f}B_{j_i}B_{j_f}\int \chi_f^*(\mathbf r_{Bb};\mathbf k_f)\left[\phi_{j_f}(\mathbf r_{A_1})\phi_{j_f}(\mathbf r_{A_2})\right]^{0*}_0  D_{m_\gamma} \left[\phi_{j_f}(\mathbf r_{A_2})\phi_{j_f}(\mathbf r_{b_1})\right]^{K}_M\,v(r_{b1}) d\mathbf r_{Cc}\, d\mathbf r_{b_1}\, d\mathbf r_{A_2}\\
&\times\int G(\mathbf r_{Cc},\mathbf r'_{Cc})\left[\phi_{j_f}(\mathbf r'_{A_2})\phi_{j_i}(\mathbf r'_{b_1})\right]^{K*}_M v(r'_{c2})\left[\phi_{j_i}(\mathbf r'_{b_2})\phi_{j_i}(\mathbf r'_{b_1})\right]^{0}_0\chi_i(\mathbf r'_{Aa};\mathbf k_i)\, d\mathbf r'_{cC}\, d\mathbf r'_{b_1}\, d\mathbf r'_{A_2}.
\end{align}
\end{widetext}
The dipole operator is defined in Eq. (\ref{eq:60}),
and the coefficients $B_{j_i},B_{j_f}$ are the BCS coherence factors $U_kV_k$,   describing the ground states of   nuclei $a$ and $B$, respectively, i.e. the two-nucleon transfer spectroscopic amplitudes associated with the reaction considered. The coordinates involved in the expressions above are defined in Fig. \ref{notes_fig1}. The effective charge is $e_{eff}=-e\,\frac{(Z_A+Z_b)}{A_A+A_b}$. 
For a numerical evaluation of Eq. (\ref{notes_eq:301}),  $T_{m_\gamma}(\mathbf k_f)$ is expanded in partial waves.

If the photon polarization is not measured, the triple differential cross section is
\begin{widetext}
\begin{align}
  \label{notes_eq:303}
  \nonumber\frac{d^3\sigma^{\gamma}_{2n}}{d\Omega_\gamma d\Omega dE_\gamma}&=\rho_f(E_f)\,\rho_\gamma(E_\gamma)\left(\left|\mathcal T^{1}(\mathbf k_\gamma,\mathbf k_f)\right|^2+\left|\mathcal T^{-1}(\mathbf k_\gamma,\mathbf k_f)\right|^2\right)\delta(E_i-E_\gamma-E_f+Q)\\
 &=\frac{\mu_i\mu_f}{(2\pi\hbar^2)^2}\frac{k_f}{k_i}\left(\frac{E_\gamma^2}{(\hbar c)^3}\right)\left(\left|\mathcal T^{1}(\mathbf k_\gamma,\mathbf k_f)\right|^2+\left|\mathcal T^{-1}(\mathbf k_\gamma,\mathbf k_f)\right|^2\right)\delta(E_i-E_\gamma-E_f+Q),  
\end{align}
\end{widetext}
where $\rho_f(E_f)$ and $\rho_\gamma(E_\gamma)$ are the heavy ion and photon phase spaces, respectively.  $Q$ is the reaction $Q$-value, and $E_i$, $E_f$ are the kinetic energies in the initial and final channels. Making use of the explicit expressions of the Wigner matrices one obtains, from the equation above the following  expression for the triple differential cross section,
\begin{widetext}
\begin{align}
  \label{notes_eq:303}
  \nonumber \frac{d^3\sigma^{\gamma}_{2n}}{d\Omega_\gamma d\Omega dE_\gamma}&=\frac{\mu_i\mu_f}{(2\pi\hbar^2)^2}\frac{k_f}{k_i}\left(\frac{E_\gamma^2}{(\hbar c)^3}\right)\left[\left(1-\frac{1}{2}\sin^2\theta_\gamma\right)\left(\left|T_1(\mathbf k_f)\right|^2+\left|T_{-1}(\mathbf k_f)\right|^2\right)+\sin^2(\theta_\gamma)\left|T_{0}(\mathbf k_f)\right|^2\right.\\
\nonumber  &+\sin^2(\theta_\gamma)\left(\cos2\phi_\gamma\,\text{Re}(T_{1}(\mathbf k_f)T^*_{-1}(\mathbf k_f))+\sin2\phi_\gamma\,\text{Im}(T_{1}(\mathbf k_f)T^*_{-1}(\mathbf k_f))\right)\\
  \nonumber  &+\frac{2}{\sqrt{2}}\cos\phi_\gamma\sin\theta_\gamma\cos\theta_\gamma\,\text{Re}\left(T_{1}(\mathbf k_f)T^*_{0}(\mathbf k_f)-T_{-1}(\mathbf k_f)T^*_{0}(\mathbf k_f)\right)\\
   &\left.+\frac{2}{\sqrt{2}}\sin\phi_\gamma\sin\theta_\gamma\cos\theta_\gamma\,\text{Im}\left(T_{1}(\mathbf k_f)T^*_{0}(\mathbf k_f)+T_{-1}(\mathbf k_f)T^*_{0}(\mathbf k_f)\right)\right]\delta(E_i-E_\gamma-E_f+Q).
\end{align}
\end{widetext}
It is also   useful to write the above expression in terms of the following dimensionless cartesian coordinates,
\begin{align}
  \label{notes_eq:304}
  x=\sin\theta_\gamma\cos\phi_\gamma;\quad y=\sin\theta_\gamma\sin\phi_\gamma;\quad z=\cos\theta_\gamma.
\end{align}
Then,
\begin{widetext}
\begin{align}
  \label{notes_eq:305}
  \nonumber \frac{d^3\sigma^{\gamma}_{2n}}{d\Omega_\gamma d\Omega dE_\gamma}&=\frac{\mu_i\mu_f}{(2\pi\hbar^2)^2}\frac{k_f}{k_i}\left(\frac{E_\gamma^2}{(\hbar c)^3}\right)\left\{\vphantom{\int\frac{1}{2}}\left(1-\tfrac{1}{2}(x^2+y^2)\right)\left(\left|T_1(\mathbf k_f)\right|^2+\left|T_{-1}(\mathbf k_f)\right|^2\right)\right.\\
  \nonumber  &+(x^2+y^2)\left|T_{0}(\mathbf k_f)\right|^2+2xy\,\text{Im}(T_{1}(\mathbf k_f)T^*_{-1}(\mathbf k_f))+(x^2-y^2)\,\text{Re}(T_{1}(\mathbf k_f)T^*_{-1}(\mathbf k_f))\\
  &\left.+\sqrt{2}\,z\,\left[x\,\text{Re}\left(T_{1}(\mathbf k_f)T^*_{0}(\mathbf k_f)-T_{-1}(\mathbf k_f)T^*_{0}(\mathbf k_f)\right)+y\,\text{Im}\left(T_{1}(\mathbf k_f)T^*_{0}(\mathbf k_f)+T_{-1}(\mathbf k_f)T^*_{0}(\mathbf k_f)\right)\right]\vphantom{\int\frac{1}{2}}\right\}\delta(E_i-E_\gamma-E_f+Q).  
\end{align}
\end{widetext}
For $E_\gamma=4$ MeV, and a deflection angle of 140$^\circ$ in the c.m. frame for the nucleus $^{114}$Sn, we obtain, from the numerical evaluation of the integral (\ref{notes_eq:301}), the following components of the $T$-matrix,
\begin{align}
  \label{notes_eq:26}
  \nonumber &T_{-1}=\left(-5.5+41.0\,i\right)\text{MeV}^{3/2} \text{fm}^{9/2},\\
  \nonumber &T_{0}=\left(-32.1-22.3\,i\right)\text{MeV}^{3/2} \text{fm}^{9/2},\\
  &T_{1}=\left(-26.1-13.7\,i\right)\text{MeV}^{3/2} \text{fm}^{9/2}.
\end{align}
The resulting radiation pattern is  displayed in Figs. \ref{notes_fig2} and \ref{notes_fig2x} in cartesian and spherical coordinates,  respectively, while in Fig. \ref{notes_fig4} we display the maximum of the angular radiation distribution for different $\gamma$-energies (for the minimum, see Fig. \ref{notes_fig2x} b), c)).

Making use of Eq. (\ref{notes_eq:300}), one has calculated the strength of the two different polarizations
\begin{align}
  \label{eq:2600}
  \left|\mathcal T^1(\mathbf k_\gamma,\mathbf k_f)\right|^2;\quad \left|\mathcal T^{-1}(\mathbf k_\gamma,\mathbf k_f)\right|^2, 
\end{align}
and the associated analyzing power 
\begin{align}
  \label{eq:3600}
    \mathcal P(\mathbf k_\gamma,\mathbf k_f)=\frac{\left|\mathcal T^{1}(\mathbf k_\gamma,\mathbf k_f)\right|^2-\left|\mathcal T^{-1}(\mathbf k_\gamma,\mathbf k_f)\right|^2}{\left|\mathcal T^{1}(\mathbf k_\gamma,\mathbf k_f)\right|^2+\left|\mathcal T^{-1}(\mathbf k_\gamma,\mathbf k_f)\right|^2},
\end{align}
and displayed in Fig. \ref{fig:pola}.

One can obtain the double differential cross section by integrating Eq. (\ref{notes_eq:303}) over all $\gamma$ angles, that is,
\begin{align}
  \label{notes_eq:306}
  \nonumber\frac{d^2\sigma^{\gamma}_{2n}}{d\Omega\,dE_\gamma}&=\int \frac{d^3\sigma}{d\Omega_\gamma\, d\Omega\,dE_\gamma}\,d\Omega_\gamma=\frac{\mu_i\mu_f}{(2\pi\hbar^2)^2}\frac{k_f}{k_i}\left(\frac{8\pi}{3}\frac{E_\gamma^2}{(\hbar c)^3}\right)\\
  &\times\left(\left|T_1(\mathbf k_f)\right|^2+\left|T_{-1}(\mathbf k_f)\right|^2+\left|T_{0}(\mathbf k_f)\right|^2\right)\delta(E_i-E_\gamma-E_f+Q).    
\end{align}
The corresponding $\gamma$-strength function for $\theta_{cm}=140^\circ$ is displayed in\footnote{It is of note that the difference with that displayed in Fig. 1 of reference \cite{Potel:21}, in particular concerning the $\gamma$ energy integrated area (5.18 $\mu$b/sr), stems from the fact that the radial value of the dipole moment was calculated in that publication, making use of the approximation $\mathbf{\hat O}=2\mathbf r_{O1}$ instead of $\mathbf{\hat O}=\mathbf r_{O1}+\mathbf r'_{O2}$ (see Eq. (\ref{eq:59}) and (\ref{eq:60})).} Fig. \ref{notes_fig5}, its $\gamma$-energy integrated area being $\approx$6.90 $\mu$b/sr.

In the reaction (\ref{abs_eq:6abs}) the system evolves between a coherent initial $\ket{\tilde i}\,(\equiv \ket{BCS(^{116}\text{Sn}(gs))}\otimes\ket{BCS(^{60}\text{Ni}(gs))}$) and a coherent final $\ket{\tilde f}\,(\equiv \ket{BCS(^{114}\text{Sn}(gs))}\otimes\ket{BCS(^{62}\text{Ni}(gs))}$) state, that is, proceeding along the levels of the corresponding pairing rotational bands (see Apps. \ref{rotgauge} and \ref{AppNernst}).

As soon as the pairing densities overlap, and because the vacuum between them is a dielectric, a (dc) voltage $V^N_J=Q_{2n}/2e_{eff}$ is created --and thus an electric field-- whereupon an alternating current of frequency $\nu^N_J=Q_{2n}/h$ becomes established. All (transferred) Cooper pair dipole moment will respond coherently to the field --according to the coherence two-nucleon transfer spectroscopic factors $B_j=((2j+1)/2)^{1/2}U_j V_j$ -- giving a ``macroscopic'' polarization to the transient composite system, which now drives the field acting as a source of directional photons \cite{Rogovin:76} of frequency $\nu^N_J$, emitted at a rate proportional to the number of Cooper pairs $\alpha_0'(=\sum_j ((2j+1)/2)^{1/2} B_j)$ quantity squared\footnote{In fact, proportional to the squared of the reduced number of Cooper pairs, i.e., $\alpha_0'(^{116}\text{Sn})\times\alpha_0'(^{60}\text{Ni})/(\alpha_0'(^{116}\text{Sn})+\alpha_0'(^{60}\text{Ni}))$).} (superradiance).

Because the members of the (gs) pairing rotational bands involved in the reaction (\ref{abs_eq:6abs}) display no damping, in particular no rotational (inhomogeneous) damping \cite{Broglia:87}, the conjugate variables $Q_{gs}(\equiv Q_{n=0})$ and $P_{gs}(\equiv P_{n=0})$ associated with these states (see p. 80 \cite{Brink:05}), fulfill the minimum value allowed by the uncertainty relations ($\Delta Q_{gs}\, \Delta P_{gs}=\hbar$), consistent with a Gaussian wave packet. It is then not surprising that the associated $\gamma$-strength function line shape is Gaussian.

\subsubsection{Change of reference frames}
Since theoretical calculations are usually worked out in the center of mass (cm) reference frame, it is useful to be able to transform the triple differential cross section (\ref{notes_eq:303}) to the laboratory (lab) frame, where the results of experimental observations are recorded. Primed (unprimed) quantities will refer to quantities measured in the lab (cm) frame of reference. Let us first establish the relationship between primed and unprimed variables (see \cite{Satchler:80} p. 300 Eq. (B.14) and \cite{Dedrick:60}, Eqs (1a) and (2)),
\begin{align}
  \label{notes_eq:102}
  &\tan\theta'=\frac{\sin\theta}{m_P/m_T+\cos\theta},\\
  &\frac{E'_\gamma}{E_\gamma}=\Gamma\left(1+\beta\cos\theta_\gamma\right),\\
  &\frac{\sin\theta_\gamma}{\sin\theta_\gamma'}=\Gamma\left(1+\beta\cos\theta_\gamma\right),
\end{align}
with
\begin{align}
  \label{notes_eq:103}
  \beta=\text{v}/c;\quad \Gamma=\frac{1}{\sqrt{1-\beta^2}},
\end{align}
$\text{v}$ being the relative velocity between the cm and lab frames, and $m_P,m_T$ are the masses of the projectile and of the target, respectively. Note that, since the azimutal angle is unchanged by the transformation, $\phi=\phi'$; $\phi_\gamma=\phi'_\gamma$.

The conservation of number particles can be expressed as
\begin{align}
  \label{notes_eq:100}
\sigma(\Omega,\Omega_\gamma ,E_\gamma)\,d\Omega d\Omega_\gamma dE_\gamma=\sigma'(\Omega',\Omega'_\gamma ,E'_\gamma)\,d\Omega' d\Omega'_\gamma dE'_\gamma,
\end{align}
where
\begin{align}
  \label{notes_eq:101}
  \nonumber &\sigma(\Omega,\Omega_\gamma ,E_\gamma)\equiv\frac{d^3\sigma(\Omega,\Omega_\gamma ,E_\gamma)}{d\Omega d\Omega_\gamma dE_\gamma};\\
  &\sigma'(\Omega',\Omega'_\gamma ,E'_\gamma)\equiv\frac{d^3\sigma(\Omega',\Omega'_\gamma ,E'_\gamma)}{d\Omega' d\Omega'_\gamma dE'_\gamma}.
\end{align}
Then,
\begin{align}
  \label{notes_eq:104}
  \sigma'(\Omega',\Omega'_\gamma ,E'_\gamma)=J\times\sigma(\Omega,\Omega_\gamma ,E_\gamma)
\end{align}
where
\begin{align}
  \label{eq:91} J=\frac{d(\cos\theta)}{d(\cos\theta')}\frac{\partial(E_\gamma,\cos\theta_\gamma)}{\partial(E'_\gamma,\cos\theta'_\gamma)}
\end{align}
is the Jacobian of the transformation, while (see \cite{Satchler:80}, Eq. (B18))
\begin{align}
  \label{notes_eq:105} \frac{d(\cos\theta')}{d(\cos\theta)}=\frac{1+(m_P/m_T)\cos\theta}{\left(1+(m_P/m_T)^2+2(m_P/m_T)\cos\theta\right)^{3/2}},
\end{align}
and (see \cite{Dedrick:60} Eq. (12))
\begin{align}
  \label{notes_eq:106}
  \frac{\partial(E_\gamma,\cos\theta_\gamma)}{\partial(E'_\gamma,\cos\theta'_\gamma)}=\frac{\sin\theta_\gamma}{\sin\theta'_\gamma}.
\end{align}
Let us apply the transformation to the reaction $^{116}$Sn +$^{60}$Ni$\to$$^{114}$Sn(gs) +$^{62}$Ni(gs), with a cm energy $E$=154.26 MeV, with the $^{114}$Sn ion  at the cm angle $\theta=140^\circ$. The velocity characterizing the transformation is that of the cm motion, $\text{v}_{cm}=\frac{m_{\text{Sn}}}{m_{\text{Sn}}+m_{\text{Ni}}}\text{v}_{lab}$. We obtain $\beta=0.059$, $\Gamma=1.003$. The maximum of the triple differential cross section for $E_\gamma=4$ MeV, $\phi_\gamma=0$, $\theta_\gamma\approx 150^\circ$ in the cm lab frame (see Fig. \ref{notes_fig2x}), corresponds to a lab energy $E'_\gamma=3.8$ MeV, and a lab polar angle $\theta'_\gamma=151.6^\circ$. The corresponding value of the triple differential cross section at the maximum is $\sigma'(\theta'_\gamma=151.6^\circ,\phi'_\gamma=0^\circ,E'_\gamma=3.8\text{ MeV})=J\times\sigma(\theta_\gamma=150^\circ,\phi_\gamma=0^\circ,E_\gamma=4\text{ MeV})= 0.63\,\mu$b/(MeV sr$^2$).

\subsection{Dipole length and orientation in Cooper pair transfer}\label{S14B}
 Let us now define the value of a given quantity, expressed as a function $f$ of the coordinates of the active neutrons, averaged over the transfer process.  We  define  the transition quantity $\langle f\rangle_{i\to f}$ averaged over the transfer process $i\to f$ as
 \begin{align}
   \label{notes_eq:1}
   \langle f\rangle_{i\to f}=\frac{|\braket{\phi_f|U f|\phi_i}|}{|\braket{\phi_f|U|\phi_i}|}=\frac{|\braket{\phi_f|U f|\phi_i}|}{|T|},
 \end{align}
 where $U$ is the transfer potential, $\phi_i$, $\phi_f$ are the initial and final states, including partial waves and neutron wavefunctions, and $T=\braket{\phi_f|U|\phi_i}$ represents the $i\to f$ transfer $T$-matrix. This is just a general schematic definition that includes the two-neutron transfer case, in which  the transfer potential acts twice, and one needs to sum over the intermediate states.

 We are  interested in the following transition lengths, $\braket{r_{O1}^2}_{i\to f}$, $\braket{r_{O2}^2}_{i\to f}$, as well as
 \begin{align}
   \label{eq:109}
   r_{2n}=\left(\braket{|\mathbf r_{O1}+\mathbf r'_{O2}|^2}_{i\to f}\right)^{1/2},
 \end{align}
 and
 \begin{align}
   \label{eq:110}
   \xi=\left(\braket{|\mathbf r_{O1}-\mathbf r'_{O2}|^2}_{i\to f}\right)^{1/2},
 \end{align}
 see also App. \ref{S1}.

 These quantities are  related, and can be expressed in terms of the amplitudes $\braket{\phi_f|U \,r^2_{O1}|\phi_i}$, $\braket{\phi_f|U \,r'^2_{O2}|\phi_i}$ and $\braket{\phi_f|U\,\mathbf r_{O1}\cdot\mathbf r'_{O2}|\phi_i}$. We have
 \begin{align}
   \label{notes_eq:2}
\nonumber   &\braket{|\mathbf r_{O1}+\mathbf r'_{O2}|^2}_{i\to f}=\left\langle r^2_{O1}+ r'^2_{O2}+2\mathbf r_{O1}\cdot\mathbf r'_{O2}\right\rangle_{i\to f}=\\
   &\frac{\left|\braket{\phi_f|U \,r^2_{O1}|\phi_i}+\braket{\phi_f|U \,r'^2_{O2}|\phi_i}+2\braket{\phi_f|U\,\mathbf r_{O1}\cdot\mathbf r'_{O2}|\phi_i}\right|}{|T|},
 \end{align}
 and
 \begin{align}
   \label{notes_eq:3}
\nonumber   &\braket{|\mathbf r_{O1}-\mathbf r'_{O2}|^2}_{i\to f}=\left\langle r^2_{O1}+ r'^2_{O2}-2\mathbf r_{O1}\cdot\mathbf r'_{O2}\right\rangle_{i\to f}=\\
   &\frac{\left|\braket{\phi_f|U \,r^2_{O1}|\phi_i}+\braket{\phi_f|U \,r'^2_{O2}|\phi_i}-2\braket{\phi_f|U\,\mathbf r_{O1}\cdot\mathbf r'_{O2}|\phi_i}\right|}{|T|}.
 \end{align}
 In order to calculate the scalar product $\mathbf r_{O1}\cdot\mathbf r'_{O2}$, it is useful to express it as
 \begin{align}
   \label{notes_eq:4}
   \mathbf r_{O1}\cdot\mathbf r'_{O2}=\frac{\sqrt{3}}{4\pi}r_{O1}r'_{O2}\left[Y^1(\hat r_{O1})Y^1(\hat r'_{O2})\right]^0_0.
 \end{align}
 The amplitudes  $\braket{\phi_f|U \,r^2_{O1}|\phi_i}$, $\braket{\phi_f|U \,r'^2_{O2}|\phi_i}$ and $\braket{\phi_f|U\,\mathbf r_{O1}\cdot\mathbf r'_{O2}|\phi_i}$ can be derived using standard partial wave decomposition and Racah algebra techniques.
 The angles $\theta_D$ and $\varphi_D$ characterizing  the dipole orientation of   (\ref{eq:60}), can be defined in a similar way.

 The Cartesian components of the dipole operator are (see Sect. \ref{S7})
 \begin{align}
   \label{notes_eq:306x}
D_x=\frac{1}{\sqrt{2}}(D_{-1}-D_1);\quad   D_y=\frac{i}{\sqrt{2}}(D_{-1}+D_1);\quad   D_z=D_0.
 \end{align}
 According to the definition of transition quantities, we can obtain the angles $\theta_D$ and $\varphi_D$ from the calculated $T$-matrices $T_1,T_{-1}$ and $T_0$, using
 \begin{align}
   \label{notes_eq:307}
   \cos\theta_D=\frac{D_z}{|\mathbf D|}=\frac{|T_0|}{\sqrt{\braket{|\mathbf r_{O1}+\mathbf r_{O2}|^2}_{i\to f}}};\quad \tan\varphi_D=\frac{|T_{-1}+T_{1}|}{|T_{-1}-T_1|}.
 \end{align}
The resulting dipole for $E_\gamma=4$ MeV, and $\theta_{cm}=140^\circ$ for the nucleus $^{114}$Sn(gs), is displayed in Fig. \ref{notes_fig3}.
\subsection{Maximum of the angular distribution}\label{AppC}
 The extremes  of the triple differential cross section  (\ref{notes_eq:303}) can be determined with the help of the relation
\begin{align}
  \label{notes_eq:136}
  \nonumber &\frac{\partial\left(\left|\mathcal T^{1}(\mathbf k_\gamma,\mathbf k_f)\right|^2+\left|\mathcal T^{-1}(\mathbf k_\gamma,\mathbf k_f)\right|^2\right)}{\partial\theta_\gamma}\\
  &=A(\phi_\gamma)\sin(2\theta_\gamma)+B(\phi_\gamma)\cos(2\theta_\gamma)=0,
\end{align}
with
\begin{align}
  \label{notes_eq:137}
  \nonumber A(\phi_\gamma)&=\left|T_{0}(\mathbf k_f)\right|^2-\frac{1}{2}\left(\left|T_1(\mathbf k_f)\right|^2+\left|T_{-1}(\mathbf k_f)\right|^2\right)+\\
  &\cos2\phi_\gamma\,\text{Re}(T_{1}(\mathbf k_f)T^*_{-1}(\mathbf k_f))+\sin2\phi_\gamma\,\text{Im}(T_{1}(\mathbf k_f)T^*_{-1}(\mathbf k_f)),
\end{align}
and
\begin{align}
  \label{notes_eq:138}
  \nonumber B(\phi_\gamma)&=\sqrt{2}\cos\phi_\gamma\,\text{Re}\left(T_{1}(\mathbf k_f)T^*_{0}(\mathbf k_f)-T_{-1}(\mathbf k_f)T^*_{0}(\mathbf k_f)\right)\\
  &+\sqrt{2}\sin\phi_\gamma\,\text{Im}\left(T_{1}(\mathbf k_f)T^*_{0}(\mathbf k_f)+T_{-1}(\mathbf k_f)T^*_{0}(\mathbf k_f)\right).
\end{align}
The values of $\theta_\gamma$ for which (\ref{notes_eq:303}) is extreme can then be obtained as a function of $\phi_\gamma$ by solving
\begin{align}
  \label{notes_eq:139}
  \theta_{max}=-\frac{1}{2}\arctan\left(\frac{B(\phi_\gamma)}{A(\phi_\gamma)}\right).
\end{align}
A maximum can be identified by ensuring that the second derivative is negative, i.e.,
\begin{align}
  \label{notes_eq:140}
  A(\phi_\gamma)\cos(2\theta_{max})-B(\phi_\gamma)\sin(2\theta_{max})<0.
\end{align}

\subsection{Gamma emission in the ($1n$) transfer reaction $^{116}$Sn +$^{60}$Ni$\to$$^{115}$Sn(gs) +$^{61}$Ni(gs) ($E_{cm}$=154.26 MeV, $\theta_{cm}=140^\circ$ for $^{115}$Sn, $D_0=13.49$ fm)}\label{Sect.14.D}
Let us now consider the reaction
\begin{equation}\label{eq_onept1}
    A+a(=b+1)\longrightarrow B(=A+1)+b+\gamma_q,
  \end{equation}
  where $\gamma_q$ is a circularly polarized photon, with polarization $q=\pm1$. 
  Let   $l_i$ and  $j_i$ be the initial orbital and total angular momenta of the transferred nucleon, while $l_f$ and  $j_f$ correspond to the final single-particle state. The total spin and magnetic quantum numbers of nuclei $A,a,B,b$ being $\{J_A,M_A\};\{J_a,M_a\};\{J_B,M_B\};\{J_b,M_b\}$. The $T$-matrix for this process can be written as,
    \begin{align}
      \label{eq_onepteq:400}
\mathcal T^{q}(\mathbf k_\gamma,\mathbf k_f)=\sum_{m_\gamma} \mathcal D_{m_\gamma q}^1(R_\gamma)\,T_{m_\gamma}(\mathbf k_f,\mathbf k_i) ,        
    \end{align}
    with
    \begin{align}
      \label{eq_onepteq:402}
      \nonumber T_{m_\gamma}(\mathbf k_f,\mathbf k_i)&=\sum_{m_i,m_f}\langle j_i\; m_i \; J_b\;M_b | J_a\, M_a\rangle\\
      &\times\langle j_f\; m_f \; J_A\;M_A | J_B\, M_B\rangle\,T^{m_i,m_f}_{m_\gamma}(\mathbf k_f,\mathbf k_i)_{1n},
    \end{align}
    and
   \begin{align}\label {eq_onepteq:403}
     \nonumber T^{m_i,m_f}_{m_\gamma}&(\mathbf k_f,\mathbf k_i)_{1n}=\int \chi_f^*(\mathbf r_{Bb};\mathbf k_f)\,\psi_{m_f}^{j_f*}(\mathbf{r}_{A1},\sigma)\,  D_{m_\gamma} \\
     &\times v(r_{b1})\, \psi_{m_i}^{j_i}(\mathbf{r}_{b1},\sigma)\,\chi_i(\mathbf r_{Aa},\mathbf k_i)\, d\mathbf r_{aA}\, d\mathbf r_{b_1},
   \end{align}
   the dipole operator being
   \begin{align}
     \label{eq_onepteq:24}
        D_{m_\gamma}=e_{eff}\sqrt{\frac{4\pi}{3}}\,r_{O1}Y^1_{m_\gamma}(\hat r_{O1}),
   \end{align}
   with an effective charge
   \begin{align}
     \label{eq_onepteq:25}
     e_{eff}=-e\,\frac{(Z_A+Z_b)}{A_A+A_b}.
   \end{align}
   The Clebsh-Gordan coefficients in Eq. (\ref{eq_onepteq:402}) account for the proper angular momentum coupling of the transferred nucleon to the core nuclei. A numerically  tractable expression can be obtained making use of standard partial wave decomposition techniques,
   \begin{widetext}
\begin{align}
  \label{eq_onepteq:28}
    \nonumber \frac{d^3\sigma^{\gamma}_{1n}}{d\Omega_\gamma d\Omega dE_\gamma}&=\frac{\mu_i\mu_f}{(2\pi\hbar^2)^2}\frac{k_f}{k_i}\left(\frac{E_\gamma^2}{(\hbar c)^3}\right)\sum_{m_i,m_f}\left[\left(1-\frac{1}{2}\sin^2\theta_\gamma\right)\left(\left|T^{m_i,m_f}_1(\mathbf k_f,\mathbf k_i)\right|^2+\left|T^{m_i,m_f}_{-1}(\mathbf k_f,\mathbf k_i)\right|^2\right)+\sin^2(\theta_\gamma)\left|T^{m_i,m_f}_0(\mathbf k_f,\mathbf k_i)\right|^2\right.\\
\nonumber  &+\sin^2(\theta_\gamma)\left(\cos2\phi_\gamma\,\text{Re}(T^{m_i,m_f}_1(\mathbf k_f,\mathbf k_i)T^{m_i,m_f*}_{-1}(\mathbf k_f,\mathbf k_i))+\sin2\phi_\gamma\,\text{Im}(T^{m_i,m_f}_1(\mathbf k_f,\mathbf k_i)T^{m_i,m_f*}_{-1}(\mathbf k_f,\mathbf k_i))\right)\\
  \nonumber  &+\frac{2}{\sqrt{2}}\cos\phi_\gamma\sin\theta_\gamma\cos\theta_\gamma\,\text{Re}\left(T^{m_i,m_f}_1(\mathbf k_f,\mathbf k_i)T^{m_i,m_f*}_{0}(\mathbf k_f,\mathbf k_i)-T^{m_i,m_f}_{-1}(\mathbf k_f,\mathbf k_i)T^{m_i,m_f*}_{0}(\mathbf k_f,\mathbf k_i)\right)\\
  &\left.+\frac{2}{\sqrt{2}}\sin\phi_\gamma\sin\theta_\gamma\cos\theta_\gamma\,\text{Im}\left(T^{m_i,m_f}_{1}(\mathbf k_f,\mathbf k_i)T^{m_i,m_f*}_{0}(\mathbf k_f,\mathbf k_i)+T^{m_i,m_f}_{-1}(\mathbf k_f,\mathbf k_i)T^{m_i,m_f*}_{0}(\mathbf k_f,\mathbf k_i)\right)\right]\delta(E_i-E_\gamma-E_f+Q).
\end{align}
\end{widetext}
In the case in which the initial beam is not polarized,  and the final polarization  is not measured, the above expression has to be multiplied by the factor
\begin{align}
  \label{eq:92}
  \frac{(2J_B+1)(2J_a+1)}{(2j_i+1)(2j_f+1)}.
\end{align}
The corresponding double differential cross section can be obtained integrating over the angles $\theta_\gamma$ and $\phi_\gamma$,
\begin{widetext}
 \begin{align}
   \label{eq_onepteq:18}
\frac{d^2\sigma^{\gamma}_{1n}}{d\Omega\,dE_\gamma}&=\frac{(2J_B+1)(2J_a+1)}{(2j_i+1)(2j_f+1)}\frac{\mu_i\mu_f}{(2\pi\hbar^2)^2}\frac{k_f}{k_i}\left(\frac{8\pi}{3}\frac{E_\gamma^2}{(\hbar c)^3}\right)\sum_{m_i,m_f}\left(\left|T^{m_i,m_f}_1(\mathbf k_f,\mathbf k_i)\right|^2+\left|T^{m_i,m_f}_{-1}(\mathbf k_f,\mathbf k_i)\right|^2+\left|T^{m_i,m_f}_{0}(\mathbf k_f,\mathbf k_i)\right|^2\right)\delta(E_i-E_\gamma-E_f+Q).    
 \end{align}
 \end{widetext}
 In Fig. \ref{fig1pt} the double differential cross section corresponding to the summed contributions of the  $^{61}$Ni and $^{115}$Sn states that are more strongly populated in the reaction $^{116}$Sn+$^{60}$Ni$\to^{115}$Sn+$^{61}$Ni,  for the summed $^{61}$Ni+ $^{115}$Sn excitation energies $E_x\lesssim 2.5$ MeV (see Fig. \ref{fig:5x} for the case of $^{61}$Ni)  is displayed.  

Within this context and that of possible  parallels with condensed matter physics, it is of note that metallic cables conducting current emit photons due to the Joule effect. Thermal radiation consists in electromagnetic radiation of wavelength in the range 0.1$\mu$m to 100$\mu$m. It includes part of the ultraviolet (UV) and all of the visible and infrared (IR) spectrum. The elements of the present section, in particular Eq. (\ref{eq_onepteq:18}) are at the basis of a study of the quantum origin of a Joule-like nuclear effect in terms of $\gamma$-emission \cite{Broglia:22}.

\section{Analysis of the $\gamma$ spectra of the 
$^{60}$Ni+$^{116}$Sn experiment with the AGATA demonstrator array}\label{S15} 
 In what follows we report on a reanalysis of the $\gamma$ spectra obtained in 
 Ref. \cite{Montanari:16}, considering also the energy range higher than 
 that of the discrete $\gamma$ lines. 

 The $\gamma$-spectra, shown in Fig. \ref{fig:g2}, 
 were obtained for the $^{60}$Ni+$^{116}$Sn reaction \cite{Montanari:16}
 in a direct kinematic experiment at E$_{\rm lab}$=245 MeV, 
 with the PRISMA spectrometer \cite{Szilner:07,Corradi:09} 
 at $\theta_{\rm lab}$=70$^{\circ}$ (detecting Ni ions)
 coupled to the AGATA Demonstrator 
 $\gamma$-array \cite{Gadea:11}. That experiment 
 had as purpose to determine the transfer strength 
 to the known excited states, to be compared with the inclusive
 data obtained in Ref. \cite{Montanari:14}. 
 It was shown that the main flux 
 in the two-neutron transfer channel is in the ground to ground state 
 transition, in agreement with microscopic calculations.
 The emitted $\gamma$ rays correspond to the known discrete lines of the 
 Ni and Sn isotopes, 
 Doppler corrected taking into account the geometry of particle and 
 $\gamma$ detectors and the velocity of both the light and 
 heavy reaction products. In the experiments performed so far with
 PRISMA one measured directly the velocity vector for the (detected) 
 light partner while the one for the heavy partner was derived assuming 
 binary kinematics. For the $^{62}$Ni and $^{114}$Sn nuclei, associated with
 the two-neutron transfer channel, the only recognizable $\gamma$-line  
 within the limited statistics,  
 corresponds to the transition from the 2$^+$ state to the 0$^+$ ground 
 state of $^{62}$Ni.

 The spectra shown in Fig. \ref{fig:g2} 
 correspond to an average distance of closest 
 approach $D_0$ $\approx$14.5 fm (see  \cite{Montanari:16}, as well as Fig. \ref{fig:7p}), 
 where the transfer probabilities of the one-and  two-neutron transfer 
 channels differ by about one order of magnitude.  
 Theory  predicts that, by lowering the bombarding energy below the Coulomb 
 barrier, the strength distribution generated by the dipole 
 oscillations maintains its shape, the energy integrated one  reflecting the 
 decrease of the cross section for the two-neutron transfer channel. 
 Calculations have been performed at 
 the bombarding energy E$_{\rm lab}$= 425.4 MeV, 
 and $\theta_{\rm lab}$=20$^\circ$, corresponding 
 to the same distance of closest approach as 
 in the direct kinematic experiment of Ref. \cite{Montanari:16}. 
 On this basis, and 
 properly taking into account the relevant experimental conditions (i.e. 
 detector solid angles, efficiencies, beam current and target thickness) 
 one estimates to observe $\approx$ 1 $\gamma$ event per day 
 in coincidence with $^{62}$Ni ions. 

 In Fig. \ref{fig:g2} one observes a distribution of events with energies 
 larger than those of the discrete $\gamma$ lines (all lying within the 
 1-2 MeV range). 
 The poor  statistics does not allow  to quantitatively  
 discuss and eventually identify the origin of the structure of the 
 event distributions, which may be due, for instance, to   
  random coincidences or to high energy $\gamma$ rays.
  The group of few events near 3-4 MeV 
 in the $^{62}$Ni spectrum is located, in average, 
 at an higher energy as compared to that of the corresponding group in 
 $^{114}$Sn, the difference between the centroids of the two groups 
 being a few hundreds of keV,   a consequence 
 of the Doppler corrections applied for the discrete $\gamma$ lines. 
 In the case in which one considers a $\gamma$ ray of e.g. $\approx$4 MeV, 
 produced by the dipole oscillation associated with the two-neutron 
 transfer channel, one should take into 
 account the fact that it would be emitted from a moving source corresponding 
 to the center of mass of the binary system. With the geometry of the 
 detectors used in \cite{Montanari:16}, the computed Doppler shift of the  
 $\approx$4 MeV $\gamma$ ray amounts to tenths 
 instead of hundredhs of keV. Such a value is too 
 small compared with the relative Doppler shift, and one does not have 
 sufficient sensitivity to fix an absolute reference, 
 having in mind that the predicted width 
 of the $\gamma$ ray distribution is of few MeV.

\subsection{Simulations}

 In building-up the structure of the measured 
 spectra, a basic role is played 
 not only by the complex convolution of processes 
 of $\gamma$-ray interactions with the Germanium crystals 
 (e.g. Compton scattering) 
 but also by the finite size of the $\gamma$ array. The  experiment reported in \cite{Montanari:16} was 
 performed with the AGATA Demonstrator, an array which at the time of the 
 measurement included four triple clusters and therefore covered only a 
 limited fraction ($\approx$ 7-8\%) 
 of the full solid angle. Thus, the reconstruction of 
 the $\gamma$-ray energy via tracking procedure was significantly affected by 
 the details of the detector geometry,   
 in particular by those events hitting the outer 
 crystals and eventually leaking out from them. 
 Moreover, one also had to take into account the fact 
 that the detection efficiency 
 depends on the $\gamma$-ray energy, which distorts the final 
 event distribution. 

 A simulation of the spectra incorporating the above 
 mentioned features has been carried out by 
 reconstructing the $\gamma$-ray energy 
 via a tracking procedure. It considers 
 the $\gamma$-array characteristics and the kinematics of the binary reaction, 
 with the $\gamma$ rays emitted by the light and heavy reaction partners, 
 the former detected in PRISMA, the latter moving at the 
 correlated kinematic angle. The program has as input the gamma energies 
 and strengths of the discrete lines experimentally observed. The event 
 distribution, also at energies higher than the last observed discrete gamma 
 lines, has been generated by implementing 
 an exponential function 
 which takes into account the (measured) efficiency dependence 
 on the $\gamma$-ray energy.  
 The results for the inelastic channel are 
 shown in Fig. \ref{fig:sim1}, where the results of the simulations have been  
 normalized to the experimental data in the region of the discrete lines. 
 It is seen that simulations follow the structure of the data  
 rather accurately, 
 indicating the correctness of the array input parameters. 
 Making use of this inelastic channel 
 ( $^{60}$Ni) parameters, simulations have been also carried out 
 for $^{62}$Ni (two-particle transfer channel), with the absolute 
 strengths adjusted to reproduce the experimental spectrum, 
 obtaining an agreement similar to that observed in connection 
 with the inelastic channel, and a solid ground to study the gamma 
 distribution generated by the predicted two-nucleon transfer 
 dipole oscillations.

 Within this context, a simulation has been carried out using as 
 input the predicted 
 gamma-strength function  and employing 
 the same tracking algorithm used previously. It was assumed for 
 simplicity that the $\gamma$-ray angular distribution is isotropic. 
 It was also assumed that the $\gamma$ rays are emitted by a 
 non-moving source, which is justified by the fact that, 
 as said before, the Doppler shift is very small compared with the 
 predicted width of the distribution (few tenths of keV as compared to few MeV).

 In Fig. \ref{fig:tracked} the initial and final distributions are shown, 
 with the ratio of output (green histogram) and 
 input (blue histogram) reflecting the average efficiency 
 of the $\gamma$-array. 
 It is seen  that the original distribution, peaked at 4 MeV 
 becomes asymmetric, its centroid shifting to  
 lower energies (about 2.5 MeV). 
 This is mainly due to the energy dependence of the detectors 
 efficiency.  A peak near the 200-300 keV region is observed
 mainly as a consequence of Compton scattering due to the 
 finite size of the array. 

 The shape of the final distribution turns out to be distorted 
 towards lower energies, overlapping with the region of discrete 
 $\gamma$ lines while preserving the exponential high energy tail. 
 In Fig. \ref{fig:tracked2} we show the spectrum of 
 $^{62}$Ni where we include the original data (black), 
 the simulation previously discussed (green) and the simulated distribution 
 of the predicted strength function (magenta). Within the limitation of the 
  poor statistics and the consequent large fluctuations in the 
 final number of counts and their location in the spectrum,  
 these results allow to 
 conclude that the event distribution of the data is not in 
 contradiction with the theoretical predictions.

\section{Conclusions}
Studied with the specific probe, namely two-nucleon transfer reactions, e.g. $^{116}$Sn+$^{60}$Ni$\to$ $^{114}$Sn(gs)+$^{62}$Ni(gs), nuclear Cooper pairs display dimensions much larger than $r_0\,((13.5$ fm/1.2 fm)$^{3}\approx10^3)$, but also than the nuclear radius ($R=r_0A^{1/3}$  $\approx 6$ fm ($A\approx116$)), and the parallel with low temperature superconductivity in metals ($(\xi/r_s)^3\approx10^{12}$, $r_s$ being the Wigner-Seitz radius), becomes clearer.

The very large value of the correlation length displayed by the partner fermions of Cooper pairs as compared to the interparticle distance in both metals and nuclei, is at the basis of the fact that successive is the main Cooper pair tunneling mechanism ($P_2\approx P_1$) through a weak link between two superconductors. Also in heavy ion reactions  between two superfluid nuclei, at energies below the Coulomb barrier.

The associated tunneling amplitude in second order of the transfer, mean field potential, results from the integration of the nuclear form factor $F(\mathbf r_{O1},\mathbf r'_{O2},\mathbf r_{Cc})$ and of the phase function $PF=\exp(iQ_{2n}t/\hbar)$, along the trajectory of relative motion. Phase function which oscillates  with time, as compared with the smooth exponential behaviour of $F$ (see Fig. \ref{fig:form}). The effect of the rapidly alternating sign of $PF$ for large values of $Q_{2n}$ as compared to the smooth behaviour with time of $F$, results in a strong cancellation of the second order amplitude. Cancellation which parallels the one found between the positive and negative values of the function $\sin\left(2\phi_0+\frac{2e}{\hbar}V_0 t\right)$ describing the alternating current (ac) Josephson effect, which averages out essentially to zero. Within this context, one can posit that the action of the $PF$ leads to an alternating single Cooper pair current of effective charged neutrons which, as the electron Cooper pairs in a Josephson weak link, radiates photons.

The fact that the photon line width of a Josephson link is essentially zero while that of the Josephson-like nuclear $\gamma$-strength function is  few MeV wide, reflects the fact that, after all, nuclei are not infinite systems, displaying recoil effects to Cooper pair transfer,  and  that the nuclear link lasts a split second.

The energy to substain the radiation is provided by the barrier biasing battery in the case of two metallic superconductors, and by the relative motion in a heavy ion collision when the $Q$-value of the reaction is negative as for example in the case of the ($1n$) process (Sect. \ref{Sect.14.D}), or where $E_\gamma>Q_{2n}$ as in the ($2n$) process (Sect. \ref{Sect.14.A}), as testified by the energy-conserving $\delta$-function $\delta(E_i-E_\gamma-E_f+Q_{2n})$ appearing in the expression of the double differential cross sections $\frac{d^2\sigma^{\gamma}_{1n}}{d\Omega\,dE_\gamma}$ and $\frac{d^2\sigma^{\gamma}_{2n}}{d\Omega\,dE_\gamma}$.

Introducing the coupling of the effectively charged neutrons (dipole approximation) to the electromagnetic field, in the second order DWBA $T$-matrix describing a single Cooper pair transferred between two superfluid nuclei, the nuclear quantities entering the Josephson-like $\gamma$-radiation effect, namely the critical current $J^N_c$, the equivalent critical bias potential $V^N_{eq}$ and the barrier resistance $R^N_{b}$, have the same dimensions as their condensed matter counterparts. While it seems natural to introduce the dipole-electromagnetic field coupling when dealing with Cooper pair tunneling through a barrier in the case of electrons, this has not been so in the case of effectively charged neutrons, not even in the case in which the transferred particles were  protons.

 Studied with its specific probe, namely Cooper pair transfer (tunneling), the parallel between weakly coupled nuclei and superconductors carried out above, allows for the use  of simple concepts like current and resistance within the nuclear framework --instead of successive transfer through virtual states, and microscopic damping in terms of selected doorway virtual states-- to advance  the understanding of pairing in nuclei, as well as to make predictions --$\gamma$-strength function-- which can be experimentally tested.


\section{Hindsight}
Let us consider a two-nucleon transfer process induced by a heavy ion reaction, and concentrate on the dominant, successive transfer contribution, that is
\begin{align}
  \label{over_eq:10}
  a(=b+2)+A\to f(=b+1)+F(=A+1)\to b+B(=A+2)
\end{align}
Making use of the notation $\alpha\equiv(a,A),\gamma\equiv(f,F)$ and $\beta\equiv(b+B)$. one can write the successive transfer amplitude in the semiclassical approximation (\cite{Broglia:04a} p. 306 Eq. (23)), as
\begin{align}
  \label{over_eq:11}
    \nonumber (a_\beta(t))_{succ}&=\left(\frac{1}{i\hbar}\right)^2\sum_{\gamma\neq\beta}
  \int_{-\infty}^{t}dt'\,\braket{\Psi_{\beta}|V_{\beta}-\braket{V_{\beta}}|\Psi_{\gamma}}_{\mathbf R_{\beta,\gamma}(t')}\\
  &\times \int_{-\infty}^{t'}dt''\,\braket{\Psi_{\gamma}|V_{\alpha}-\braket{V_{\alpha}}|\Psi_{\alpha}}_{\mathbf R_{\gamma,\alpha}(t'')},
\end{align}
where the quantal $cm$ coordinate $\mathbf r_{\beta\gamma}=(\mathbf r_{\beta}+\mathbf r_\gamma)/2$ should be identified with the average classical coordinate,  i.e. $\mathbf r_{\beta\gamma}\to \mathbf R_{\beta\gamma}=(\mathbf R_{\beta}+\mathbf R_\gamma)/2$ which, together with   $\mathbf v_\beta=\mathbf {\dot R}_\beta$ and similar for the channel $\gamma$, are assumed to describe the motion of the centers of the wavepackets, and satisfy the classical equations $m_\beta \mathbf {\dot v_\beta}=-\pmb \nabla_\beta U(\mathbf R_\beta)$. The functions $\Psi$ describe the structure of the nuclei and,  e.g. $\Psi_\alpha=e^{-i\frac{E_\alpha}{\hbar}t}\psi_\alpha$ while $\psi_\alpha=\psi^a(\xi_a)\psi^A(\xi_A)$ is the product of the intrinsic BCS wavefunctions describing the structure of nuclei $a$ and $A$, $\xi$ being the corresponding intrinsic variables. One can rewrite (\ref{over_eq:11}) as,
\begin{align}
  \label{over_eq:12}
    \nonumber (a_\beta(t))_{succ}&=\left(\frac{1}{i\hbar}\right)^2\sum_{\gamma\neq\beta}
  \int_{-\infty}^{t}dt'\,\braket{\Psi_{\beta}|V_{\beta}-\braket{V_{\beta}}|\psi_{\gamma}}_{\mathbf R_{\beta,\gamma}(t')}\\
  &\times \int_{-\infty}^{t'}dt''\,\braket{\psi_{\gamma}|V_{\alpha}-\braket{V_{\alpha}}|\Psi_{\alpha}}_{\mathbf R_{\gamma,\alpha}(t'')}  e^{i\frac{E_\gamma}{\hbar}(t''-t')},
\end{align}
where
\begin{align}
  \label{over_eq:18} \exp\left(i\frac{E_\gamma}{\hbar}(t''-t')\right)=\cos\left(\frac{E_\gamma}{\hbar}(t''-t')\right)+i\sin\left(\frac{E_\gamma}{\hbar}(t''-t')\right).
\end{align}
Assuming the quasiparticle excitation energies $E_\gamma$ to be much larger than the reaction $Q-$value, the periodic functions in (\ref{over_eq:18}) oscillate so rapidly, that the outcome of the integration in (\ref{over_eq:11}) amounts to nothing unless $t'\approx t''$. In other words and further assuming that the matrix elements (formfactors) are smooth functions of time along the trajectories of relative motion one can write $ \sum_{\gamma\neq \beta}\exp\left(i\frac{E_{\gamma}}{\hbar}(t''-t')\right)\approx \frac{1}{\Delta E}\int d E \exp\left(i\frac{E}{\hbar}(t''-t')\right)\approx\frac{\hbar}{\Delta E}\times\delta(t''-t')$, where $1/\Delta E$ is the average density of levels of the two-quasiparticle states.  Making use of Eq. (\ref{over_eq:9}), one can write Eq. (\ref{over_eq:12}) as,
\begin{align}
  \label{over_eq:13}
  \nonumber &(a_\beta(t))_{succ}=\left(\frac{1}{i\hbar}\right)^2 \int_{-\infty}^{t}dt'\braket{\psi_{\beta}|(V_{\beta}-\braket{V_{\beta}})|\psi_{\gamma}}_{\mathbf R_{\beta,\gamma}(t')}\\
  &\times \frac{\hbar}{ \Delta E}\, {}_{\mathbf R_{\gamma,\alpha}(t')}\braket{\psi_{\gamma}|V_{\alpha}-\braket{V_{\alpha}}|\psi_{\alpha}}\,e^{i\frac{Q_{2n}}{\hbar}t'},    
\end{align}
where $(Q_{2n}/\hbar)=\omega^N_J$ is the nuclear parallel to the angular Josephson frequency, while $Q_{2n}/(2\times e_{eff})=V^N$ is the nuclear equivalent to the weak link direct current bias. The quantity $e_{eff}$ is the effective charge of a nucleon which, for the reaction $^{116}{}_{50}$Sn+$^{60}_{28}$Ni$\to$ $^{114}$Sn+$^{62}$Ni is $e_{eff}=-e((50+28)/(114+62))=-e\times0.44$. It is of notice that the fact that (\ref{over_eq:11}), and thus (\ref{over_eq:13}) has the right gauge phasing, is  in keeping with the fact that the two-particle reaction mechanism at the basis of (\ref{over_eq:12}) has been worked out fully quantum mechanically in a particle conserving, gauge invariant representation.

Defining
\begin{align}
  \label{over_eq:14}
     T_{\beta\alpha}=\left((V_{\beta}-\braket{V_{\beta}})\ket{\psi_{\gamma}}_{\mathbf R_{\beta,\gamma}(t')}\,\frac{\hbar}{\Delta E}\, {}_{\mathbf R_{\gamma,\alpha}(t')}\bra{\psi_{\gamma}}(V_{\alpha}-\braket{V_{\alpha}})\right)      
\end{align}
one obtains
\begin{align}
  \label{over_eq:15}
   \nonumber   (a_\beta(t))_{succ}&=\left(\frac{1}{i\hbar}\right)^2
  \int_{-\infty}^{t}dt'\braket{\psi_{\beta}|T_{\beta\alpha}(t')|\psi_{\alpha}}\\
  &\times(\cos\omega_J^N t'+i\sin\omega_J^N t').    
\end{align}
The argument of the time integral can thus be viewed as the product of the two-nucleon transfer amplitude, modulated by the phase function $\exp(i\omega_J^N t')$, e.g. by the $\sin\omega_J^N t'$ periodic function. It implies that the two-particle transfer amplitude rises rather fast in the first quarter of the first period for then to decrease somewhat  slower for the next quarter, after which $\sin\omega_J^N t'$ changes sign. From there on the flux describes a pair of nucleons moving back to the original configuration. That is, a pair being transferred successively to the initial state. Until the next sign change makes the pair move in the original direction, i.e. towards the final configuration. In other words, one is in the presence of a transient and thus exponentially attenuated  alternating supercurrent of effective charged pair of neutron carriers and frequency $\omega_J^N$, which parallels the (ac) Josephson effect, the expression (\ref{over_eq:11})  containing from the start, the appropriate gauge phasing.

Inserting the dipole operator Eq. (\ref{eq:60}) in (\ref{over_eq:11}), the argument of (\ref{over_eq:15}) becomes
\begin{align}
  \label{over_eq:16}
  \braket{\psi_\beta|T_{\beta\alpha}(t')|\psi_\alpha}\mathbf D_{m_\gamma}(\cos\omega_J^N t'+i\sin\omega_J^N t').
\end{align}
Consequently, there would be emission of a $\gamma$-ray of energy $\hbar\omega_J^N$, each time a Cooper pair tunnels through the junction. The second order DWBA $T$-matrix amplitude integrated as a function of the relative center of mass coordinate is equivalent to (\ref{over_eq:11}), integrated as a function of time. One  then expects to observe  the (ac) nuclear Josephson effect (Fig. \ref{abs_fig:19}), when the tunneling (transfer) time for a Cooper pair ($\hbar/|Q_{2n}|$) \cite{Hara:71}, is shorter than the collision time ($\tau_{coll}\approx0.6\times 10^{-21}$ s),  the situation for $Q_{2n}=1.307$ MeV ($\hbar/Q_{2n}\approx0.5\times10^{-21}$ s) being marginal, the $T$-matrix amplitude displaying  one quarter of a cycle.


Within the above scenario and that of a coarse grained approximation one can, arguably, describe the nuclear parallel to the (dc) Josephson effect as an (ac) one with $Q_{2n}\approx0.1$ MeV (in which case $\left(\frac{Q_{2n}}{\hbar}\right)^{-1}=0.7\times10^{-20}$ s, an order of magnitude longer than $\tau_{coll}$).

Inserting in (\ref{over_eq:11})  the square of the absolute value of the relative distance between the partner nucleons of the transferred Cooper pair instead of the dipole moment, the argument of (\ref{over_eq:15}) becomes
\begin{align}
  \label{over_eq:17}
\braket{\psi_\beta|T_{\beta\alpha}(t')|\psi_\alpha}|\mathbf r_{O1}-\mathbf r_{O2}|^2(\cos\omega_J^N t'+i\sin\omega_J^N t').  
\end{align}
Extending the integration over $t'$ to $+\infty$ (equivalent for the $T$-matrix to do similarly for the relative center of mass coordinate), one can determine the Cooper pair mean square radius. In the case of the reaction $^{116}$Sn+$^{60}$Ni$\to$ $^{114}$Sn+$^{62}$Ni at $E_{cm}=154.26$ MeV ($(D_0)_c=13.49$ fm). A numerical difference which can be used to ascribe an error to the theoretical estimate of the correlation length, that is $\xi\approx12\pm1.5$ fm. 
\section{Acknowledgments}
This work was performed under the auspices of the U.S. Department of Energy by Lawrence Livermore National Laboratory under Contract No. DE-AC52-07NA27344. This work is part of the I+D+i project with Ref. PID2020-114687GB-I00, funded by MCIN/AEI/10.13039/501100011033.


\begin{appendix}
\section{Rotations in gauge space}\label{rotgauge}
The BCS wavefunction does not have a fixed number of particles and reflects a privileged orientation in gauge space (Fig. \ref{abs_figI}). It can be viewed as the intrinsic state of a pairing rotational band. Its expression in the laboratory system is,
\begin{align}
  \label{abs_eq:16}
  \ket{BCS(\phi)}_{\mathcal K}=\mathcal G^{-1}(\phi)\,\ket{BCS(0)}_{\mathcal K'}=\prod_{\nu>0}\left(U_\nu+V_\nu e^{-i2\phi} a^\dagger_\nu a^\dagger_{\tilde\nu}\right)\ket{0}.
\end{align}
It is obtained  by rotating the intrinsic BCS wavefunction  through an angle $\phi$ in gauge space making use of the inverse of the operator
\begin{align}
  \label{abs_eq:17}
  \mathcal G(\phi)=e^{-i\phi},
\end{align}
and of the relation
\begin{align}
  \label{abs_eq:19}
  a^{\dagger}_\nu = \mathcal G^{-1}(\phi)\,a^{'\dagger}_\nu\, \mathcal G(\phi)=e^{i\phi}a^{'\dagger}_\nu.
\end{align}

The state described by the BCS wavefunction can be viewed as an axially symmetric rotor in the two-dimensional (gauge) space, where the particle number operator $\hat N=\sum_\nu a^{\dagger}_\nu a_\nu$ plays the role of the angular momentum operator in the case of rotations in 3D-space. The state (\ref{eq:52}) is the $N$th member of a pairing rotational band (see \cite{Bohr:64,Bohr:76,Bohr:75,Hogassen:61,Bes:66,Papenbrock:22}, and references therein).

The gauge angle $\phi$ and the particle number operator satisfy the commutation relations $[\phi,\hat N]=i$ and, in the number representation
\begin{align}
  \label{abs_eq:20}
  \phi=i\frac{\partial}{\partial N}.
\end{align}
The Hamiltonian for the BCS pairing problem is,
\begin{align}
  \label{abs_eq:21}
  H=H_0-\lambda \hat N,
\end{align}
where $H_0$ includes the kinetic energy of the nucleons and the pairing interaction, and $\lambda$ is a Lagrange multiplier which is used to fix the average number of nucleons. Physically it is the Fermi energy and is determined by measuring the energy change of the system when adding or substracting particles. For example the change in energy of the nucleus when an even number $\delta N$ of nucleons is added is
\begin{align}
  \label{abs_eq:22}
  \delta\braket{E}\approx\lambda\delta N.
\end{align}
This is in keeping with the fact that the time derivative of the gauge angle is given by
\begin{align}
  \label{abs_eq:23}
  \dot\phi=\frac{i}{\hbar}[H,\phi]=\frac{1}{\hbar}\frac{\partial H}{\partial N}=\frac{1}{\hbar}\lambda.
\end{align}
Thus the combination
\begin{align}
  \label{abs_eq:24}
  2\hbar\dot\phi=2\lambda=\delta\braket{E}
\end{align}
has the physical meaning of the change in energy of the nucleus when a pair ($\delta N$) of nucleons is added to it. Taking the derivative with respect to $N$
\begin{align}
  \label{abs_eq:25}
  \frac{\partial \dot\phi}{\partial N}=\frac{1}{\hbar}\frac{\partial \lambda}{\partial N}=\frac{1}{\hbar}\frac{\partial^2H}{\partial N^2}=\frac{\hbar}{\mathcal J}.
\end{align}
This equation defines the ``pairing moment of inertia'' $\mathcal J$ describing rotation in gauge space. It can also be written as
\begin{align}
  \label{abs_eq:26}
  \frac{\mathcal J}{\hbar^2}=\frac{\partial N}{\partial \lambda}.
\end{align}
If $E(N_0)$ is the energy of a nucleus with $N_0$ nucleons and if the pairing moment of inertia is approximately constant with $N$, then Eq. (\ref{abs_eq:25}) can be integrated to give the energy of a pairing rotational band
\begin{align}
  \label{abs_eq:27}
  E(N)\approx E(N_0)+\lambda(N-N_0)+\frac{\hbar^2}{2\mathcal J}(N-N_0)^2.
\end{align}
In the case of the pairing rotational band (see Fig. \ref{abs_figII}) associated with the ground state of the Sn-isotopes, and making use of the single $j$-shell model expression, in which case 
\begin{align}
  \label{abs_eq:28}
\frac{\hbar^2}{2\mathcal J}=\frac{G}{4}=\frac{25}{4N_0}=0.092\text{ MeV};\quad(N_0=68)  ,
\end{align}
one can write
\begin{align}
  \label{abs_eq:29}
  E(N)=E(68)+8.124\text{ MeV}\times(N-68)+0.092\text{ MeV}\times(N-68)^2.
\end{align}
Thus, for example (Eq. (\ref{abs_eq:5})),
\begin{align}
  \label{abs_eq:30}
  E(70)-E(68)\approx 16.248\text{ MeV}+\mathcal O(\hbar^2/2\mathcal J)
\end{align}
and
\begin{align}
  \label{abs_eq:31}
  \dot\phi\approx\frac{8.124\text{ MeV}}{\hbar}\approx 12.2\text{ ZHz}
\end{align}
essentially corresponds to the rotational frequency in gauge space associated with the Coriolis force (Eq. (\ref{abs_eq:21})) felt by the system $^{120}$Sn in the intrinsic frame of reference $\mathcal K'$.

It is of notice that $G\alpha_0'=\Delta'$. In the nuclear case $\Delta'\approx1.5$ MeV and $\alpha_0'\approx 6$. Thus $G\approx0.3$ MeV and $\left(\hbar^2/2\mathcal J\right)\times 4\approx0.3 $ MeV(0.3 MeV/16.248 MeV$\approx 2$\%). In the case of metallic superconductors, $\Delta\approx1.4$ meV (lead), while the number of Cooper pairs participating in the BCS condensate is a macroscopic quantity ($\approx10^{20}$), and the rotational term $(G/4)\times 4$ can be considered negligible. As a consequence $V=(\lambda_1-\lambda_2)/e$, and the energy difference for the transfer of a Cooper pair is $\hbar\omega_J=2e\times V\approx2(\lambda_1-\lambda_2)=2\times\hbar(\dot\phi_1-\dot\phi_2)$ which, for all purposes can be used as an identity. Within this context, because the phase  of a quantum state evolves in time as $\exp\left(i\frac{E}{\hbar}t\right)$, and the tunnel current arises from a transition between a state in which a pair is on one side of the junction and a state in which it is on the other side of it, the relative phase function of the two states will oscillate at an angular frequency $\omega_J$ corresponding to the energy difference $\hbar\omega_J=2(\lambda_1-\lambda_2)$. Consequently, and in keeping with the fact that supercurrents are lossless, each time a Cooper pair tunnels through the junction a $\gamma$-ray of this energy will be emitted (See e.g. \cite{Tinkham:96}).
\section{Nuclear pairing rotational bands and Nernst theorem}\label{AppNernst}
The classical expression of the heat capacity of a mole of many solids --heat required to rise it by 1K-- is constant (about 3$R$, where $R$ is the gas constant) for temperatures far from absolute zero (Dulong-Petit law), a result associated with the connection of heat capacity and lattice vibrations of the solid. Conspicuous deviations from this law are observed at low temperatures, close to absolute zero (-273.15 $^\circ$C).

The total energy of a substance can be obtained by measuring this specific heat. However, to calculate the free energy from the total energy, a further assumption is needed, known as Nernst theorem or third law of thermodynamics. Its significance for low temperature physics is testified  by the following formulations. One says that, while absolute zero can be approached to an arbitrary degree, it can never be reached. The other says that at absolute zero temperature the entropy becomes zero.

The total energy is then the sum of the thermal energy and of the zero point energy ($1/2 h \nu_0$ per degree of freedom), a quantity which does not play a significant role at  ``high'' ($kT/h\nu_0\geq 1)$ temperatures.

The existence of the zero point energy must make itself felt in the behaviour of matter because it can not be taken out of --given away from-- the substance, and therefore cannot take part in energy exchanges. Thus, with Nernst assumption one can identify the range of low temperature physics as that in which the zero point energy is the dominant part in the energy content of the substance\footnote{Planck was one of the first to recognize the third law and thus zero point energy, as the most important manifestation of the quantum principle.}.

Within this context the process (\ref{abs_eq:6abs}) between the corresponding members of the ground state (gs) rotational bands, i.e.
\begin{align}
  \label{eq:104}
    ^{116}\text{Sn}+^{60}\text{Ni}\to^{114}\text{Sn (gs)}+^{62}\text{Ni (gs)}\quad(Q_{2n}=1.307\text{ MeV}),
\end{align}
can be viewed as paradigmatic examples of Nernst hypothesis. 

The pairing Hamiltonian describing the motion of independent fermions, e.g. of like nucleons, interacting through a pairing force of constant matrix elements $G$, can be written as the sum of a deformed ($\alpha_0=\braket{BCS|\sum_{\nu>0}P_\nu^\dagger|BCS}$; $\Delta=G\alpha_0$) mean field term in gauge space which has as lowest energy solution the $\ket{BCS}$ state, and a fluctuating term leading to pairing vibrations with energy $\geq 2\Delta$, and pairing rotations which admits a zero energy solution ($\hbar\omega_1''=0$)
\begin{align}
  \label{eq:105}
  \ket{1''}=\frac{\Lambda_{1''}}{2\Delta}\left(\hat N-N_0\right)\ket{0''},
\end{align}
which restores gauge invariance (see e.g. \cite{Brink:05} Sect. 4.2 p. 75 and App. I, Sect. I.4 p. 330). In other words, a RPA solution where $\ket{0''}$ is the vacuum of the RPA Hamiltonian and in which $\hat N$ is the number of particles operator written in terms of quasiparticles, $N_0=\sum_{\nu}2V_\nu^2$ is the mean field (BCS) average number of particles, while the prefactor ($\Lambda_{1''}/2\Delta$) is the zero point amplitude of the mode, that is ($\frac{\hbar^2}{2D_{1''}}\frac{1}{\hbar\omega_{1''}}$), where $D_{1''}$ stands for the mass parameter (pairing moment of inertia $\mathcal J$, Eq. (\ref{abs_eq:25}) of the mode). Because $\Lambda_{1''}\to\infty$ ($\hbar\omega_{1''}=0$), this can only happen if in Eq. (\ref{eq:105}) ($\hat N-N_0)\ket{0''}\to0$, implying that particle number conservation is restored in the RPA approximation. Because $\hbar \omega_{1''}=\hbar\sqrt{\frac{C_{1''}}{D_{1''}}}$ and $D_{1''}(=\mathcal J)$ is finite, the above result implies that the restoring force $C_{1''}=0$. In other words, all orientations in gauge space have the same energy, and the one phonon state (\ref{eq:105}) results from rotations of the intrinsic deformed $\ket{BCS}$ state of divergent amplitudes,   that is, fluctuations of $\phi$ over the whole $0-2\pi$ range of values (Eq. \ref{eq:52}), the resulting states 
\begin{align}
  \label{eq:106}
      \ket{N}\sim\int_0^{2\pi} d\phi\,e^{-iN\phi}\,\ket{BCS_{\mathcal K'}(\phi)}\sim \left(c'_\nu P^\dagger_\nu\right)^{N/2}\ket{0''},
\end{align}
where $c'_\nu=U_{nu'}/V_{\nu'}$, are the members of the pairing rotational band (Eq. (\ref{eq:104})), whose energy is all zero point energy. As the system is in its ground state, it is also at absolute zero temperature that is, the system characterized by Eq. (\ref{eq:105}) has zero entropy. It is then not surprising that the associated $\gamma$-strength function line shape is Gaussian (Fig. \ref{notes_fig5}).   
\section{Degrees of freedom of rotations in 3D-space}\label{AppH}
The orientation of a body in three-dimensional space involves three angular variables, such as the Euler angles, $\omega=\phi,\theta,\psi$, and three quantum numbers are needed in order to specify the sate of motion. The total angular momentum $I$ and its component $M=I_z$ on a space-fixed axis (laboratory system of reference) provide two of these quantum numbers. The third may be obtained by considering the components of $\mathbf I$ with respect to an intrinsic (body-fixed) coordinate system with orientation $\omega$ with respect to the laboratory system. One calls $K$ the projection of $\mathbf I$ on the corresponding 3-axis  ($z'$-axis) (Fig. \ref{abs_figI}). The unsymmetrized wavefunction associated with a member of the corresponding rotational band can be written as \cite{Bohr:75},
\begin{align}
  \label{abs_eq:8}
  \Psi_{KIM}=\left(\frac{2I+1}{8\pi^2}\right)^{1/2}\Phi_K(q)\mathcal D_{MK}^I(\omega),
\end{align}
where $\Phi_K(q)$ is the intrinsic state (e.g. Nilsson state \cite{Nilsson:55}) which, for simplicity one denotes $\ket{K}$.

An axially symmetric quadrupole deformation can be characterized by the intrinsic electric quadrupole moment
\begin{align}
  \label{abs_eq:9}
  \nonumber eQ_0&=\braket{K|\int \rho_e(\mathbf r') r'^2(3\cos^2\theta'-1)d\tau'\,|K}\\
  &=\left(\frac{16\pi}{5}\right)^{1/2}\braket{K|\mathcal M(E2,\nu=0)|K}.
\end{align}
The prime coordinates refer to the intrinsic (body-fixed) system, and $\mathcal M(E2,\nu)$ denotes the components of the electric quadrupole tensor relative to the intrinsic system.

The $E2$ moments $\mathcal M(E2,\mu)$ referred to the laboratory axis can be written as 
\begin{align}
  \label{abs_eq:10}
  \mathcal M(E2,\mu)=\sum_\nu \mathcal M(E2,\nu)\mathcal D^2_{\mu\nu}(\omega)=\left(\frac{5}{16\pi}\right)^{1/2}e Q_0\mathcal D^2_{\mu0}(\omega).
\end{align}
The $E2$-matrix elements between two members of a rotational band ($\Delta I\leq2$) can be evaluated  by employing the relations
\begin{align}
  \label{abs_eq:11}
  \nonumber  \int_0^{2\pi}\sin\theta & d\theta\int_0^{2\pi}d\phi\int d\psi \mathcal D^{I*}_{MM'}(\omega)\mathcal D^{I_1}_{M_1M_1'}(\omega)\\
  &=\frac{8\pi^2}{2I+1}\delta(I,I_1)\delta(M,M_1)\delta(M',M_1'),
\end{align}
and
\begin{align}
  \label{abs_eq:12}
  \nonumber \sum_{M_1 M_2}&\braket{I_1\,M_1\,I_2\,M_2|I\,M}\mathcal D^{I_1}_{M_1M_1'}(\omega)\mathcal D^{I_2}_{M_2M_2'}(\omega)\\
  &=\braket{I_1\,M'_1\,I_2\,M'_2|I\,M'}\mathcal D^{I}_{MM'}(\omega),
\end{align}
leading to
\begin{align}
  \label{abs_eq:13}
  \braket{KI_2||\mathcal M (E2)||KI_1}=(2I_1+1)^{1/2}\braket{I_1\,K\,2\,0|I_2\,K}\left(\frac{5}{16\pi}\right)^{1/2}eQ_0,
\end{align}
and
\begin{align}
  \label{abs_eq:14}
  \nonumber B(E2;&KI_1\to K I_2)=(2I_1+1)^{-1}| \braket{KI_2||\mathcal M (E2)||KI_1}|^2\\
  &=\frac{5}{16\pi} e^2Q_0^2\braket{I_1\,K\,2\,0|I_2\,K}^2.
\end{align}
The integral (\ref{abs_eq:11}) and summation (\ref{abs_eq:12}) result in a Clebsch-Gordan coefficient and a rotation function, implying angular momentum conservation in the quadrupole electromagnetic decay along a quadrupole rotational band. An important consequence of symmetry relations,  as the factor $5/16\pi$ (phase space relation) together with  the distribution of the single-particle orbits in the intrinsic Nilsson state $(\rho_e(\mathbf r)=\sum_{i\in\pi}|\varphi_i(\mathbf r)|^2$; $(i\equiv[Nn_3\Lambda\Omega])$) determines the associated $BE2$-values. In other words, the physics at the basis of the large $BE2$-values between members of a rotational band  is to be found in the spontaneous symmetry breaking of rotational invariance taking place in the intrinsic system. The $\omega$-projection provides the elements to express the Nilsson predictions in the laboratory reference frame, where they can be measured.
  \section{(ac) Josephson effect}\label{AppB}
  Each of the two weakly coupled superconductors can be viewed as a rotor which defines a privileged orientation  in the two dimensional gauge space. The associated deformations are measured by the order parameters $\alpha_i=(\alpha_0')_ie^{-2i\phi_i}$ $(i=1,2)$ (Fig. \ref{fig:7} (E)).

  The zero point frequency of the rotors is defined by the chemical potential of the superconductor,
  \begin{align}
    \label{eq:37}
    \dot\phi_i=\frac{1}{\hbar}\frac{\partial H}{\partial N_i}=\frac{1}{\hbar}\mu_i.\quad(i=1,2).
  \end{align}
  In gauge space, number of particles play the role of angular momentum in 3D-space . Let us for simplicity neglect the second term in (\ref{eq:14}) and consider
  \begin{align}
    \label{eq:35}
    \delta\dot\phi=\dot\phi_1-\dot\phi_2=\frac{1}{\hbar}(\mu_1-\mu_2)=\frac{e}{\hbar}V.
  \end{align}
  Thus\footnote{Here we follow \cite{Feynman:63} Vol. 3.}, 
  \begin{align}
    \label{eq:38}
    \delta\phi(t)=\phi_0+\frac{e}{\hbar}\int V(t)\,dt.
  \end{align}
  In the case in which the junction is biased by a dc voltage $V_0$ one can write (Eqs. (\ref{eq:16}), (\ref{special_eq:7})),
  \begin{align}
    \label{eq:39}
    J=J_c\sin\left(2\phi_0+\frac{2e}{\hbar}V_0 t\right).
  \end{align}
  For a bias $V_0\approx1$ mV, one obtains for the prefactor\footnote{Making use of $\hbar=6.58\times10^{-16}$eV$\times$s and $V_0=1$mV, one obtains for the prefactor of $t$ in (\ref{eq:39}) $\omega_J=2eV_0/\hbar=3\times10^{12}$ s$^{-1}$=3THz.} of $t$ in Eq. (\ref{eq:39}), $3\times10^{12}$ s$^{-1}$ (3 THz) (=$\omega_J$, angular frequency). Consequently, the function $\sin\left(2\phi_0+(2eV_0/\hbar)t\right)$ oscillating very rapidly between positive and negative values averages out to essentially zero. A way to observe this high frequency ac is by detecting the microwave photons of frequency $\nu_J=2eV_0/h$ emitted by the junction.

  Another way to observe the ac Josephson current is by applying, in addition to the dc voltage a weak ($V_1\ll V_0$) voltage at a very high frequency to the junction, namely
  \begin{align}
    \label{eq:40}
    V=V_0+V_1\cos\omega t
  \end{align}
  in which case (\ref{eq:38}) becomes
  \begin{align}
    \label{eq:41}
    \delta\phi(t)=\phi_0+\frac{e}{\hbar}V_0t+\frac{e}{\hbar}V_1\int\cos\omega t\,dt=
    \phi_0+\frac{e}{\hbar}V_0 t+\frac{e}{\hbar}
    V_1\frac{\sin\omega t}{\omega}
  \end{align}
  Thus\footnote{This is in keeping with the fact that for $\Delta x\ll1$, $\sin(x+\Delta x)=\sin x+\Delta x\cos x$ $(\sin\Delta x\approx\Delta x, \cos\Delta x\approx 1)$.}
  \begin{align}
    \label{eq:42}
    \sin2\delta\phi(t)=\sin\left(2\phi_0+\frac{2e}{\hbar}V_0 t\right)+\frac{2eV_1}{\hbar\omega}\sin\omega t\cos\left(2\phi_0+\frac{2e}{\hbar}V_0 t\right).
  \end{align}
  The first term coincides with (\ref{eq:39}) and is zero in average. Concerning the second, and assuming $\omega=2eV_0/\hbar=\omega_J(\equiv2\pi\nu_J$, where $\nu_J=K_JV$ is the Josephson frequency, $K_J=2e/h$ being the Josephson constant, inverse of the fluxon, the flux quantum), one finds
  \begin{align}
    \label{eq:43}
    \left(\frac{V_1}{V_0}\right)\left(\cos2\phi_0(\sin\nu_Jt)\times(\cos\nu_Jt)-(\sin2\phi_0)\times(\sin^2\nu_Jt)\right). 
  \end{align}
  While the first term of (\ref{eq:43}) averages to zero, the second one does not. The practical way to implement the modulation described in Eqs. (\ref{eq:40}) and (\ref{eq:43}) is to bathe the Josephson junction in microwave radiation (Fig. \ref{fig:B1}).
  Within this context, one can calculate   the supercurrent with the help of (\ref{eq:42}),
  \begin{align}
    \label{eq:44}
    \nonumber     J=&J_c\sin2\delta\phi(t)\\
    &=J_c\times \text{Im}\left(\exp\left(i\left(2\phi_0+\frac{2e}{\hbar}V_0 t+\frac{2eV_1}{\hbar\omega}\sin\omega t\right)\right)\right).
  \end{align}
  Making use of the expansion
  \begin{align}
    \label{eq:45}
    \exp\left(i z\sin\alpha\right)=\sum_{k=-\infty}^\infty J_k(z)\cos k\alpha+\sum_{k=-\infty}^\infty J_k(z)\sin(k\alpha)
  \end{align}
  and the parity relations of the Bessel functions
  \begin{align}
    \label{eq:46}
    J_k(z)=(-1)^k J_{-k}(z),
  \end{align}
  it is possible to write the total current as,
  \begin{align}
    \label{eq:47}
    \nonumber I&=J+\frac{V_0}{R}=J_1\sum_{k=-\infty}^\infty (-1)^k J_k\left(\frac{2eV_1}{\hbar\omega}\right)\\
&\times\sin\left(2\phi_0+\frac{2e}{\hbar}V_0 t-k\omega t\right)+\frac{V_0}{R_b},    
  \end{align}
  where one has added the shunt current $V_0/R_b$ to the superconducting one. It is of notice that this is the dc part of the current since the {\it{sin}} part averages to zero, unless
  \begin{align}
    \label{eq:48}
    V_0=\frac{k\hbar\omega}{2e},\quad k=0,\pm 1, \pm 2,\dots
  \end{align}
  Then, the supercurrent has the dc component
  \begin{align}
    \label{eq:49}
    J=J_c (-1)^k J_k\left(\frac{2eV_1}{\hbar\omega}\right)\sin2\phi_0.
  \end{align}
  Summing up, the dc current increases linearly with $V_0$ as $V_0/R_b$, except when the potential is an integer multiple of $\hbar\omega/2e$, where the dc current suddenly jumps to the value given by (\ref{eq:49}). These are called Shapiro steps, and the result of an experiment is shown in Fig. \ref{fig:B2}.
  \section{Relation between bias and Josephson frequency}\label{AppE}
  The rotational frequency in gauge space of a superconductor or of a superfluid nucleus is defined by the relation
  \begin{align}
    \label{eq:99}
    \dot\phi=\frac{i}{\hbar}\left[H,\phi\right]=\frac{1}{\hbar}\frac{\partial\phi}{\partial\mathcal N}=\frac{1}{\hbar}\lambda,
  \end{align}
  the identification of the chemical potential $\lambda$ with the rotational angular frequency in gauge space is implicit in the description of the (mean field) pair correlation by the BCS Hamiltonian,
  \begin{align}
    \label{eq:100}
    H'=H-\lambda\mathcal N.
  \end{align}
  The last term (Coriolis force) can be associated with a transformation to a rotating reference system in which $\phi$ is stationary (body fixed frame)
  \begin{align}
    \label{eq:101}
    (\dot\phi)'=\frac{i}{\hbar}\left[H',\phi\right]=\frac{i}{\hbar}\left[H,\phi\right]-\frac{i}{\hbar}\lambda\left[\mathcal N,\phi\right]=\dot\phi-\frac{\lambda}{\hbar}=0.
  \end{align}
  While there is no reference frame for the gauge angle of an isolated superconductor, the gauge angle of one superconductor may be referred to that of another,
  \begin{align}
    \label{eq:102}
    \hbar\dot\phi_{rel}=\hbar(\dot\phi_l-\dot\phi_r)=\lambda_l-\lambda_r=e\times V_0.
  \end{align}
  Thus, the energy difference of a Cooper pair on one or the other side of the weak link is
  \begin{align}
    \label{eq:103}
    \Delta E_J=2\hbar\dot\phi_{rel}=2\hbar(\dot\phi_l-\dot\phi_r)=2(\lambda_l-\lambda_r)=2e\times V_0
  \end{align}
  The Josephson frequency is defined as
  \begin{align}
    \label{eq:54}
    \nu_J=\frac{2eV_0}{h},
  \end{align}
  the quantity
  \begin{align}
    \label{eq:93}
    V_0=\frac{h\nu_J}{2e},
  \end{align}
  being the biasing potential, while the energy difference of a Cooper pair on one and the other side of the junction and thus the energy of the emitted photons is
  \begin{align}
    \label{eq:94}
    h\nu_J=\hbar\omega_J=2eV_0.
  \end{align}
  For a bias potential $V_0=1$ mV, $h\nu_J=2$ meV, and\footnote{Making use of $h=4.136\times10^{-15}$ eV$\times s$ one obtains $\nu_J=2\times$ meV/(4.136$\times10^{-15}$ eV$\times$s)=0.484$\times10^{12}$ Hz.}
  \begin{align}
    \label{eq:95}
    \nu_J=0.484\times10^{12}\text{ Hz (THz)}.
  \end{align}
  It is of note that the Cooper pair binding energy is $\approx2\Delta$ which e.g. in the case of lead ($\Delta\approx1.4$ meV) amounts to 2.8 meV. Furthermore, and in keeping with (\ref{eq:95}) and text following Eq. (\ref{eq:16}), the range of frequencies one is dealing with is that of microwaves of extreme high frequency to the far infrared.
  
  Similarly, in the case of the reaction\footnote{$Q_{2n}=BE(^{114}\text{Sn})+BE(^{62}\text{Ni})-BE(^{116}\text{Sn})-BE(^{60}\text{Ni})$=1516.83 MeV-1515.53 MeV=1.307 MeV.} (\ref{abs_eq:6abs}),
  \begin{align}
    \label{eq:96}
    h\nu^N_J=2e_{eff}V_0^N=Q_{2n}=1.307\text{ MeV}
  \end{align}
  and ($e_{eff}=0.443e$)
  \begin{align}
    \label{eq:97}
    V_0^N=1.469\text{ MV}.
  \end{align}
  Thus\footnote{$\nu_J^N=1.307$ MeV/(4.136$\times10^{-15}$ eV$\times$ s)=0.316$\times 10^{21}$ Hz.}
  \begin{align}
    \label{eq:98}
    \nu_J^N=\frac{Q_{2n}}{h}=0.316\times10^{21}\text{ Hz (ZHz)}.
  \end{align}
  In the nuclear case, typical values of the Cooper pair binding energy are found in the range 2.4-3.0 MeV ($\Delta\approx1.2-1.5$ MeV), and (\ref{eq:98}) corresponds to gamma-rays. 
  \section{Relation between normal and abnormal density}\label{AppA}
  From the relation (\ref{eq:1}) defining the correlation length one can view the quantity $\pi\Delta$ as the Cooper pair binding energy.
  For a bias implying an energy $E<\pi\Delta$, one cannot talk about single-particle, fermionic motion, the entities describing the system being Cooper pairs. The bias thus acts on the center of mass of these entities (see Fig. \ref{fig:A1} (a)), which will move in a coherent fashion, leading  in superconductor, to supercurrents of carrier $q=2e$. Going back again the expression (\ref{eq:1}), the quantity $\hbar/(\pi\Delta)$ can be viewed as the effective period for the back and forth ($\ell=0$) motion of the partners of the Cooper pair (Fig. \ref{fig:A1} (b)). In this case thus, it is the intrinsic motion of the fermionic partners of the Cooper pairs one refers to. Being this a virtual motion, it is not observable. For bias such that the center of mass of Cooper pairs move with the critical velocity
  \begin{align*}
    \text{v}_c=\frac{\hbar s_c}{m}=\frac{1}{m}\frac{\hbar}{\xi}=\frac{\pi\Delta}{m\text{v}_F},
  \end{align*}
  implying energies of the order of the depairing energy
  \begin{align*}
    m\text{v}_c\text{v}_F\approx\pi\Delta,
  \end{align*}
  and thus critical violation of time-reversal invariance, the system moves into the regime of independent fermionic motion, that is, S-Q regime of carriers of charge $q=e$.
\section{Single fermion tunneling probability}\label{AppG}
For particles of momentum $k=(2M\times E)^{1/2}/\hbar$ incident from the left on a barrier of thickness $w$ and height $V_0$  (see Fig. \ref{fig:barrier}) and assuming $\kappa w\gg1$ ($\kappa=(2M(V_0-E))^{1/2}/\hbar$), the tunneling probability  is
\begin{align}
  \label{special_eq:34}
  T\approx\left(\frac{4\kappa k}{k^2+\kappa^2}\right)^2e^{-2\kappa d}. 
\end{align}
\subsection*{Electrons}
Assuming the tunneling electron moves close to  the Fermi energy (e.g. $\kappa_F\approx 1.18$\AA$^{-1}$, Nb) and that $w\approx 1$nm, $T\approx e^{-2\kappa d}\approx10^{-10}$. Thus the probability for two electrons to tunnel simultaneously is $10^{-20}$, as nearly impossible as no matter.
\subsection*{Nucleons}
In the case of the reaction $^{116}$Sn+$^{60}$Ni$\to^{115}$Sn+$^{61}$Ni at $E_{cm}=154.26$ MeV ($(D_0)_c=13.49$ fm), $w\approx 2.24$ fm, and $\kappa\approx 0.62$ fm$^{-1}$, corresponding to a typical nucleon separation energy of 8 MeV.  Thus $2\kappa w\approx 2.78$ and $P_1=T\approx7.5\times 10^{-2}$. As a consequence, the probability for two nucleons to tunnel simultaneously is $P_2\approx P_1^2\approx 6\times 10^{-3}$, and thus quite small.

  \section{Assessing the validity of the parallel with BCS superconductivity (\cite{Bardeen:57a};\cite{Bardeen:57b}) at the basis of nuclear BCS (\cite{Bohr:58})}\label{AppC}
  \subsection{Successive and simultaneous transfer}
  The order parameter of the phenomenon of spontaneous symmetry breaking in gauge space associated with BCS condensation can be written as
  \begin{align}
    \label{eq:50}
    \alpha_0=\sum_{\nu,\nu'>0}\braket{BCS(N+2)|a^\dagger_\nu a^\dagger_{\tilde \nu}|BCS(N)}=\sum_{\nu>0}U_\nu(N)V_\nu(N+2),
  \end{align}
  where $U_\nu=U'_\nu$, $V_\nu=V'_\nu e^{-2i\phi}$, $U'_\nu$ and $V_\nu'$ real,
  but equally well as,
  \begin{widetext}
  \begin{align}
    \label{eq:51}
    \nonumber \alpha_0&=\sum_{\nu,\nu'>0}\braket{BCS(N+2)|a^\dagger_\nu |\nu'}\braket{\nu'|a^\dagger_{\tilde \nu}|BCS(N)}=\sum_{\nu,\nu'>0}\braket{BCS(N+2)|a^\dagger_\nu \alpha^\dagger_{\nu'}|BCS(N+1)}\braket{BCS(N+1)|\alpha_{\nu'}a^\dagger_{\tilde \nu}|BCS(N)}\\
 &=\sum_{\nu,\nu'>0}\braket{BCS(N+2)|V_{\nu}(N+2)\alpha_{\tilde\nu} \alpha^\dagger_{\nu'}|BCS(N+1)}\braket{BCS(N+1)|\alpha_{\nu'}U_\nu(N)\alpha^\dagger_{\tilde \nu}|BCS(N)}=\sum_{\nu>0}V_\nu(N+2)U_\nu(N),
  \end{align}
  \end{widetext}
  where the inverse quasiparticle transformation $a^\dagger_\nu=U_\nu\alpha^\dagger_\nu+V_\nu\alpha_{\tilde \nu}$ was used in dealing with the intermediate states. The above
relations  testify to the fact that nucleons moving in time reversal states around the Fermi energy and described by $\ket{BCS}=\prod_{\nu>0}\left(U_\nu+V_\nu a^\dagger_\nu a^\dagger_{\tilde \nu}\right)\ket{0}$ are equally well pairing correlated (entangled), when undergoing successive than simultaneous transfer.

  Furthermore, because the binding energy of the Cooper pairs\footnote{Due to the very important role zero point pairing fluctuations play in nuclei, one can talk about $U,V$ factors both in connection with superfluid nuclei as well as with pairing vibrational nuclei, in which case one refers to $U^{eff},V^{eff}$. These effective occupation numbers are closely related to the backwardsgoing QRPA amplitudes, the so called $Y$-amplitudes, $(V^{eff}_\nu)^2=2Y_{pv}^2(j_\nu)/(j_\nu+1/2)$, $(U^{eff}_\nu)^2=1-(V^{eff}_\nu)^2$, where $pv$ labels either the pair addition or pair removal modes \cite{Potel:13b}.} is much smaller than the Fermi energy $(2\Delta/\epsilon_F\ll1)$ and the pairing strength is conspicuously smaller than the value of the mean field around the nuclear surface region, successive is the dominant two-nucleon transfer mechanism. It then does not seem surprising that the absolute cross section associated with pair transfer may be comparable to that of one-particle transfer under similar kinematic conditions\footnote{For example: \cite{Cavallaro:17} $d\sigma(^9\text{Li} (d,p)^{10}\text{Li} (1/2^-))/d\Omega|_{\theta_{max}}\approx0.8$ mb/sr, as compared to \cite{Tanihata:08}  $d\sigma(^{11}\text{Li} (p,t)^{9}\text{Li} (1/2^-))/d\Omega|_{\theta_{max}}\approx1 $ mb/sr, \cite{Fortune:94} $^{10}$Be$(t,p)^{12}$Be(gs) $(\sigma=1.9\pm0.5$ mb, $4.4^\circ\leq\theta_{cm}\leq54.4^\circ)$ as compared to \cite{Schmitt:13} 
$^{10}$Be$(d,p)^{11}$Be($1/2^+$) $(\sigma=2.4\pm0.013$ mb, $5^\circ\leq\theta_{cm}\leq39^\circ)$ in the case of light nuclei around closed $(N=6)$ shell, and \cite{Bassani:65} $^{120}$Sn$(p,t)^{118}$Sn(gs) $(\sigma=3.024\pm0.907$ mb, $5^\circ\leq\theta_{cm}\leq40^\circ)$ as compared to \cite{Bechara:75} $^{120}$Sn$(d,p)^{121}$Sn($7/2^+$) $(\sigma=5.2\pm0.6$ mb, $2^\circ\leq\theta_{cm}\leq58^\circ)$.}.

  Nonetheless, one has to be somewhat careful in using these results in connection with the parallel one may try to establish with the similarity existing between the critical (dc) Josephson current of Cooper pairs ($S$-$S$), and the single electrons current ($S$-$Q$) observed by Giaever, setting a (dc) bias $2\Delta/e$ to the weak link. This is because the superfluid-quasiparticle ($S$-$Q$) current observed by Giaever implies, in the nuclear case, in particular in the case of heavy ion reactions between superfluid nuclei, both the gs$\to$gs single-particle transfer, but also the transfer to quasiparticle states of energy $\lesssim2\Delta$. Furthermore, because in the condensed matter case one is talking of transfer of Cooper pairs or of single electrons between two superconductors, something non-operative when one of the two interacting systems in the nuclear reaction under consideration is a proton, a deuteron or a triton. Again, due to the important role ZPF associated with  pairing vibrations play in nuclei, the clear cut distinction between S and N systems found in condensed matter, is blurred in the nuclear case (within this context see Fig. \ref{fig:D4} and associated discussion; see also Sect. \ref{Se4.1}). 
  \subsection{Coherent state and phase space}
Let us consider a  $(p,t)$ process connecting the ground state of two  superfluid nuclei, e.g.  Sn-isotopes. The unique property of the $\ket{N+2}\to\ket{N}$ transition induced by the $(p,t)$ reaction between members of the rotational band (gs$\to$gs transition), is due to the fact that
  \begin{align}
    \label{eq:52}
    \ket{N}\sim\int d\phi\,e^{iN\phi}\,\ket{BCS(\phi)},
  \end{align}
  where
  \begin{align}
    \label{eq:53}
    \nonumber    \ket{BCS}&=\prod_{\nu>0}\left(U_\nu+V_\nu P^\dagger_\nu\right)\ket{0}_F\\
    &=\sum_{n=0,1,2\dots}\frac{\left(c_\nu P^\dagger_\nu\right)^n}{n!}\ket{0}_F=\exp\left(c_\nu P^\dagger_\nu\right)\ket{0}_F,
  \end{align}
  is a coherent state violating pair of particles conservation. In the previous relation $c_\nu=V_\nu/U_\nu$, $P^\dagger_\nu=a_\nu^\dagger a_{\tilde \nu}^\dagger$ and $n=N/2$ (number of pairs), have been used.

  The relevance of the above result expresses itself in the microscopic description of superfluid nuclei through the fact that, for example, to  reproduce the experimental cross section $\sigma_{exp}(\text{gs}\to\text{gs})=2492\pm 374\mu$b associated with the reaction $^{116}$Sn$(p,t)^{114}$Sn (\cite{Guazzoni:04}), one needs to have the correct pairing strength  ($\text{v}_p^{eff}=\text{v}_p^{bare}+\text{v}_p^{ind}$) and (effective) phase space. This is in keeping with the fact that the contribution to the pairing gap associated with $\text{v}_p^{bare}$ and $\text{v}_p^{ind}$ are about equal, and that this last one arises from the exchange of  two-quasiparticle ($2qp$) vibrations calculated in QRPA. The experimental result $\sum_i\sigma_{exp}(\text{gs}\to2qp(i))=2012\mu$b$(1.3 \text{MeV}\leq E_{2qp}(i)\leq 4.14$ MeV, 60 states) together with the theoretical one $\sigma_{th}(\text{gs}\to\text{gs})=2078\mu$b, testify to the phase space consistency used to describe the $N=66,64$ members of the ground state pairing rotational band of the Sn-isotopes.
  \subsection{Cooper pair and single-particle transfer in heavy ion collisions between superfluid nuclei}
  From the value of the Josephson critical current $I_c=(\pi/4)I_N$, where $I_N=V_{eq}/R=\left(\frac{2\Delta}{e}\right)\frac{1}{R_b}$, one can conclude that the $S$-$S$ Cooper pair current is almost equal ($(\pi/4)\approx0.8$)   to the $S$-$Q$ single electron current resulting from the unbinding of Cooper pair partners ($\approx2\Delta$) by giving a momentum $q\approx\hbar/\xi$ to the Cooper pair center of mass, implying a critical breaking of time reversal invariance. The same can be done by forcing Cooper pair partners to be at a distance\footnote{This equivalence can be qualitatively shown through $\delta(p^2/2m_e)\approx2\Delta$, $\frac{p_F\delta p}{m_e}=\text{v}_F\delta p=2\Delta$, $\delta p\delta x\approx \delta p\xi\approx \hbar$, i.e. $\delta p\approx\hbar/\xi$, $\xi\approx(\hbar \text{v}_F/2\Delta)$, (please note the $\approx$ symbol in the indeterminacy relation).}$\gtrsim (D_0)_c$. 
  \subsection{Correlation length, nuclear dimensions and moment of inertia: further evidence of the distortion operated on the Cooper pairs by the mean field}
  For a superfluid system held in a non spherical, e.g. ellipsoidal domain, of dimension large as compared to the coherence length $\xi$, the moment of inertia approaches that of the irrotational flow. At the antipodes with the rigid moment of inertia, in the sense that, seen from the rotating frame of reference, this last one is at rest, the velocity lines of the irrotational one being ellipses (see Fig. \ref{fig:C1}).

  In nuclei, while the observed moments of inertia are smaller than the rigid one by a factor of 2, they are about a factor of 5 greater than the irrotational moment of inertia\footnote{\cite{Bohr:75} pp 75, 83, 278, 397 and 674.}. A fact which testifies to the special role pairing plays in nuclei, constrained as it is by both spatial confinement ($R_0<\xi$) and quantization (mixing of odd and even parity states $Y_{l0}(\theta=180^\circ)\sim(-1)^l;Y_{l0}(0^\circ)=1$). In other words, constrained by the strong distortion to which nuclear Cooper pairs are subject.
  \subsection{Specific probe of Josephson-like nuclear (ac) effect}
  Being the order parameter of the superconducting phase the $\ket{BCS}$ expectation value of $P^\dagger$, one should be able to calculate, within a 10\% error, the absolute two-nucleon transfer cross section ($\sim|T|^2$) in trying to assess the validity of the parallel made between low-temperature superconductors and open-shell nuclei. In particular regarding subtle consequences of BCS, like the Josephson effect. Not only this, but also to match with specific nuclear structure conditions those at the basis of the solid state phenomenon.

  The transfer of a pair of electrons through a Josephson junction is free of dissipation, aside from fulfilling $\xi\gg w$. Consequently, the bombarding conditions in a heavy ion reaction between superfluid nuclei aimed at creating a transient Josephson junction, should be such that $E_{cm}\lesssim E_B$ (no dissipation) and, at the same time, also fulfill $D_0\lesssim(D_0)_c$ (superfluid phase). Otherwise, one will be close\footnote{As already stated, because of the important role pairing vibrations play in nuclei and in nuclear reactions, in particular in the transition from the S-S to the S-Q regimes, the condensed matter effects of such transitions are in the nuclear case conspicuously attenuated.} to or in an uncorrelated single-(quasi) particle phase in which case the absolute two-nucleon transfer cross section becomes $\sim|T|^4$, precluding the qualification of Josephson-like junction. See however Sect. \ref{Se4.1}.

\section{Finite versus infinite systems}\label{AppF}    
Linear momentum is conserved in  infinite systems, while translational invariance is spontaneously broken in finite ones. A fact which underscores  the central role the  center of mass plays in the microscopic description of  finite, quantum many-body systems.

Let us illustrate this by looking at a nucleus. That is, shine on it a beam of photons and record the frequencies it absorbs. The outcome is remarkable. Nuclei  absorb essentially all of the dipole electromagnetic energy at a single frequency, through the vibration of protons against effectively charge carrying neutrons. As a consequence,  the center of mass of the system remains at rest.

Let us now illuminate a heavy nucleus with a beam of another heavy species, at a bombarding energy such that the two systems enter, only at the distance of closest approach, in weak contact. Being two leptodermic systems, when they  separate  they have to reconstruct the overlapping surface area and, in the process, give back the original surface energy, eventually returning to the initial kinetic energy of relative motion. The center of mass of each of the systems plays again a central role in defining the  radii of curvature, the associated reduced radius and thus the ion-ion potential, let alone the trajectory of relative motion and the distance of closest approach along it.

Similarly, concerning the abnormal density of the two colliding superfluid nuclei. The associated Cooper pairs can be viewed as correlated pairs of nucleons, each described in terms of single-particle wavefunctions referred to the center of mass of the corresponding mean field potential.
During the weak leptodermic contact, the abnormal densities do similarly in gauge space to what the normal ones do in three dimensional space, giving rise to a weakly coupled superconducting-superconducting junction,  interacting through    the reduced pairing gap. Interaction which allows effectively charged nuclear  Cooper pairs, referred to the center of mass of one system, to move through the junction, one nucleon at a time, to eventually become referred to the center of mass of the other superconducting system. The part of the $T$-matrix associated with the radial formfactor mediating such process,  can be viewed as the conductance, i.e. the inverse of the resistance ($1/R_b$), of the nuclear junction.  Weighted with the corresponding two-nucleon transfer spectroscopic amplitudes (coherence factors)  modulus squared, it determines the ground state-ground state ((gs)-(gs)) absolute differential cross section. Quantity which, for a range of bombarding energies ($E_{cm}$) smaller than that of the Coulomb barrier (and thus essentially free from absorption), and associated distances of closest approach $D_0\lesssim (D_0)_c$  is,  within a factor of two, equal to the incoherent summed one-quasiparticle absolute transfer cross section (Fig. \ref{fig:2}).


Because the collision time of the heavy ion reaction  associated    with the value of $E_{cm}$ leading to $D_0\approx(D_0)_c$ implies that the center of mass of the transferred Cooper pair has moved with about the critical momentum $\hbar/\xi$, one can talk of the  critical current of a single effectively charged nuclear Cooper pair. And because  of the selfconsistency found between $E_{cm},\tau_{coll},(D_0)_c$ and $\hbar/\xi$, one can view the associated transient Josephson-like junction to be stationary, as far as the Cooper pair transfer is referred (adiabatic picture).
 


\section{Correlation distances between weakly coupled many-body systems}\label{AppD}
\subsection{Selected examples}
\textbf{a}) The neutron stripping of $^9$Be on $^{12}$C ($^{12}$C($^{9}$Be$(1p_{3/2}),^8$Be)$^{13}$C$^*(2s_{1/2};3.086\text{ MeV})$) at low bombarding energies (see Fig. \ref{fig:D1}) displays an astrophysical factor $S^*(\sim E_{cm}\times\sigma)\approx117\pm8$ MeV barns at $E_{cm}\approx$ 1 MeV ($D_0\approx35$ fm)\footnote{\cite{Broglia:04a} Fig. 37, p. 467.}. Processes in which a single nucleon is transferred in grazing collisions between heavy ions, where the two nuclei have a small overlap play, as a rule (well matched conditions) an important role. This is because the associated formfactors have a longer range than the formfactors for inelastic scattering\footnote{\cite{Broglia:04a} Fig. 1, p. 298.} (Fig. \ref{fig:D2}), and the reason why transfer is, for spherical or little deformed systems, the main source of depopulation of the entrance channel. As a consequence  single-particle formfactors or better, their tail, are the basic elements needed to calculate the imaginary part of the optical potential acting between the ions.

\textbf{b})  The transition density of the soft $E1$-dipole mode (PDR) of $^{11}$Li, in which the halo neutrons oscillate out of phase with respect to the nucleons of the core ($^9$Li), is associated with a dipole moment\footnote{\cite{Broglia:19}.} $d= e_{eff}r$, with $r\approx9$ fm (Fig. \ref{fig:D3}). A quantity which can be determined by measuring the $B(E1)$ strength function of the PDR. Similarly, the $(p,p')$ or $(d,d')$ cross sections.

\textbf{c}) In the reaction $^{116}$Sn+$^{60}$Ni$\to^{114}$Sn+$^{62}$Ni between ground states, the main contribution is associated with successive transfer, in which the partner nucleons of the transferred Cooper pair are, in the intermediate channel $^{115}$Sn+$^{61}$Ni one on each nucleus. For a bombarding energy $E_{cm}=154.26$ MeV ($(D_0)_c=13.49$ fm) their relative distance is estimated to be $\approx10.56$ fm (see Fig. \ref{abs_figA}). Distance which can be probed through the measurement of the Josephson-like (ac) dipole radiation. Arguably  this can also be done studying the quasielastic process $^{120}$Sn+$p\to^{118}$Sn+$t\to^{120}$Sn+$p$, as schematically described in Fig. \ref{fig:D4}.

\subsection{Macroscopic estimates of the dipole moment from GDR, PDR and soft $E1$ modes}\label{AppI2}
The Thomas-Reiche-Kuhn (TRK) dipole energy weighted sum rule can be written as \cite{Bohr:75}
\begin{align}
  \label{eq:61}
  S(E1)=14.8\frac{NZ}{A}e^2\text{fm}^2\,\text{MeV}
\end{align}
In the case of the giant dipole resonance of $^{208}_{86}$Pb$_{126}$, which essentially carries 100\% of it,
\begin{align}
  \label{eq:62}
  S(E1)=735.2\,e^2\text{ fm}^2\text{ MeV}.
\end{align}
Making use of the expression $\hbar\omega_{GDR}=80/A^{1/3}$ MeV one obtains
\begin{align}
  \label{eq:63}
  \hbar\omega_{GDR}\approx 13.5\text{ MeV},
\end{align}
and the associated dipole moment can be estimated from the relation,
\begin{align}
  \label{eq:64}
  d\approx\left(\frac{S(E1)}{\hbar\omega_{GDR}}\right)^{1/2}\approx\left(54.5 \,e^2\text{ fm}^2\right)^{1/2}\approx7.4\,e\text{ fm}=e_{eff}r.
\end{align}
With the help of the associated effective neutron and proton charges
\begin{align}
  \label{eq:65}
  (e_{eff}/e)=\left\{\begin{array}{c}
		-Z/A=-82/208=-0.39=e_N/e, \\ 
		N/A=126/208=0.61=e_Z/e,
                \end{array} \right.
\end{align}
one obtains
\begin{align}
  \label{eq:66}
  r=\frac{d}{e_{eff}}=\left\{\begin{array}{c}
		-19\text{ fm}=r_N, \\ 
		12.1\text{ fm}=r_Z,
                \end{array} \right.  
\end{align}
the average value being\footnote{Of notice that (\ref{eq:67}) represents the summed coherent contributions of the many $p-n$ dipoles when the nucleus absorbs  (emits) a $\gamma$-ray of energy (\ref{eq:63}) (see for example right hand side of Fig. 1.5 of reference \cite{Bertsch:05}; see also Sect. 3.3 as well as p. 86 Eq. (3.15) and Fig. 3.1 of this reference). From  mean field approximation one can estimate the summed neutron, proton (dipole) ($p-h$) phase space to be of the order of 20-25. Thus $\bar r/23\approx0.7$ fm.}
\begin{align}
  \label{eq:67}
  \bar r=\frac{|r_N|+|r_Z|}{2}\approx 15.6\text{ fm},    
\end{align}
to be compared with the nuclear radius $R_0=7.1$ fm.

A similar estimate for the PDR assuming its centroid is at $\hbar\omega_{GDR}\approx 6$ MeV and carrying $\approx20$\% EWSR, one obtains
\begin{align}
  \label{eq:68}
  d^2=\frac{0.2 S(E1)}{6\text{ MeV}}=24.5\,e^2\text{ fm}^2,
\end{align}
that is $d\approx5$ fm, and
\begin{align}
  \label{eq:69}
  r=\frac{d}{e_{i}}=\left\{\begin{array}{c}
		-12.3\text{ fm}=r_N \\ 
		8.2\text{ fm}=r_Z
                \end{array} \right.    
\end{align}
leading to
\begin{align}
  \label{eq:70}
  \bar r=10.3\text{ fm}.
\end{align}
Let us repeat similar estimates for $^{120}_{50}$Sn$_{70}$. They lead to
\begin{align}
  \label{eq:71}
  &S(E1)=421\,e^2\text{ fm}^2\text{ MeV},\\
  \nonumber&\hbar\omega_{GDR}\approx 16.4\text{ MeV},
\end{align}
and
\begin{align}
  \label{eq:72}
  d=\left(\frac{S(E1)}{\hbar\omega_{GDR}}\right)^{1/2}\approx 5.1\text{ fm}.
\end{align}
The effective charges being
\begin{align}
  \label{eq:73}
  (e_{eff}/e)=\left\{\begin{array}{c}
		-0.43=e_N/e, \\ 
		0.57=e_Z/e,
                \end{array} \right.      
\end{align}
leads to
\begin{align}
  \label{eq:74}
  r=\frac{d}{e_{eff}}=\left\{\begin{array}{c}
-11.9\text{ fm}=r_N \\ 
		8.9\text{ fm}=r_Z,
                \end{array} \right.        
\end{align}
and thus to
\begin{align}
  \label{eq:75}
  \bar r=10.4\text{ fm}
\end{align}
to be compared with the radius\footnote{Similarly to the case of $^{208}$Pb, the dipole phase space is in the present case $\approx 16-18$, and $\bar r/17\approx 0.6$ fm.} $R_0=5.9$ fm. Concerning the PDR one obtains, under the assumption of 20\% EWSR and $\hbar\omega_{PDR}\approx 8$ MeV, $d=3.2\,e$ fm and
\begin{align}
  \label{eq:76}
  \bar r=6.5\text{ fm}.
\end{align}
In the case of the halo nucleus $^{11}_3$Li$_8$ one finds
\begin{align}
  \label{eq:77}
  S(E1)\approx32.3\,e^2\text{ fm}^2\text{ MeV}.
\end{align}
The soft $E1$-mode (PDR) carries $\approx8\%$ of the EWR and its centroid is found at an energy of $\approx0.75$ MeV. Consequently
\begin{align}
  \label{eq:78}
  d\approx\sqrt{4.3}\,e\text{ fm}=2.1\,e\text{ fm}.
\end{align}
In the present case we are only interested in $r_N\approx\left|2.1/(-0.27)\right|$ fm $\approx8$ fm. As seen from Fig. \ref{fig:D3}, this quantity is very similar with that associated with the (absolute value) maximum, of the halo component of the transition density ($\approx 9$ fm). In this case, the dipole phase space is rather small ($\approx 2$ and $\bar r_N/2\approx 4$ fm).
\section{Josephson frequency}\label{AppJx}
\subsection{Superconductors}
The energy of the emitted photon associated with the  tunneling of a Cooper pair across a (dc) barrier biased Josephson junction is
\begin{align}
  \label{eq:79}
  \Delta E=2eV
\end{align}
 ($\Delta E=\hbar\omega_J=h\nu_J$). Thus,
\begin{align}
  \label{eq:80}
  \nu_J=\frac{2Ve}{h}\left(\omega_J=\frac{2Ve}{\hbar}\right)
\end{align}
In keeping with the fact that
\begin{align}
  \label{eq:81}
  \dot\phi=\frac{\mu}{\hbar},
\end{align}
where $\mu$ is the chemical potential of the superconductor, the energy difference (\ref{eq:79}) forces the weak link coupled superconductors gauge phases to rotate with different angular velocities,
\begin{align}
  \label{eq:82}
  \dot\phi_{rel}=\left(\dot\phi_1-\dot\phi_2\right)=\frac{\mu_1-\mu_2}{\hbar}\quad\left(\phi_{rel}=\phi_1-\phi_2\right).
\end{align}
Because
\begin{align}
  \label{eq:83}
  \mu_1-\mu_2=eV,
\end{align}
and
\begin{align}
  \label{eq:84}
  \phi_{rel}=\frac{Ve}{\hbar}
\end{align}
one can write
\begin{align}
  \label{eq:85}
  \omega_J=2\dot\phi_{rel}=2\frac{\mu_1-\mu_2}{\hbar}=\frac{2Ve}{\hbar}
\end{align}
It is of note that  the factor of 2 in front of
$\phi_{rel}$, emerges from the fact the coupling Hamiltonian between the two rotors in gauge space arises from the exchange of pairs, that is, $P_1 P_2=e^{2i\phi_1}P'_1e^{-2i\phi}P'_2$ (see e.g. \cite{Brink:05} p. 357 and refs. therein). That is,
\begin{align}
  \label{eq:86}
  H_{coupl}\sim e^{2i\phi_1}e^{-2i\phi_2}e^{i\phi_0/2}+\text{h.c.}\sim \cos\left(2(\phi_1-\phi_2)+\phi_0\right).
\end{align}
Assuming a dc bias
\begin{align}
  \label{eq:87}
  \phi_{rel}=\dot\phi_{rel}t=\frac{\mu_1-\mu_2}{\hbar}t
\end{align}
and
\begin{align}
  \label{eq:88}
  \dot N_1\sim\sin\left(2e V t+\phi_0\right)
\end{align}
\subsection{Nuclei}
The energy associated with  Cooper pair tunneling is
\begin{align}
  \label{eq:89}
  &\Delta E=\hbar\omega_J^N=Q_{2n}=2V^N\,e_{eff},
\end{align}
where
\begin{align}
 &\omega_J^N=\frac{Q_{2n}}{\hbar}=2\dot\phi_{rel}^N=2\frac{\mu_1-\mu_2}{\hbar}=\frac{2V^N e_{eff}}{\hbar},
\end{align}
and
\begin{align}
  \label{eq:90}
  V^N=\frac{Q_{2n}}{2e_{eff}}.
\end{align}
The energy of the associated $\gamma$-rays is taken from the relative motion (see $\delta$-function $\delta(E_i-E_\gamma-E_f+Q)$) Eq. (\ref{notes_eq:306}).

\section{Wick's theorem}\label{AppJ}
The relation (\ref{special_eq:18}) is in contrast with that expected for the  normal state of metals as can be seen by calculating (\ref{special_eq:16}).  Better, the two-particle density matrix, which we write in its general, non-local form, as
\begin{align}
  \label{special_eq:19}
  \Gamma(r_1r_2;r'_1r'_2)=\braket{\Psi|\hat\psi^\dagger(\mathbf r'_2)\hat\psi^\dagger(\mathbf r'_1)\hat\psi(\mathbf r_1)\hat\psi(\mathbf r_2)|\Psi}.
\end{align}
Assuming $\ket{\Psi}$ to be a single determinant,
\begin{align}
  \label{special_eq:20}
  \Gamma(r_1r_2;r'_1r'_2)=\rho(r_1,r'_1)\rho(r_2,r'_2)-\rho(r_1,r'_2)\rho(r_2,r'_1)
\end{align}
where
\begin{align}
  \label{special_eq:21}
  \rho(r,r')=\braket{\Psi|\hat\psi^\dagger(\mathbf r')\hat\psi(\mathbf r)|\Psi}=\sum_{i=1}^N\psi_i(\mathbf r)\psi^*_i(\mathbf r')
\end{align}
Thus
\begin{align}
  \label{special_eq:22}
  \lim_{|\mathbf r -\mathbf r'|\to\infty}\Gamma(r_1r_2;r'_1r'_2)=0,
\end{align}
as
\begin{align}
  \label{special_eq:23}
  \lim_{|\mathbf r -\mathbf r'|\to\infty}\rho(r,r')=0.
\end{align}

The rule for calculating the expectation value of a product of creation and annihilation operators is that it must be split up into pairs of operators in all possible ways, and each pair replaced by its expectation value. Within this context (\ref{special_eq:19}) becomes,
\begin{align}
  \label{special_eq:25}
  \nonumber \Gamma(r_1r_2;r'_1r'_2)&=\frac{1}{2}\rho(r_1,r'_1)\rho(r_2,r'_2)\\
  &-\frac{1}{2}\rho(r_1,r'_2)\rho(r_2,r'_1)+\frac{1}{2}\chi^*(r_2',r_1')\chi(r_2,r_1). 
\end{align}
Similar to Hartree-Fock (HF), BCS is a mean field theory. However, while a state represented by a single determinant like $\ket{HF(gs)}$ can be specified by  its one-particle density matrix (\ref{special_eq:21}), and all higher order density matrices can be calculated once this is known, the quasiparticle vacuum is not. The ground state $\ket{BCS}$ of a superconductor is not completely determined by its density matrix, but depends also on its \textit{pairing function} $\chi$ (abnormal density), which, in configuration space, can be written as
\begin{align}
  \label{special_eq:26}
  \chi(r,r')=\braket{\Psi|\hat\psi(r')\hat\psi(r)|\Psi},
\end{align}
in analogy to (\ref{special_eq:21}) for the density matrix. In configuration space it is a function that goes to zero if $r$ and $r'$ are far apart, and the range over which it is significantly different from zero is known as the correlation length. In momentum space it has matrix elements
\begin{align}
  \label{special_eq:27}
  \chi(\mathbf k\uparrow,-\mathbf k\downarrow)=U_k V_k.
\end{align}
Let us now write the appropriate  singlet pairing function associated with the $\ket{BCS}$ state,
\begin{align}
  \label{special_eq:28}
  \nonumber \chi(r_1,r_2)&=\braket{BCS|\hat\psi_\downarrow(\mathbf r_2)\hat\psi_\uparrow(\mathbf r_1)|BCS}\\
  &\sim\sum_k U_k V_ke^{i\mathbf k\cdot(\mathbf r_1-\mathbf r_2)}e^{i2\,\mathbf s\cdot\mathbf R-2i\phi},
\end{align}
where $\mathbf R=(\mathbf r_1+\mathbf r_2)/2$ is the center of mass coordinate of the Cooper pair. The BCS pair function corresponds to a state in which each pair has (CM) momentum $2\hbar s$ and phase $-2\phi$. On performing the Fourier sum over $\mathbf k$ one finds that the internal part of the pair function corresponds to a spherical symmetric bound $s$-state. The radial $k$-value is of the order of $k_F$, and the mean square radius of the state is of the order of
\begin{align}
  \label{special_eq:29}
  \xi=\frac{\hbar\text{v}_F}{\pi\Delta},
\end{align}
which can be interpreted as the inverse of the range of $k$-values near the Fermi surface over which the product $U_kV_k$ is significant. The two-particle density matrix thus displays the following property
\begin{align}
  \label{special_eq:30}
  \lim_{\substack{|r-r'|\to\infty\\(r_2,r_1),(r_2',r_1')<\xi}}\Gamma(r_1r_2;r'_1r'_2) =\chi^*(r_2',r_1')\,\chi(r_2,r_1).
\end{align}
Namely, it is large whenever the pair of points $r_1$ and $r_2$ and $r_1'$ and $r_2'$ are within a correlation length, however far apart the pair of points $(r_2',r_2)$ , $(r_1',r_1)$, $(r_2',r_1)$ , $(r_1',r_2)$ are from each other. This ODLRO property results in the two-particle matrix having a large eigenvalue. The corresponding eigenvector is the pairing function,
\begin{align}
  \label{special_eq:31}
  \int\int \Gamma(r_1r_2;r'_1r'_2) \chi(r_1',r_2')\,d\mathbf r'_1\,d\mathbf r'_2=a N \chi(r_1,r_2),
\end{align}
where $N$ is the number of pairs of particles participating in the condensate, $a$ being a number of the order of unity.

\section{Effective dipole}\label{S1}
 The intrinsic (i.e., referred to the center of mass $O$, see Fig. \ref{notes_fig1}) electric dipole moment is,
\begin{align}\label {eq202}
\mathbf D=\sum_i\mathbf r_{Oi}\,q_i.
\end{align}
The sum in the expression above extends to all the nucleons in the system, and the coordinates $\mathbf r_{Oi}$ are measured with respect to the center of mass $O$. Although we don't write the creation operator  explicitly, we assume that the vector potential $\mathbf A$ creates a photon of energy $E_\gamma$.  The charges $q_i$ are equal to $e$ (positive electron charge) for protons, and zero for neutrons. Let us write separately the contributions of the two nuclei $a$ and $A$, extend the sum only to the protons, and rewrite the coordinate vectors,
\begin{align}\label {eq204}
\mathbf D=e\sum_{i\in b,\pi}(\mathbf r_{Ob}+\mathbf r_{bi})+e\sum_{i\in A,\pi}(\mathbf r_{OA}+\mathbf r_{Ai}). 
\end{align}
We now make use of the approximation of considering that the cores $b$ and $A$ remain in their ground states $\Psi^0_{A}$ and $\Psi^0_{b}$ during the transfer process. Because of parity conservation,
 \begin{align}\label {eq205}
\sum_{i\in b,\pi} \int |\Psi^0_{b}(\pmb\xi_b)|^2 \mathbf r_{bi} \,d\pmb\xi_b=\sum_{i\in A,\pi} \int |\Psi^0_{A}(\pmb\xi_A)|^2 \mathbf r_{Ai} \,d\pmb\xi_A=0,
 \end{align}
and
 \begin{align}\label {eq206}
\nonumber &e\mathbf r_{Ob}\sum_{i\in b,\pi} \int |\Psi^0_{b}(\pmb\xi_b)|^2\,d\pmb\xi_b=eZ_b\mathbf r_{Ob},\\
 &e\mathbf r_{OA}\sum_{i\in A,\pi} \int |\Psi^0_{A}(\pmb\xi_A)|^2\,d\pmb\xi_A=eZ_A\mathbf r_{OA}.
\end{align}
In the context of the process we are interested in, we can thus adopt the ``effective'' dipole operator
\begin{align}\label {eq203}
\mathbf D=Z_A\mathbf r_{OA}+Z_b\mathbf r_{Ob}=(Z_A+Z_b)\mathbf r_{OQ}+Z_A\mathbf r_{QA}+Z_b\mathbf r_{Qb},
\end{align}
where the point $Q$ is the center of mass of the system formed by the two cores $A$ and $b$ (see Fig. \ref{notes_fig1}). Note that the two last terms in Eq. (\ref{eq203}), i.e.
\begin{align}\label {eq210}
\mathbf d_{brem}=Z_A\mathbf r_{QA}+Z_b\mathbf r_{Qb},
\end{align}
 correspond to the dipole operator associated with the system formed by the two nuclei $A$ and $b$ alone, without the two transferred neutrons, and is responsible for the bremsstrahlung radiation of the colliding charged cores. Making use of the definition of the center of mass,
 \begin{align}\label {eq207}
\mathbf r_{OQ}&=-\frac{\mathbf r_{O1}+\mathbf r_{O2}}{A_A+A_b},\\
\mathbf r_{QA}&=\frac{A_b}{A_A+A_b}\mathbf r_{bA},\\
\mathbf r_{Qb}&=-\frac{A_A}{A_A+A_b}\mathbf r_{bA},
 \end{align}
 where  $A_A,A_b$ are the mass numbers of $A$ and $b$, respectively. We can thus rewrite the bremsstrahlung dipole in terms of the $b$--$A$ coordinate $\mathbf r_{bA}$,
 \begin{align}\label {eq211}
\mathbf d_{brem}=e\,\frac{Z_AA_b-Z_bA_A}{A_A+A_b}\mathbf r_{bA}.
 \end{align}
Finally, if we define
 \begin{align}\label {eq208}
&\mathbf d_1=-e\frac{(Z_A+Z_b)}{A_A+A_b}\mathbf r_{O1},\\
&\mathbf d_2=-e\frac{(Z_A+Z_b)}{A_A+A_b}\mathbf r_{O2},
\end{align}
we can write
 \begin{align}\label {eq209}
\mathbf D=\mathbf d_1+\mathbf d_2+\mathbf d_{brem}.
\end{align}
The dipole operator can thus be decomposed in a bremsstrahlung term $\mathbf d_{brem}$ (which we are not going to consider here), plus a neutron contribution for each one of the transferred neutrons. It is interesting to note that the bremsstrahlung term vanishes if the two cores are identical or have the same charge-to-mass ratio (see Eq. (\ref{eq211})).


\end{appendix}

	\begin{center}
		\begin{table}
			\begin{tabular}{|c|c|c|c|}
				\hline
              $D_0$(fm)	& $E_{cm}$ (MeV) & $(E_B-E_{cm})$ (MeV)  & $\left(\frac{4}{\pi}\right)^2$ $\left(\frac{\sigma_{2n}}{\sigma_{1n}}\right)$\\
              \hline
				13.12	& 158.63 &-1.03 & 1.14\\
              13.49	& 154.26 & 3.34 & 0.57\\
              13.70	& 151.86 & 5.74 & 0.59\\
              13.81	& 150.62 & 6.98 & 0.46\\
              14.05	& 148.10 & 9.50 & 0.27\\
              14.24	& 146.10 & 11.50 & 0.22\\
              14.39	& 145.02 & 12.58 & 0.18\\
				\hline
			\end{tabular}
			\caption{In columns 1)--4) the distance of closest approach ($D_0$), the center of mass bombarding energy ($E_{cm}$), the difference between this quantity and the energy of the Coulomb barrier ($E_B=157.60$ MeV) \cite{Montanari:14}, and the ratio between the experimental value of the two- and one-particle transfer absolute cross sections multiplied by the factor $(4/\pi)^2$, are given.}\label{tab:1}
		\end{table}
	\end{center}

 \begin{figure}[h]
	\centerline{\includegraphics*[width=8cm,angle=0]{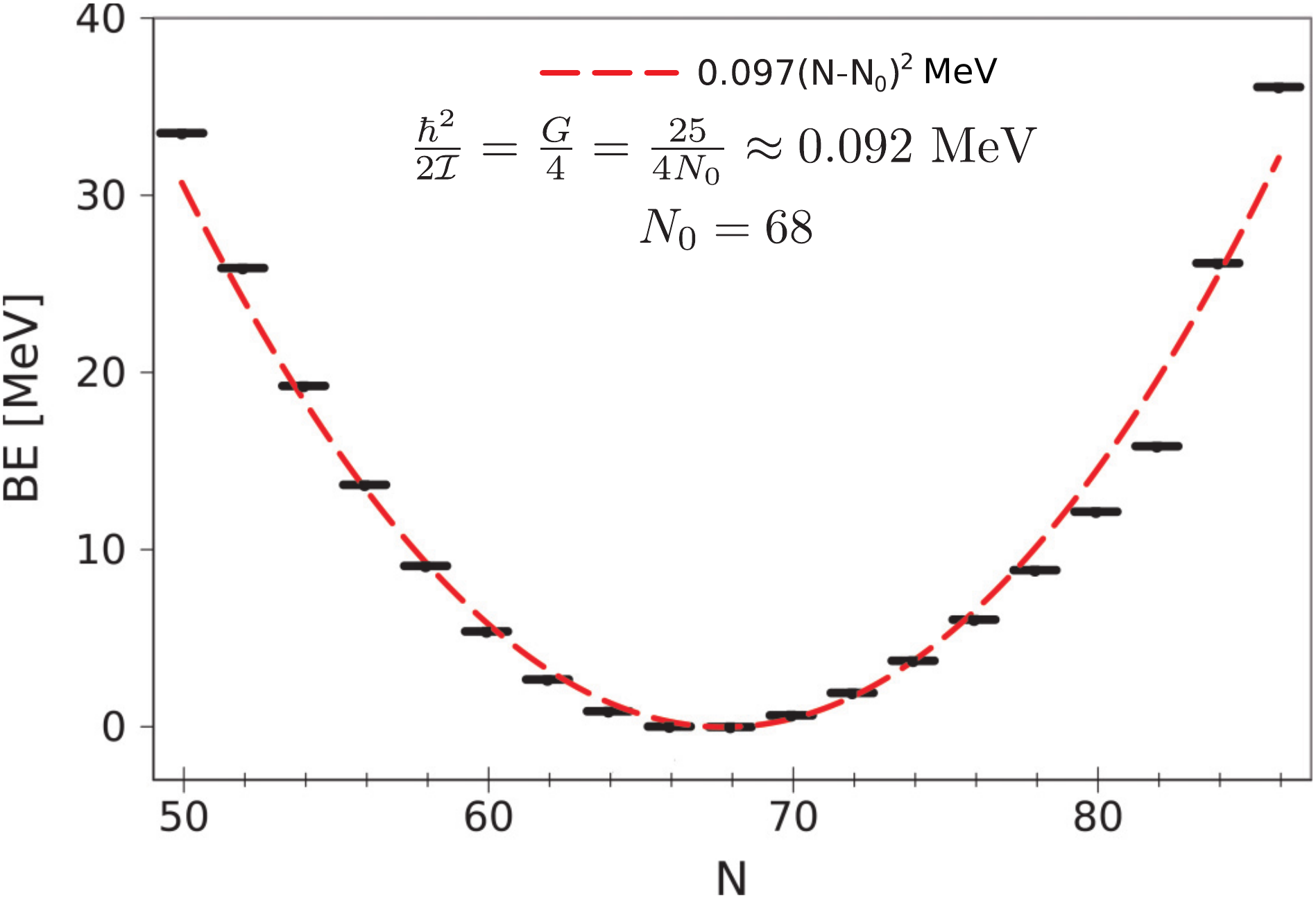}}
	\caption{Pairing rotational band associated with the ground state of the Sn-isotopes. The heavy drawn horizontal lines  display the energies calculated making use of the expression $BE=BE(^{50+N}$Sn$_N$)-8.124$N$+46.33, obtained by subtracting  the contribution of the single nucleon addition to the nuclear binding energy obtained by a linear fitting of the binding energies of the whole Sn-chain. The minimum corresponding to $N=68$ ($^{116}$Sn) has been arbitrarily shifted to make it coincide with the $BE=0$ value (for more details see \cite{Potel:13b}).}
    \label{abs_figII}  
  \end{figure}

    \begin{figure}[h]
	\centerline{\includegraphics*[width=7cm,angle=0]{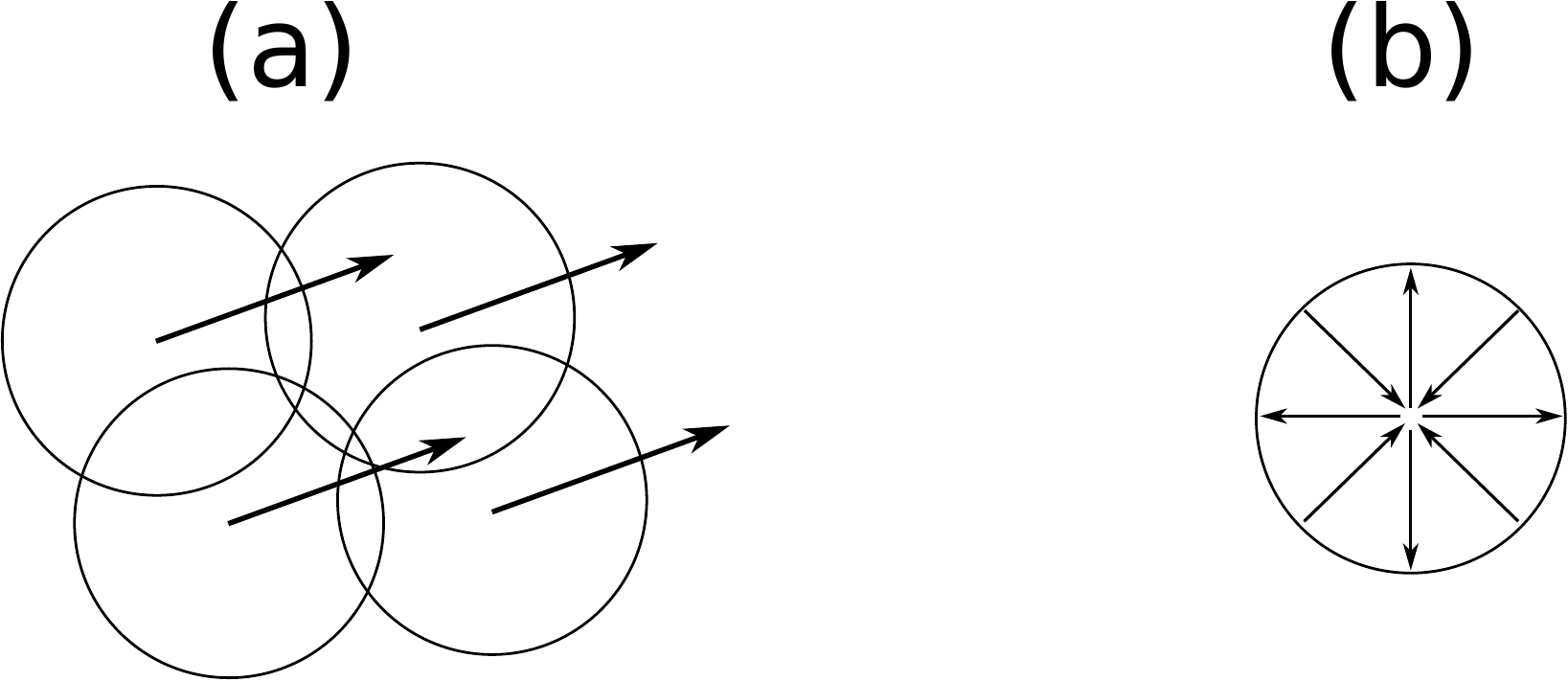}}
	\caption{(a) Schematic of a  supercurrent, for an applied bias leading to a momentum acting on the center of mass of Cooper pairs $k<1/\xi$.\; (b) intrinsic, fermionic motion of the Cooper pair.}
    \label{fig:A1}  
\end{figure} 
 \begin{figure}
  	\centerline{\includegraphics[width=7cm]{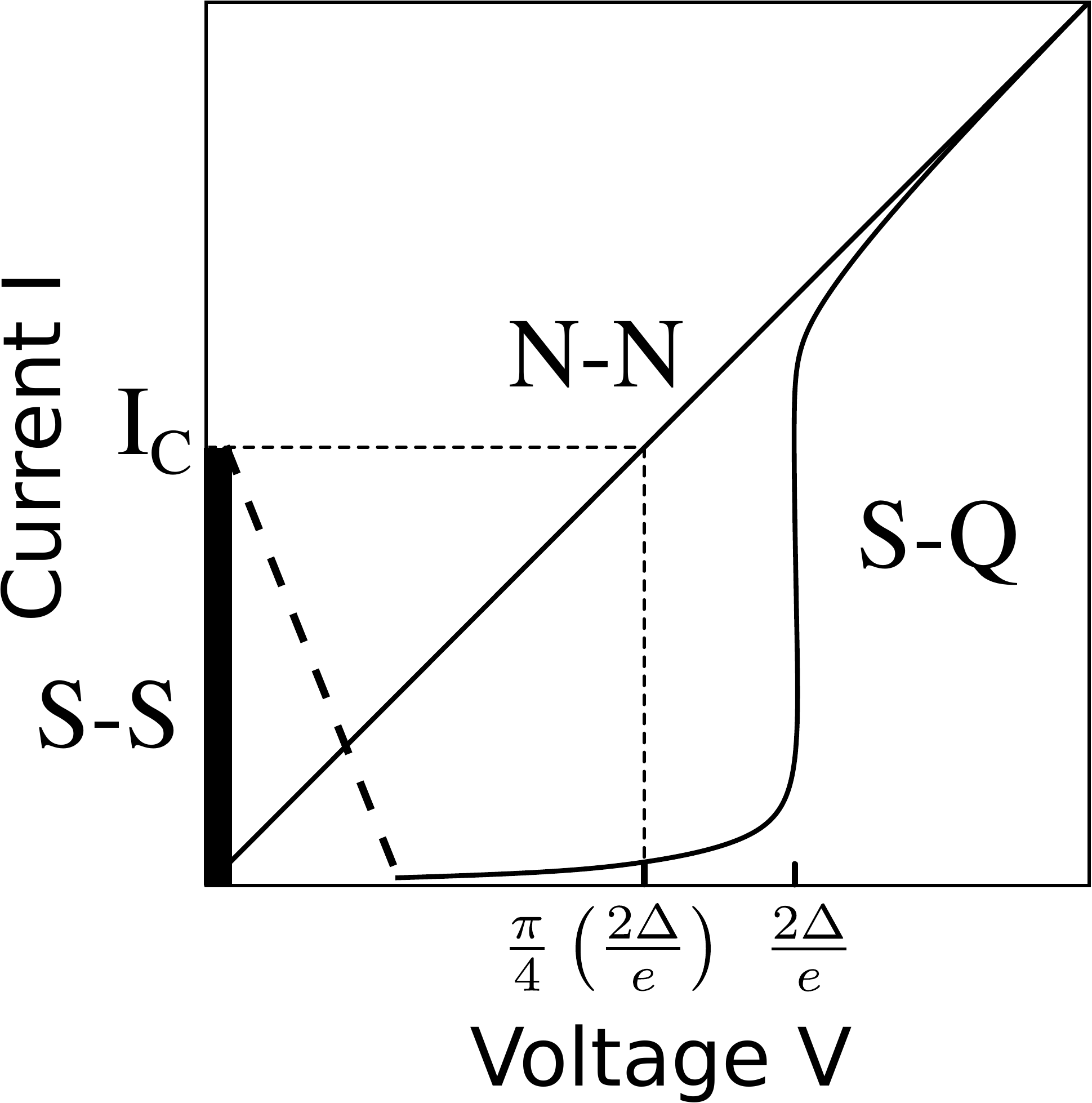}}
  	\caption{Weak link: Current-voltage characteristic curves for tunneling between two metals across a barrier (thin oxide layer of 10--30\AA) \cite{Poole:95}. It is assumed that both metals are  equal and that they can become superconductors below a critical temperature $T_c$ (e.g. Al-Al$_2$O$_3$-Al (1.140 K), Sn-SnO$_\text{x}$-Sn (3.722 K)).  \textbf{N-N} indicates the  current resulting from the \textit{tunneling of single electrons} between the metals, both of them in the \emph{normal phase} at $T>T_c$. \textbf{S-S} labels the (super) current associated with the \textit{tunneling of Cooper pairs} through the unbiased link ($V=0$, no potential drop) at $T<T_c$. That is, the (dc) Josephson effect described by the relation $I=I_c\sin\Delta\phi$ ($\Delta\phi=2(\phi_2-\phi_1)$). If a direct current source is attached to a Josephson junction, the relative phase $\Delta\phi$ adjusts to accommodate both the magnitude and sign of the current and, in the steady state, a zero-voltage current flows. As the current is changed, an instantaneous voltage $V$ appears to readjust the phase, $\delta(\Delta\phi)=(2e/\hbar)V\delta t$ to a new equilibrium value. Once equilibrium is reached, the current continues to flow and the voltage vanishes. The largest zero-voltage current which can flow $(I_c)$ is reached when $\Delta\phi=\pm\pi/2$. For $I>I_c$ the $I-V$ characteristic becomes a typical quasiparticle \cite{Giaver:73} tunneling curve, the intermediate behaviour (bold dashed line) being governed by the external circuitry. Within this context \textbf{S-Q} implies  {\emph {single-electron tunneling}} between two superconductors when the link is subject to a potential difference (direct bias)
$V\gtrsim 2\Delta/e$, in which case Cooper pairs are broken and quasiparticles (\textbf{Q}) become excited. The critical Josephson current is equal to $\pi/4$ ($\approx$ 80\%) of the \textbf{N-N} normal (single electron) current at the so called equivalent voltage $V_{equiv}=2\Delta/e$.}\label{fig:1}
  \end{figure}

\begin{figure}[h]
	\centerline{\includegraphics*[width=8cm,angle=0]{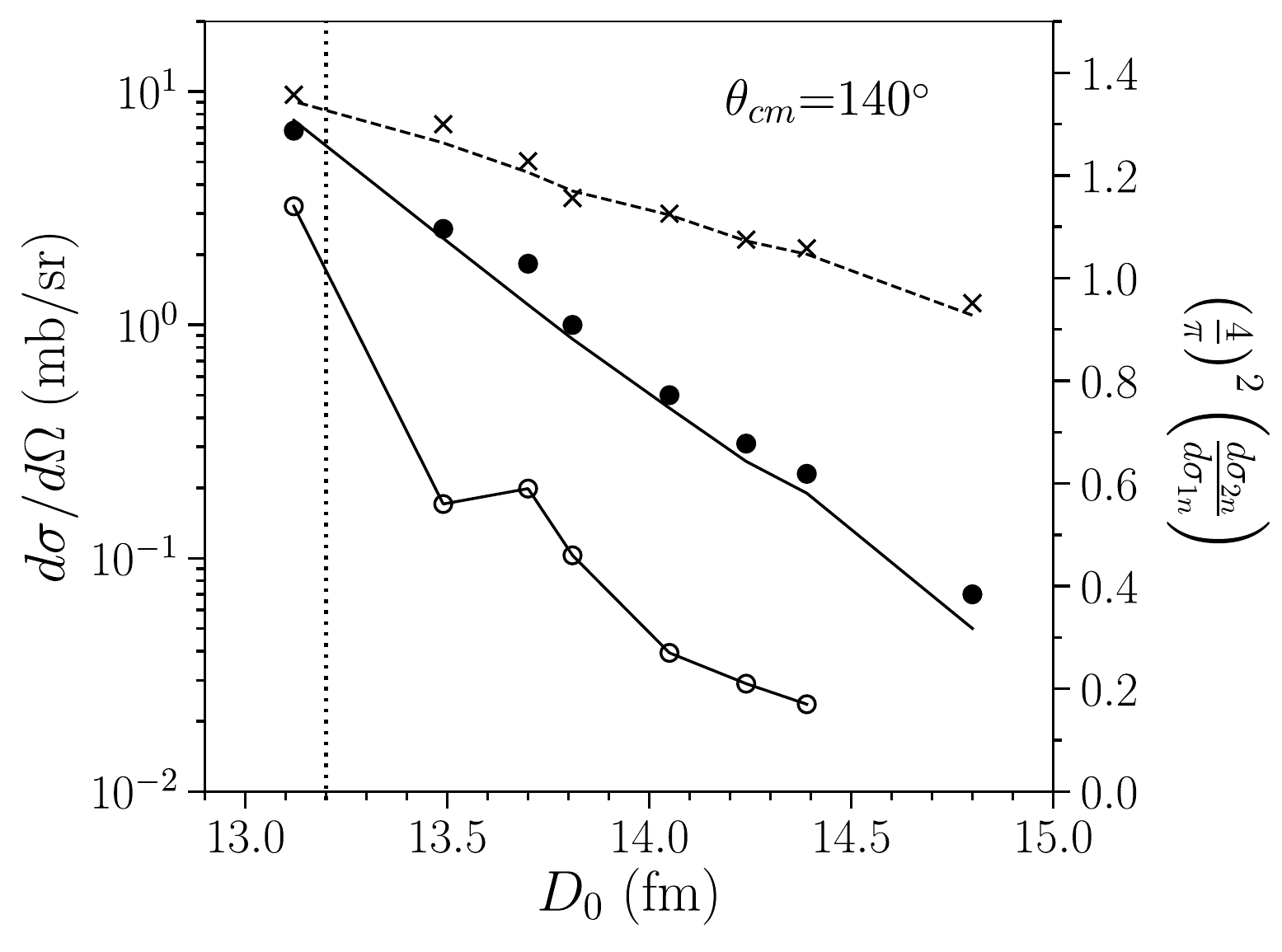}}
	\caption{Absolute differential cross sections associated with the reaction $^{116}$Sn+$^{60}$Ni$\to^{115}$Sn+$^{61}$Ni,  $\left.d\sigma_{1n}/d\Omega\right|_{\theta_{cm}=140^\circ}$ (exp. (x), th. (dashed line)), and $^{116}$Sn+$^{60}$Ni$\to^{114}$Sn+$^{62}$Ni, $\left.d\sigma_{2n}/d\Omega\right|_{\theta_{cm}=140^\circ}$ \cite{Montanari:14} (exp. (solid dot), th. (continuous line)), displayed as a function of the distance of closest approach $D_0$, for eight of the twelve bombarding energies for which the experiments were carried out. The ratio between the above experimental cross sections multiplied by $(4/\pi)^2$ (open circles) is also shown. The factor $4/\pi$ enters the relation $\frac{4}{\pi}\left(I_c/I_N\right)=1$, where $I_c$ is the critical Josephson (super) current (of Cooper pair carriers) and $I_N$ the normal (of single-electron carriers) current at the equivalent potential $V_{equiv}=2\Delta/e$ (see Fig. \ref{fig:1} as well as text; see also caption to Fig. \ref{fig:7p}). The vertical dotted line at  $D_0=13.2$ fm indicates the distance of closest approach associated with $E_{cm}=E_B=157.60$ MeV (Coulomb barrier).}
    \label{fig:2}  
\end{figure} 
\begin{figure*}
	\centerline{\includegraphics*[width=10cm,angle=0]{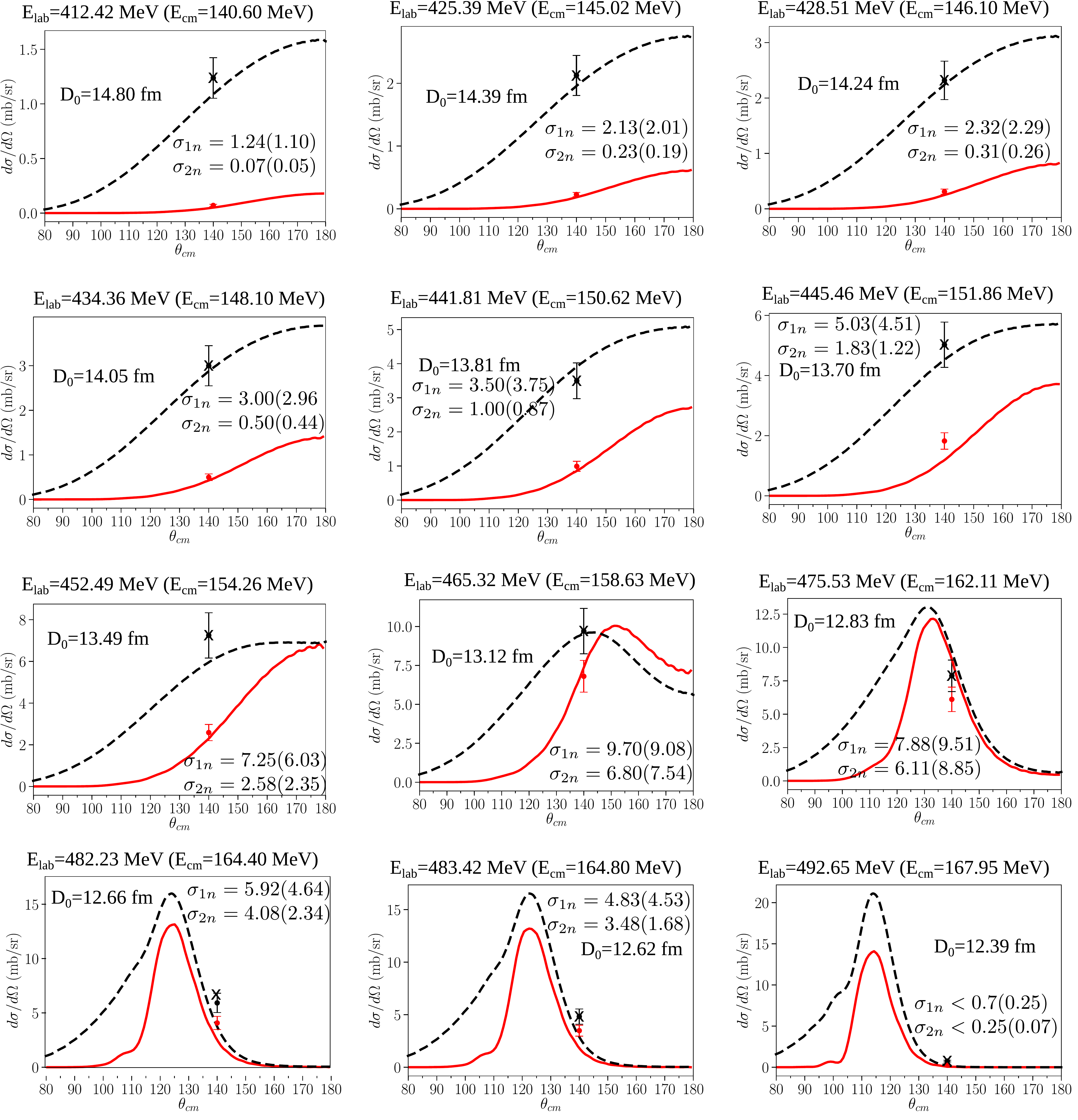}}
	\caption{Angular distribution associated with the reactions $^{60}$Ni+$^{116}$Sn$\to^{62}$Ni+$^{114}$Sn and $^{60}$Ni+$^{116}$Sn$\to^{61}$Ni+$^{115}$Sn, calculated making use of the optical potential of ref. \cite{Montanari:14} (see also ref. \cite{Broglia:04a} p. 111), as well as the spectroscopic amplitudes collected in Table 1 of \cite{Montanari:14}, in comparison with the experimental data (\cite{Montanari:14}) to which an error of 15\% has been assigned. Also given are the one- and two-nucleon transfer cross sections in mb \cite{Montanari:14} (those in parentheses are the theoretical estimates calculated as discussed in the text).  While the calculations of the two-neutron transfer reaction corresponds to a single transition populating the ground state of both the residual ($^{62}$Ni) and the outgoing ($^{114}$Sn) nucleus (inverse kinematics), that of one-particle transfer is inclusive concerning the residual system ($^{115}$Sn, $^{61}$Ni). In fact, the reported theoretical absolute differential cross sections, result from the incoherent sum of $E_{qp}\leq2.5$ MeV  (see also Table I of ref. \cite{Montanari:14}; see also \cite{Lee:09}). This is in keeping with the fact that the one-particle transfer channel corresponds, in the case of condensed matter, to Giaever's \cite{Giaver:73} S-Q ($q=e$) current. In the present case, the bias for the superfluid-normal phase transition is provided by the strain  forcing pairs to tunnel across a barrier of ``width'' (distance of closest approach) $D_0\gtrsim(D_0)_c$ ($\approx$13.5 fm). Thus, the quasiparticle states in the energy range $0\leq E_j\leq2\Delta$ are expected to play, in the nuclear case, the role of $Q$ in Giaver's observations. Within this context, it is of notice that the two-neutron transfer amplitude in the gs$\to$gs ($2n$ transfer) reaction studied, is proportional to the number of Cooper pairs, in a similar way in which the maximum value of the Josephson current is proportional to $\Delta_1\Delta_2/(\Delta_1+\Delta_2)\sim\alpha_1\alpha_2/(\alpha_1+\alpha_2)\sim\alpha$, assuming the two superconductors to be equal.
}\label{fig:7p}
  \end{figure*}

\begin{figure}[h]
	\centerline{\includegraphics*[width=8cm,angle=0]{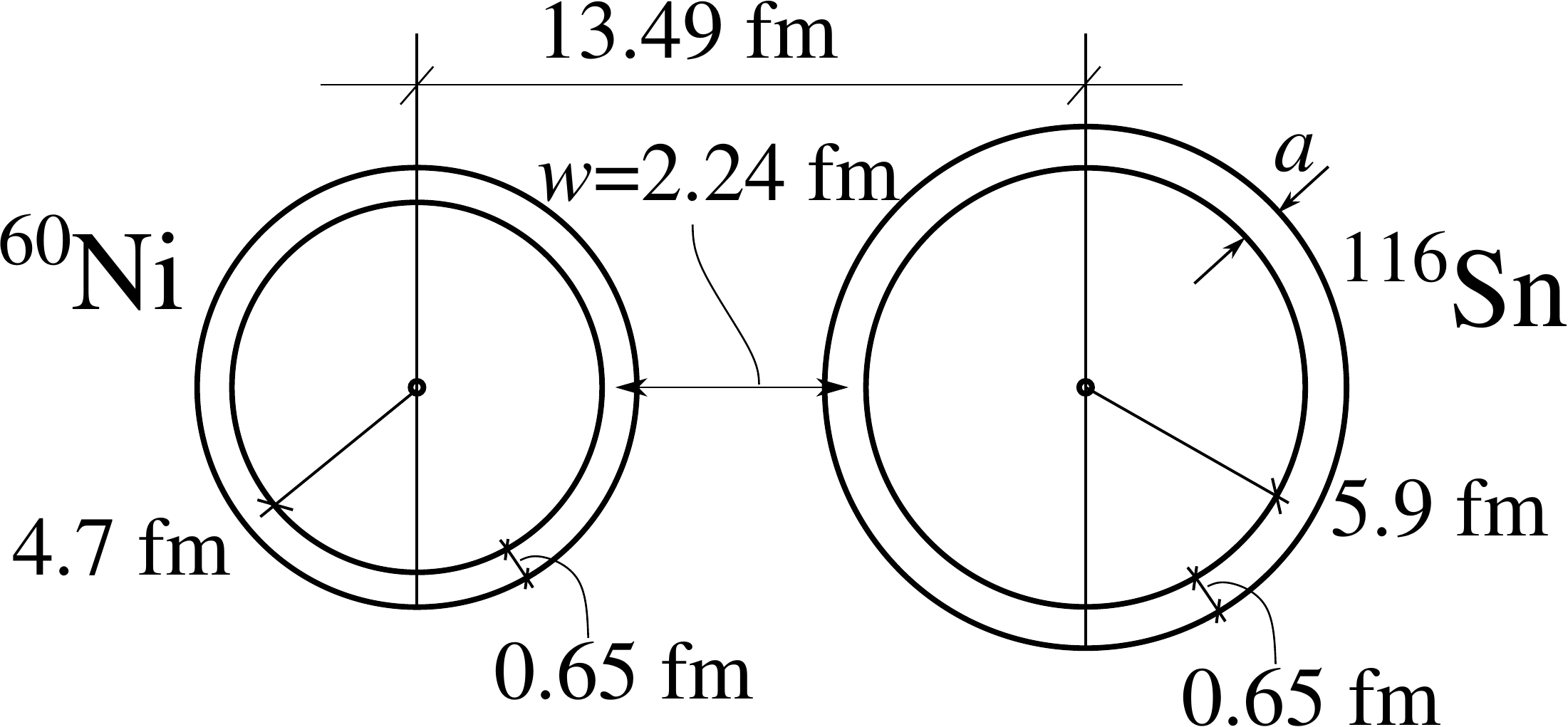}}
	\caption{Schematic representation of the relative position, at the critical distance of closest approach $(D_0)_c(=13.49$ fm), of the two superfluid nuclei $^{116}$Sn of radius $R_0=5.9$ fm and $^{60}$Ni ($R_0=4.7$ fm), corresponding to  a bombarding energy of $E_{cm}=154.26$ MeV and $\theta_{cm}=140^\circ$, which defines a situation in which the distance between the nuclear surfaces (Josephson-like barrier width) is  $w=(D_0)_c-(R_0(^{116}\text{Sn})+(1/2)a)+(R_0(^{60}\text{Ni})+(1/2)a))=2.24$ fm.}
    \label{fig:3}  
\end{figure} 

\begin{figure}[h]
	\centerline{\includegraphics*[width=8cm,angle=0]{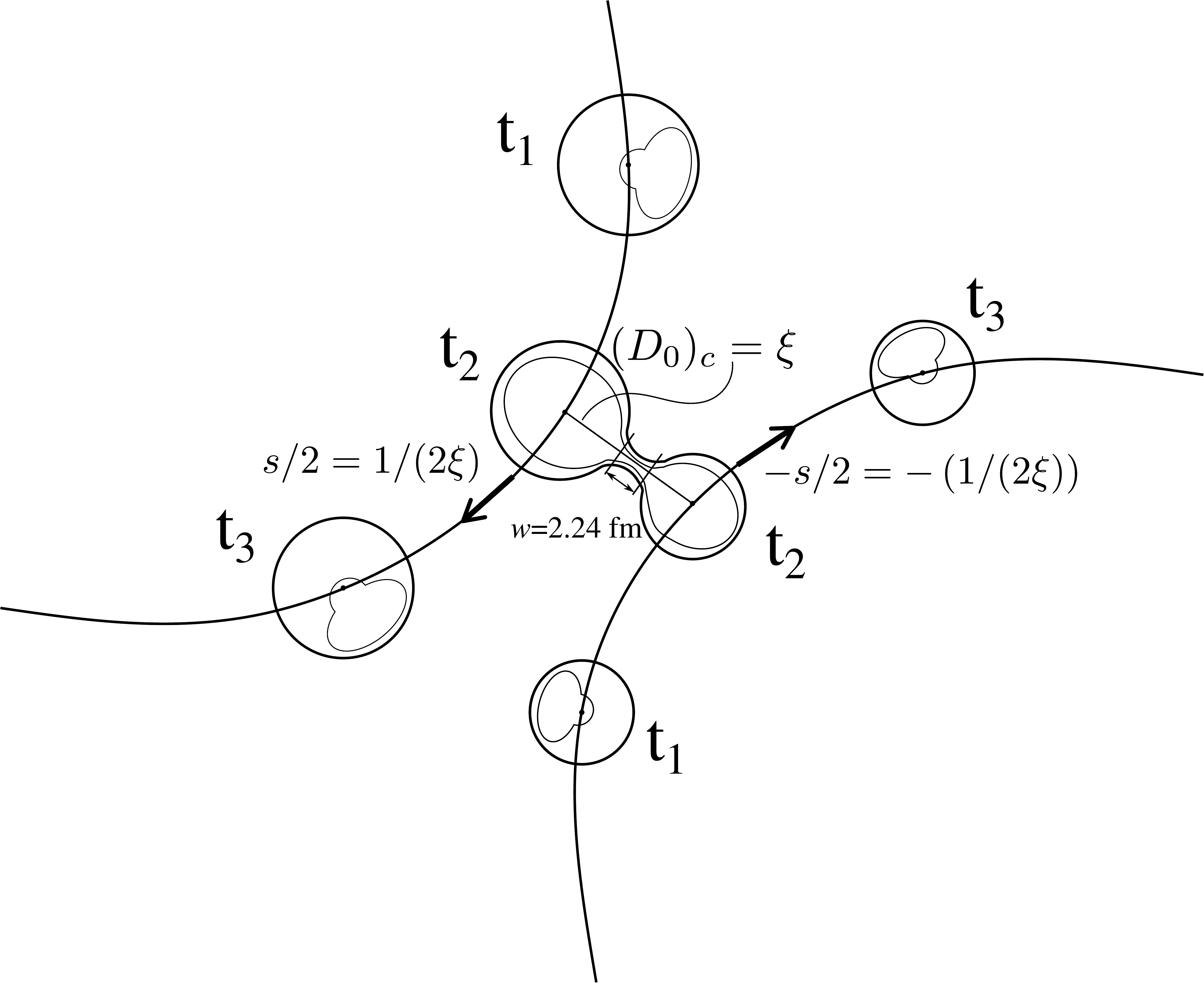}}
	\caption{Schematic representation of the relative motion of the reaction $^{116}$Sn+$^{60}$Ni$\to^{114}$Sn+$^{62}$Ni in the center of mass system, under the kinematic conditions $E_{cm}=154.26$ MeV, $\theta_{cm}=140^\circ$, implying $(D_0)_c=13.49$ fm. The ions are in the entrance (exit) channel at time $t_1(t_3)$, while the transfer --mainly successive-- process in which each of the partner neutrons of the Cooper pair is in a different nucleus takes place at time $t_2$. Indicated are the critical distance of closest approach  $(D_0)_c$ the relative distance between the nuclear surfaces (width of the Josephson-like barrier; $w=2.24$ fm) and the momentum $\pm\hbar/(2\xi)$ acting on the pair partners and thus $\hbar s=\hbar/\xi$ on the center of mass motion of the Cooper pair being transferred (critical momentum). The thin lines describe for ($t_1$,$t_3$) and ($t_3$,$t_1$) the relative distribution of the associated density for one particle fixed at $r_1=R$, as a function of the coordinate of the other partner nucleon. For the transfer situation ($t_2$,$t_2$), where each partner nucleon is in a different nucleus, it represents the intrinsic Cooper pair wavefunction modulus squared $|\varphi_q(\mathbf r)|^2$ ($\mathbf r=\mathbf r_1-\mathbf r_1$).}
    \label{fig:4}  
\end{figure}

\begin{figure}[h]
	\centerline{\includegraphics*[width=8cm,angle=0]{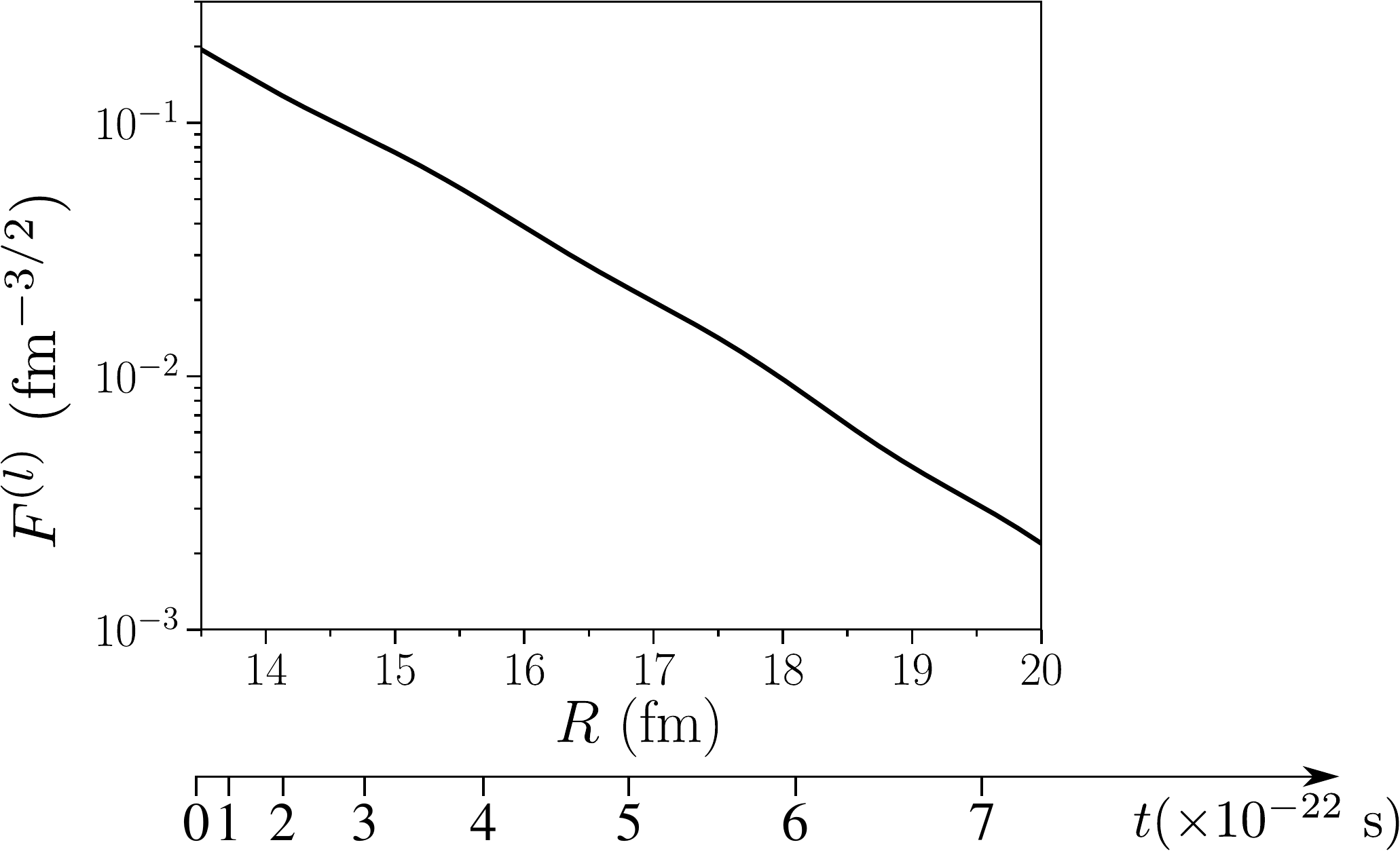}}
	\caption{Local formfactor of the ($2n$) transfer reaction (\ref{abs_eq:6abs}), defined according to Eq. (\ref{eq:112}) as a function of the distance $R$ between the centers of mass of the $^{61}$Ni and $^{115}$Sn nuclei. Also shown are the associated times, according to the classical trajectory followed by the heavy ions during the collision process.}
    \label{fig:form}  
\end{figure}

   \begin{figure}[h]
	\centerline{\includegraphics*[width=5cm,angle=0]{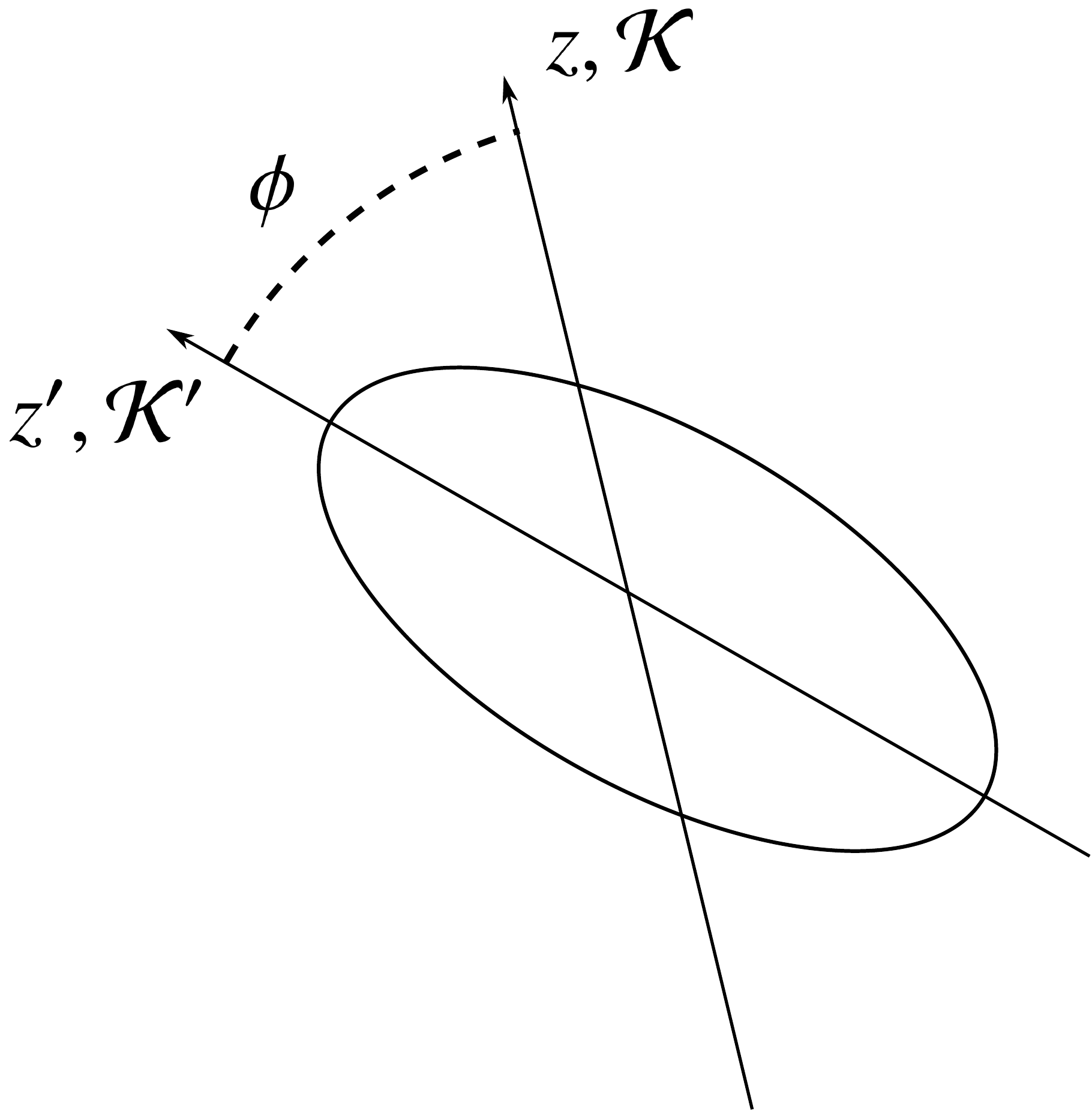}}
	\caption{Schematic representation of a deformed system in gauge space defining a privileged orientation $z'$ in the two-dimensional space and thus an intrinsic, body-fixed, coordinate system of reference $\mathcal K'$, making an angle $\phi$ with the laboratory frame of reference $\mathcal K$.}
    \label{abs_figI}  
\end{figure} 
\begin{figure}[h]
	\centerline{\includegraphics*[width=7cm,angle=0]{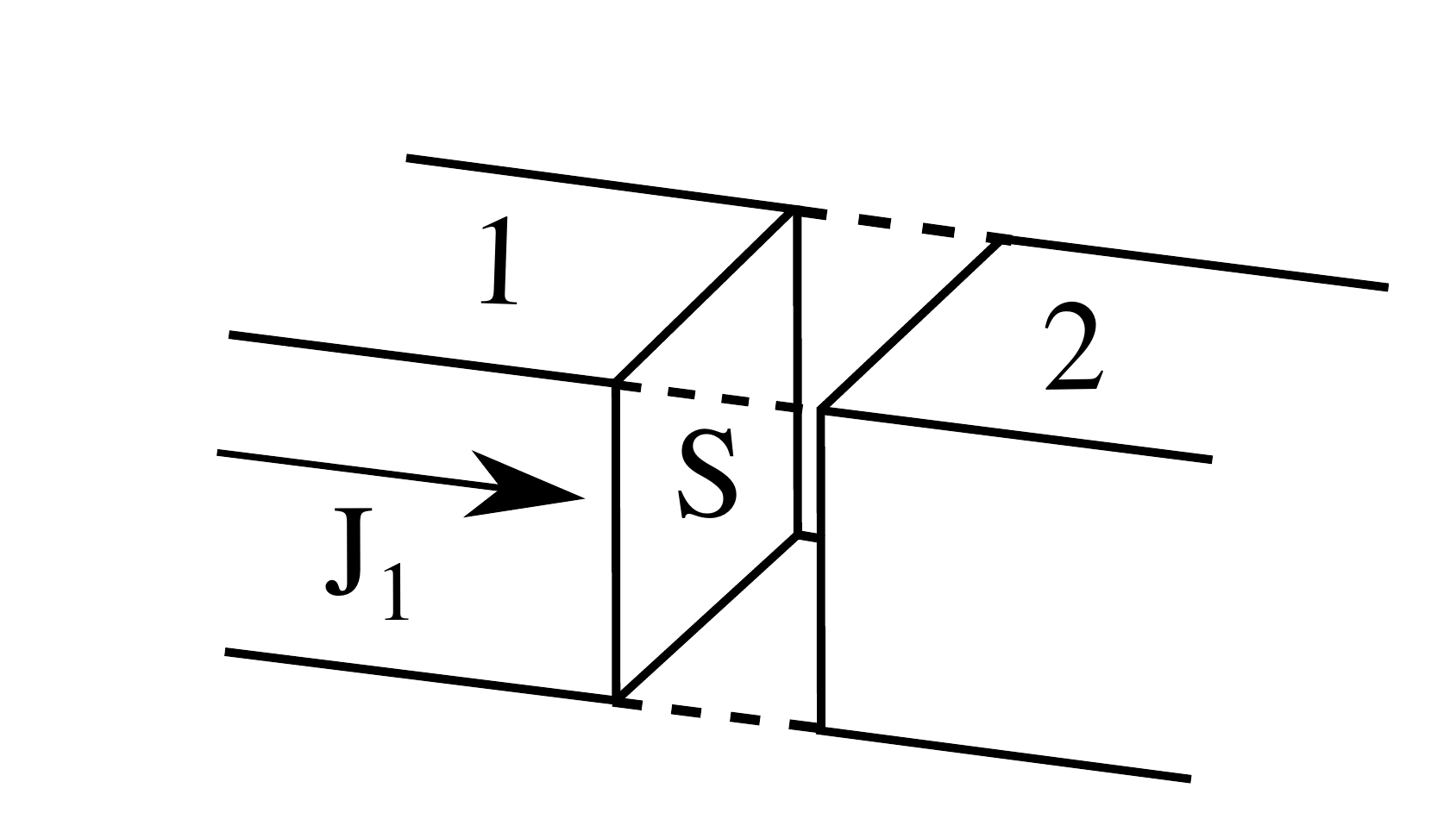}}
	\caption{Schematic representation of the Josephson junction between superconductors 1 and 2. The critical supercurrent corresponds to $J_c=\frac{\pi}{4}\frac{V_{eq}}{R_b}=\frac{\pi}{4}\frac{V}{\Omega S}=\frac{\pi}{4}\frac{A}{S}$, where $V,\Omega, A$ stands for Volt, Ohm and Ampere and the unit surface $S$ stands for square meter ($m^2$). The dashed lines define the (oxide layer) barrier region.}
    \label{fig:6}  
\end{figure} 

 \begin{figure}[h]
	\centerline{\includegraphics*[width=8cm,angle=0]{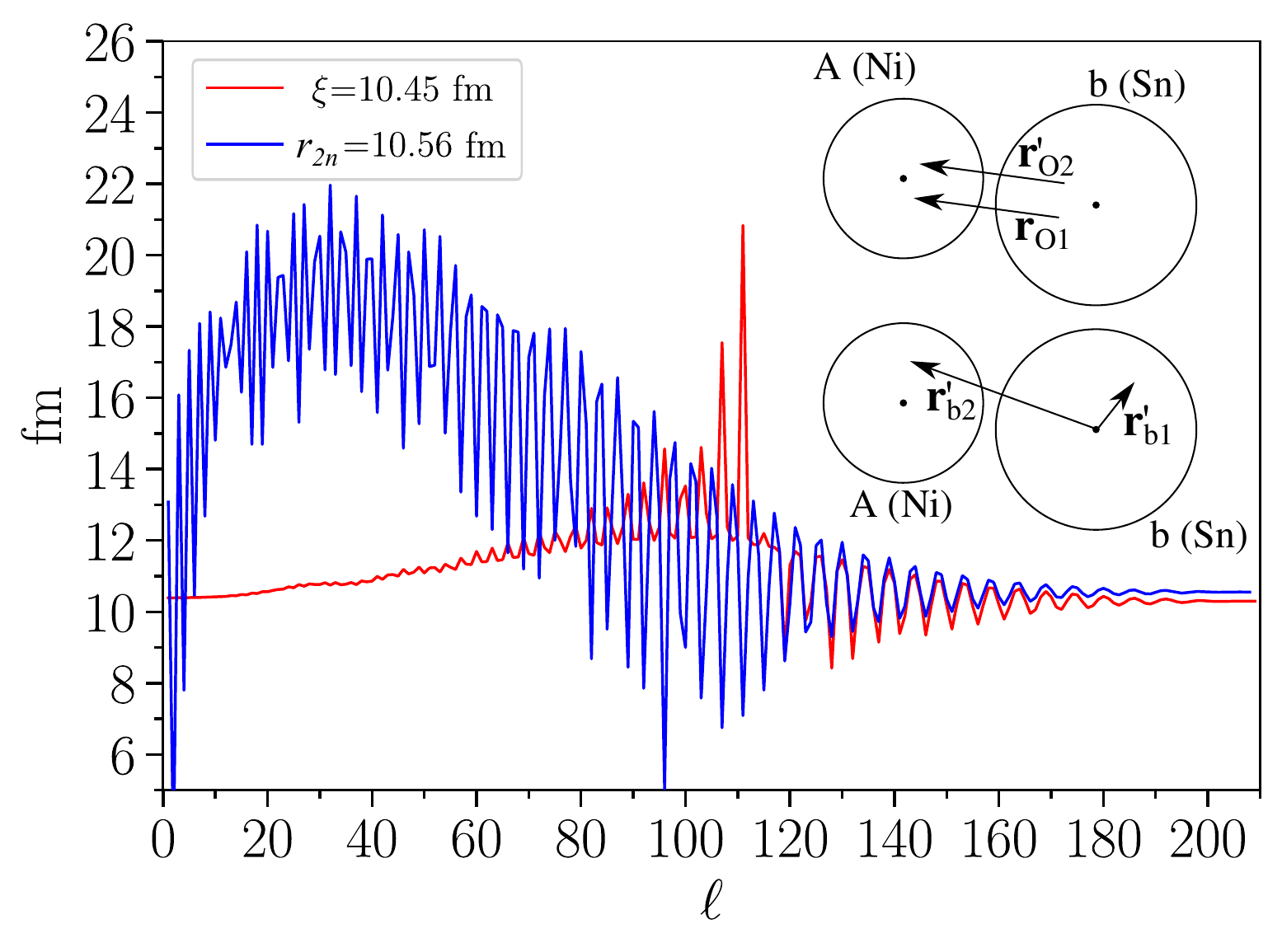}}
	\caption{The modulus of the radial component of the dipole moment $r_{2n}$ (see Eq. (\ref{eq:108})) associated with the successive tunneling of the partner nucleons (both referred to the cm of the Sn +Ni system) of the transferred Cooper pair (color online, blue continuous curve; see Sect. \ref{S14B}), and the correlation length $\xi$ (Eq. (\ref{eq:110})) associated with the transfer of the first partner nucleon from Sn(b) to Ni(A), both partners referred to the cm of Sn, as a function of the partial waves $(0\leq \ell\leq 210)$. In the inset, schematic representation of the coordinates  used in the numerical calculation of $\braket{|\hat O_i|^2}/|T|$, where $\braket{{\;}}$ stands for the average value obtained by inserting the corresponding function modulus squared in the $T$-matrix.}
    \label{abs_figA}  
\end{figure} 
  \begin{figure}[h]
	\centerline{\includegraphics*[width=8.5cm,angle=0]{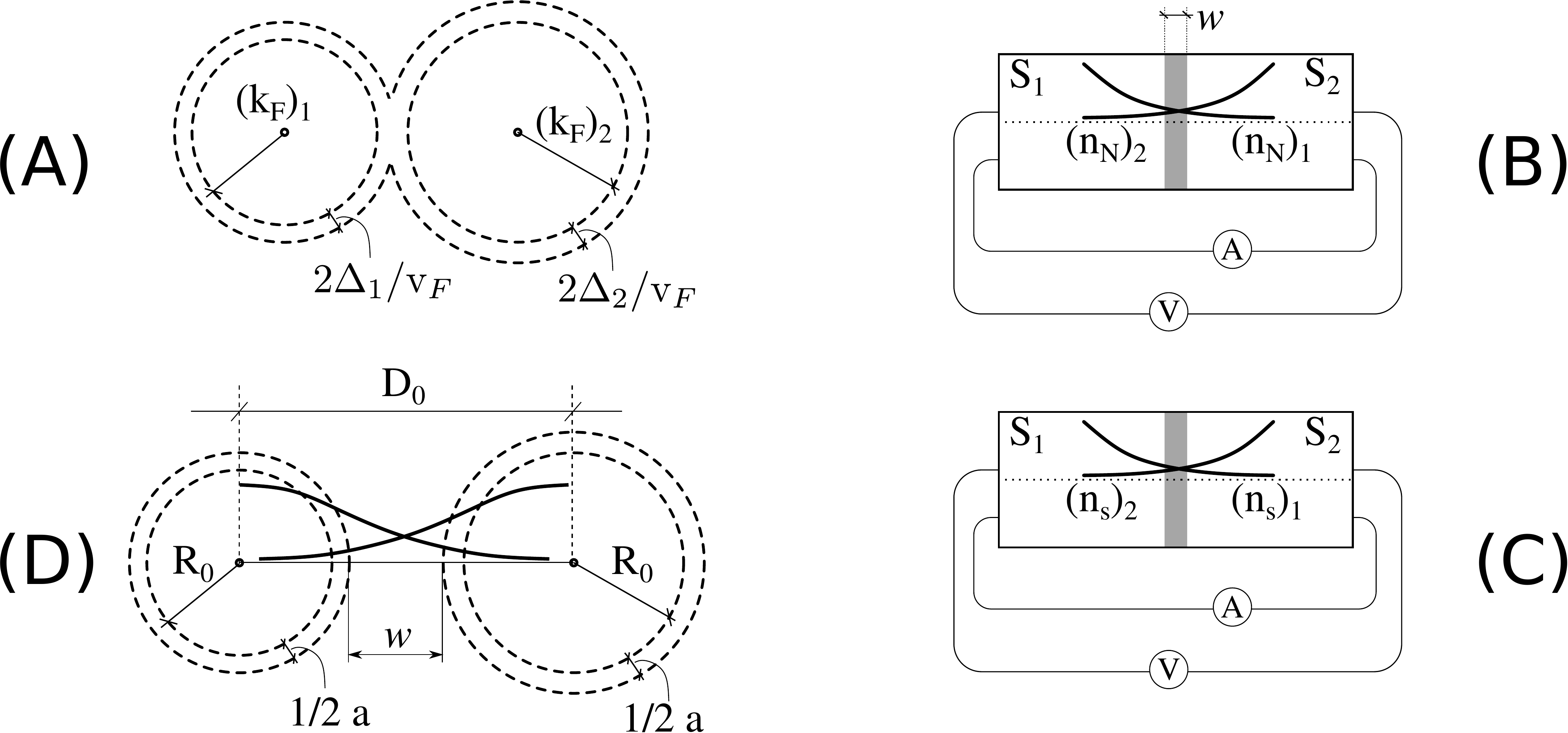}}
\caption{Schematic  of a weak link between two superconductors (S) labeled 1 and 2. \textbf{(A)} The Fermi surface (dashed curve) of the two superfluid nuclei $S_1$ and $S_2$ and associated momentum region around it of thickness $2\Delta_1/\text{v}_F$ and $2\Delta_2/\text{v}_F$. \textbf{(B)} Josephson junction of width $w\,(\approx 1-3$ nm). Also indicated is the overlap of the tails of the normal densities.  \textbf{(C)} Same as (B) but indicating the overlap of the abnormal densities $(n'_s=\alpha_0'/\mathcal V)$. \textbf{(D)} Similar to (A) but for the case of a transient Josephson-like junction associated with a heavy ion collision between two superfluid nuclei ($R_0=1.2A^{1/3}$ fm), at the distance of closest approach. A situation has been selected in which the bombarding energy is smaller than the Coulomb barrier and corresponds to a distance of closest approach $D_0\approx(D_0)_c$, in which case, the overlap between the normal densities has been indicated, while the one associated with the abnormal densities expected is to be similar to this  one.}
    \label{fig:7}  
  \end{figure}

 \begin{figure}[h]
	\centerline{\includegraphics*[width=8cm,angle=0]{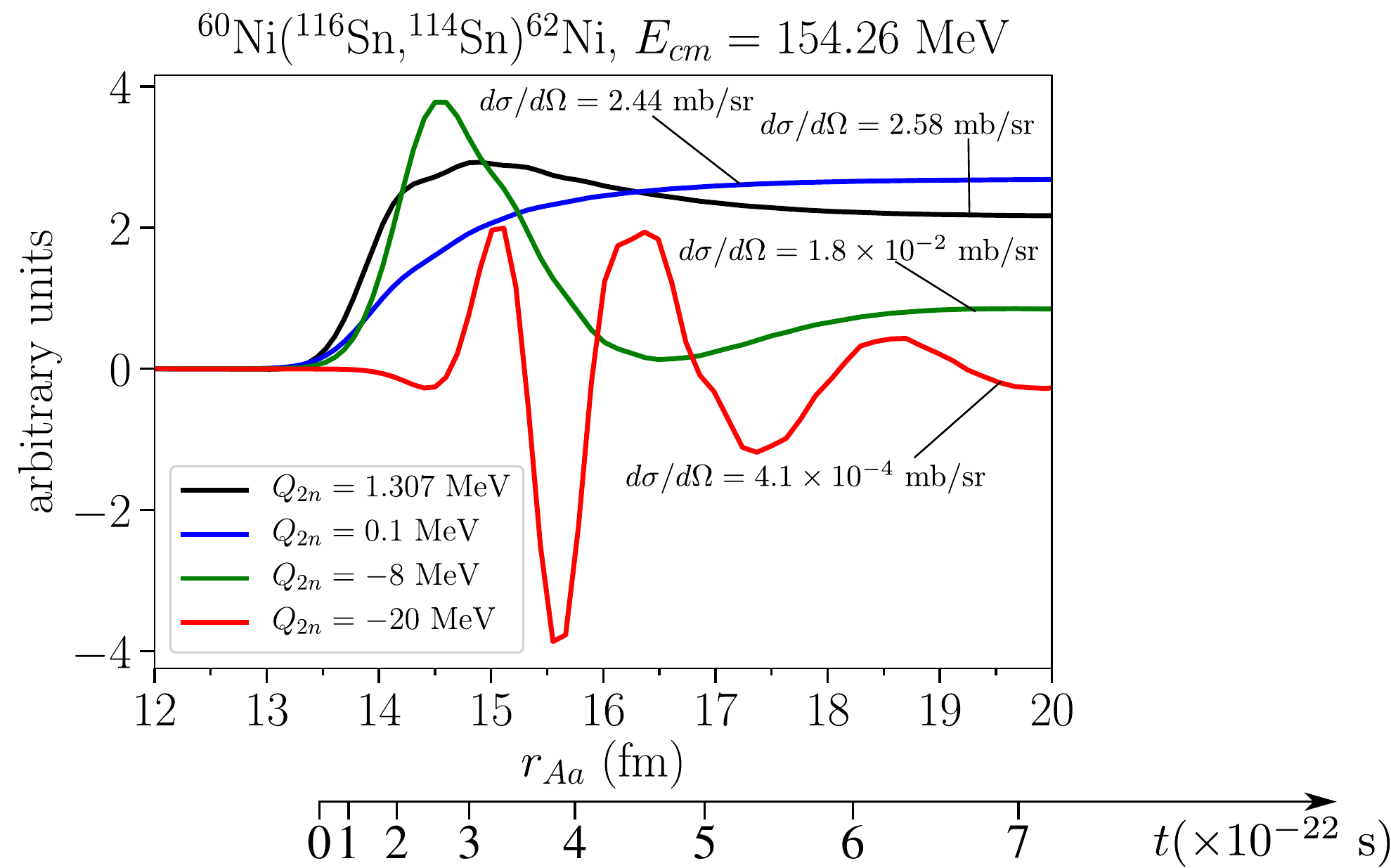}}
	\caption{Cumulative integral of the $T$-matrix as a function of the center of mass distance $r_{Aa}$ of the  two interacting nuclei $^{116}$Sn and $^{60}$Ni for the reaction $Q$-value 1.307 MeV, and for other three selected $Q$-values.  The associated absolute differential cross sections $\left.d\sigma_{2n}/d\Omega\right|_{\theta_{cm}=140^\circ}$ are  displayed. The timescale obtained with the help of the  hyperbola describing the trajectory of relative motion is also shown. Within this scenario, the quantal Cooper pair transfer indicates that for the process under consideration, $\tau_{coll}\approx 0.6-0.7\times 10^{-21}$ s.}
    \label{abs_fig:19}  
\end{figure} 
\begin{figure}
	\centerline{\includegraphics*[width=7cm,angle=0]{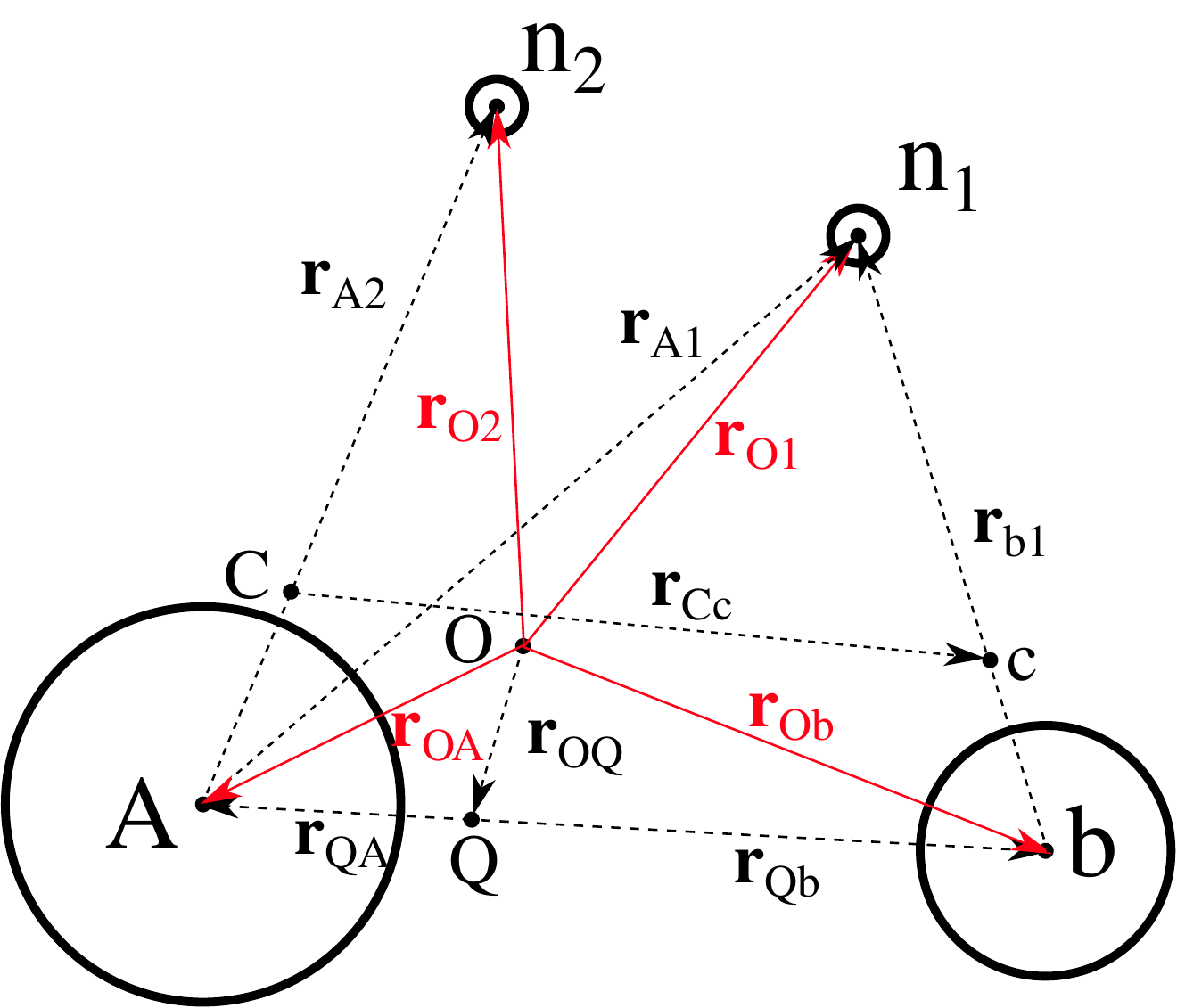}}
	\caption{Schematic representation of the two cores $A,b$ and two neutrons $n_1,n_2$ involved in the transfer process $a(\equiv b+2)+A\to b+B(\equiv A+2)$. The intrinsic electric dipole operator $\mathbf D$ is referred to the center of mass $O$. Also shown is the center of mass $Q$ of the $A,b$ system.}\label{notes_fig1}
\end{figure}

\begin{figure}
	\centerline{\includegraphics*[width=7cm,angle=0]{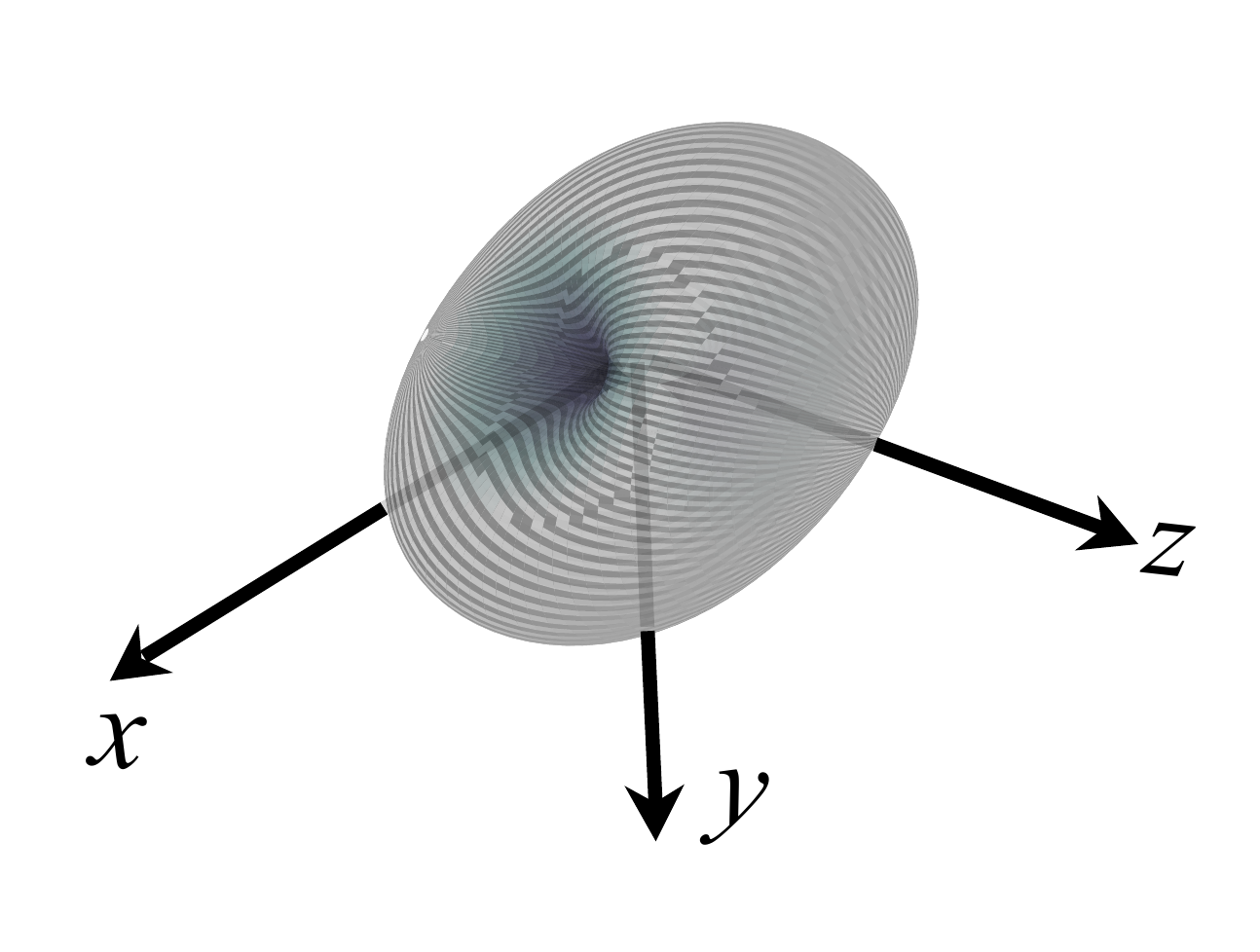}}
	\caption{Angular radiation pattern for $E_\gamma=4$ MeV, $E_{cm}=154.26$ MeV  and $\theta_{cm}=140^\circ$. The radius $r(x,y,z)$ of the surface is proportional to the triple differential cross section (\ref{notes_eq:303}).}\label{notes_fig2}
  \end{figure}
  \begin{figure}
	\centerline{\includegraphics*[width=9cm,angle=0]{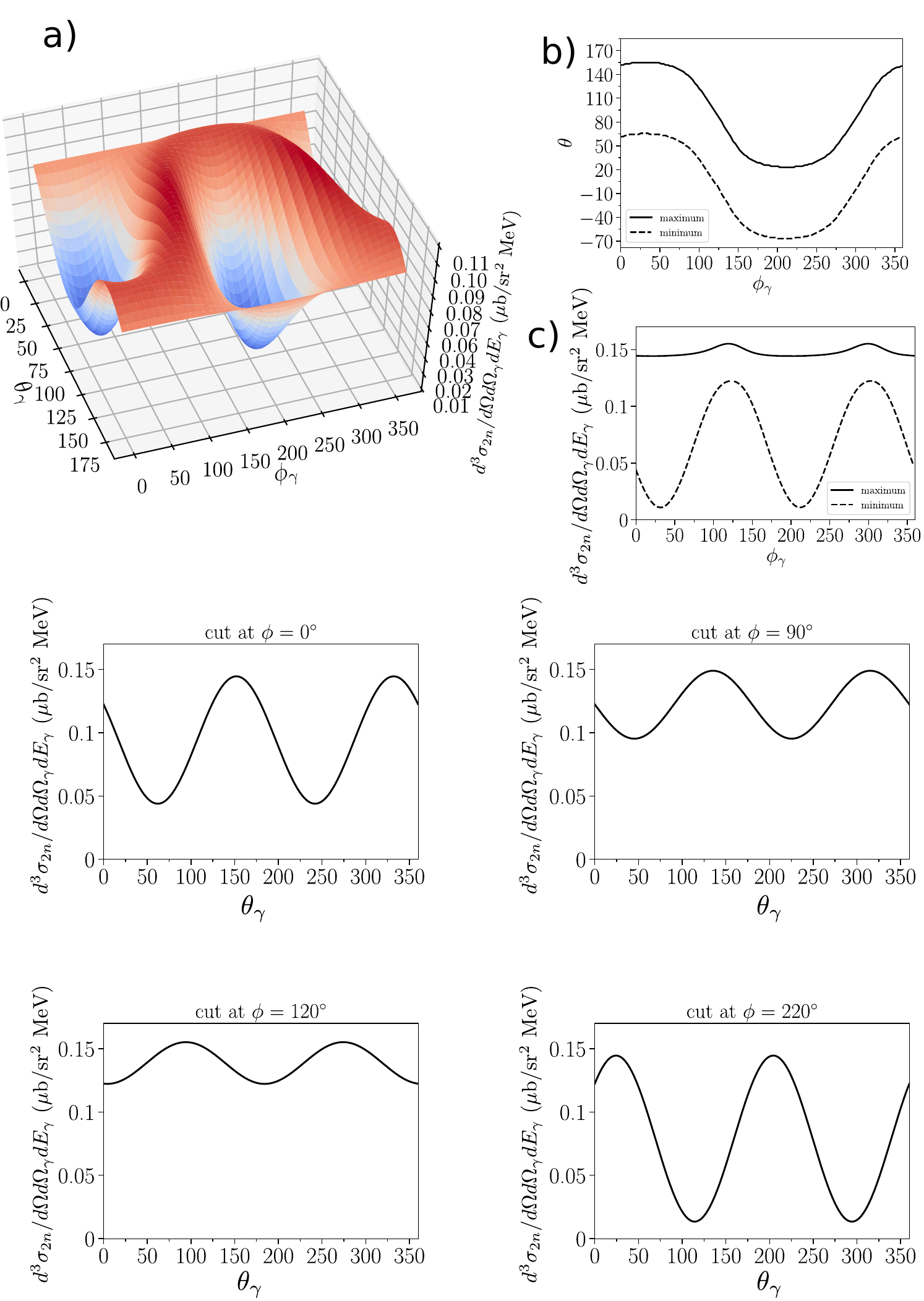}}
	\caption{a) Angular radiation pattern for $E_\gamma=4$ MeV, $E_{cm}=154.26$ MeV and $\theta_{cm}=140^\circ$ as a function of $\theta_\gamma$ and $\phi_\gamma$ (see Eq. (\ref{notes_eq:303})). b) The polar angle $\theta$ 
 for which the $\gamma$-radiation strength has its maximum (continuous curve) and its minimum (dashed curve) as a function 
 of the azimutal angle $\phi_{\gamma}$  and c) the associated values of the 
 triple differential cross section. We also show cuts of the 3-dimensional figure at constant values of $\phi_\gamma=0^\circ,90^\circ,120^\circ$ and $220^\circ$.}\label{notes_fig2x}
\end{figure}    

\begin{figure}
	\centerline{\includegraphics*[width=9cm,angle=0]{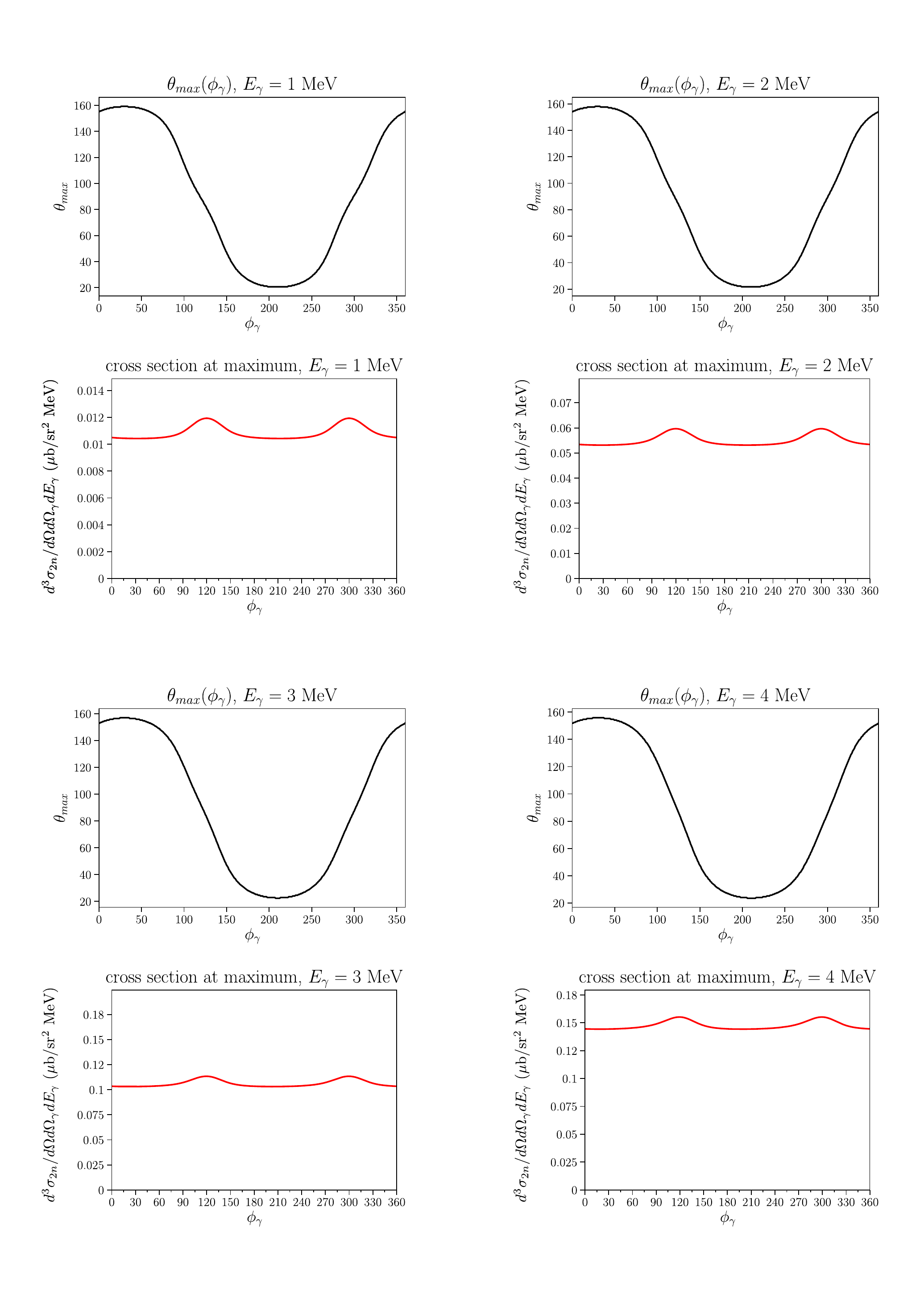}}
	\caption{Making use of Eq. (\ref{notes_eq:139}), the angle $\theta_{max}$ for which the radiation is strongest has been determined as a function of $\phi_\gamma$ for four different $\gamma$ energies (black curves). Also shown (red curves) is the value of the differential cross section (\ref{notes_eq:303}) for the corresponding maximum.}\label{notes_fig4}
\end{figure}
\begin{figure}
	\centerline{\includegraphics*[width=6cm,angle=0]{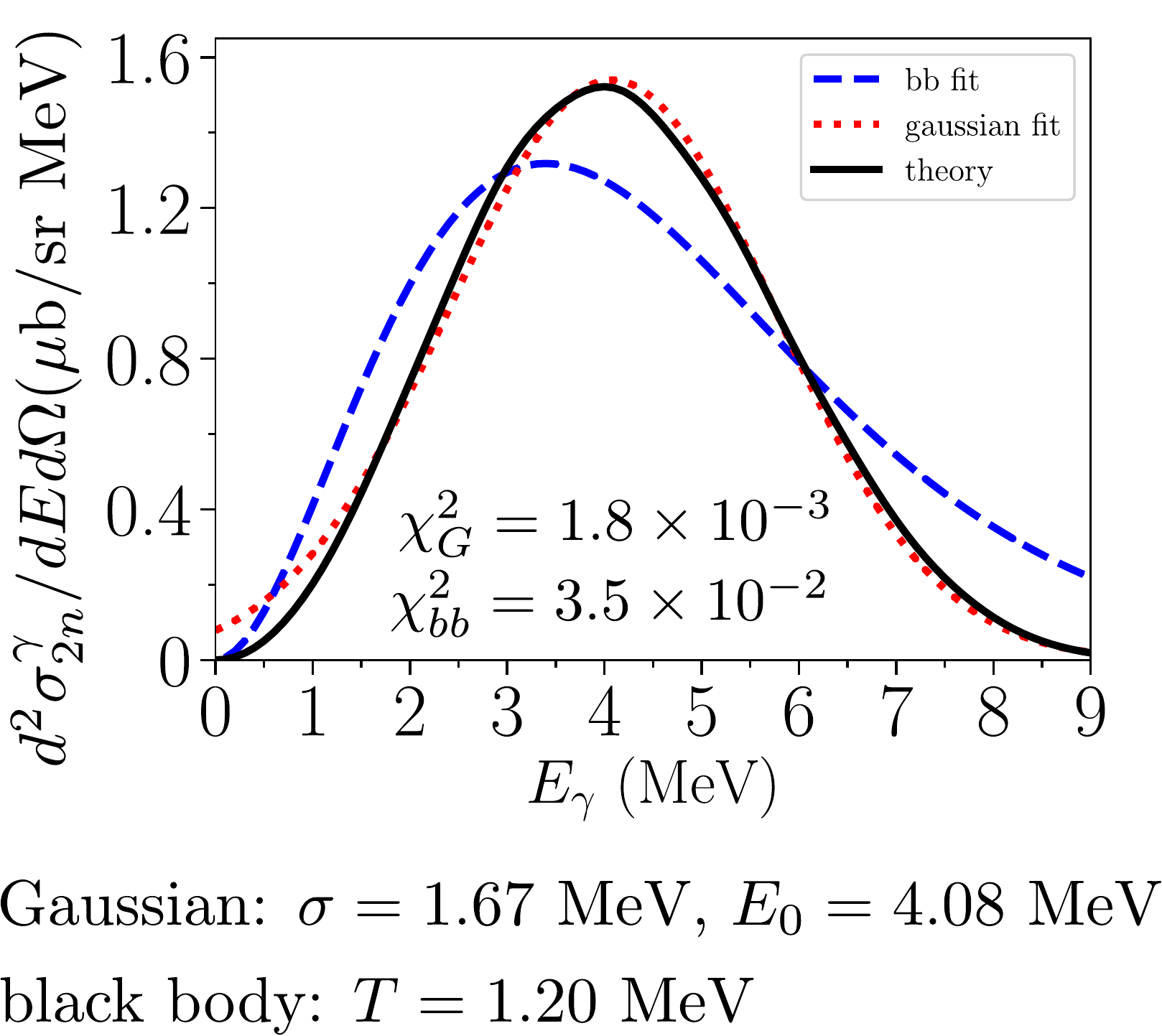}}
	\caption{(Continuous black line) double differential cross section (\ref{notes_eq:306}) for $\theta_{cm}=140^\circ$ as a function of the energy of the $\gamma$ ray. Integrating over the $\gamma$ ray energy, one obtains $(d\sigma/d\Omega)_\gamma=\int (d^2\sigma/dE_\gamma d\,\Omega) \,dE_\gamma=6.9\,\mu\text{b}/\text{sr}$. The corresponding absolute two-nucleon transfer differential cross section is 2.35 mb/sr, to be compared with the experimental value 2.58 ($\pm 0.39$) mb/sr \cite{Montanari:14}. The ratio (6.90 $\mu$b/sr) / (2.35 mb/sr) $\approx2.94\times10^{-3}$ corresponds to the number of photons emitted per cycle \cite{Potel:21}. Also shown are the $\chi^2$ fits and corresponding parameters of Gaussian (red dotted line) and blackbody (blue dashed line) line shapes.}\label{notes_fig5}
\end{figure}

\begin{figure}
	\centerline{\includegraphics*[width=9cm,angle=0]{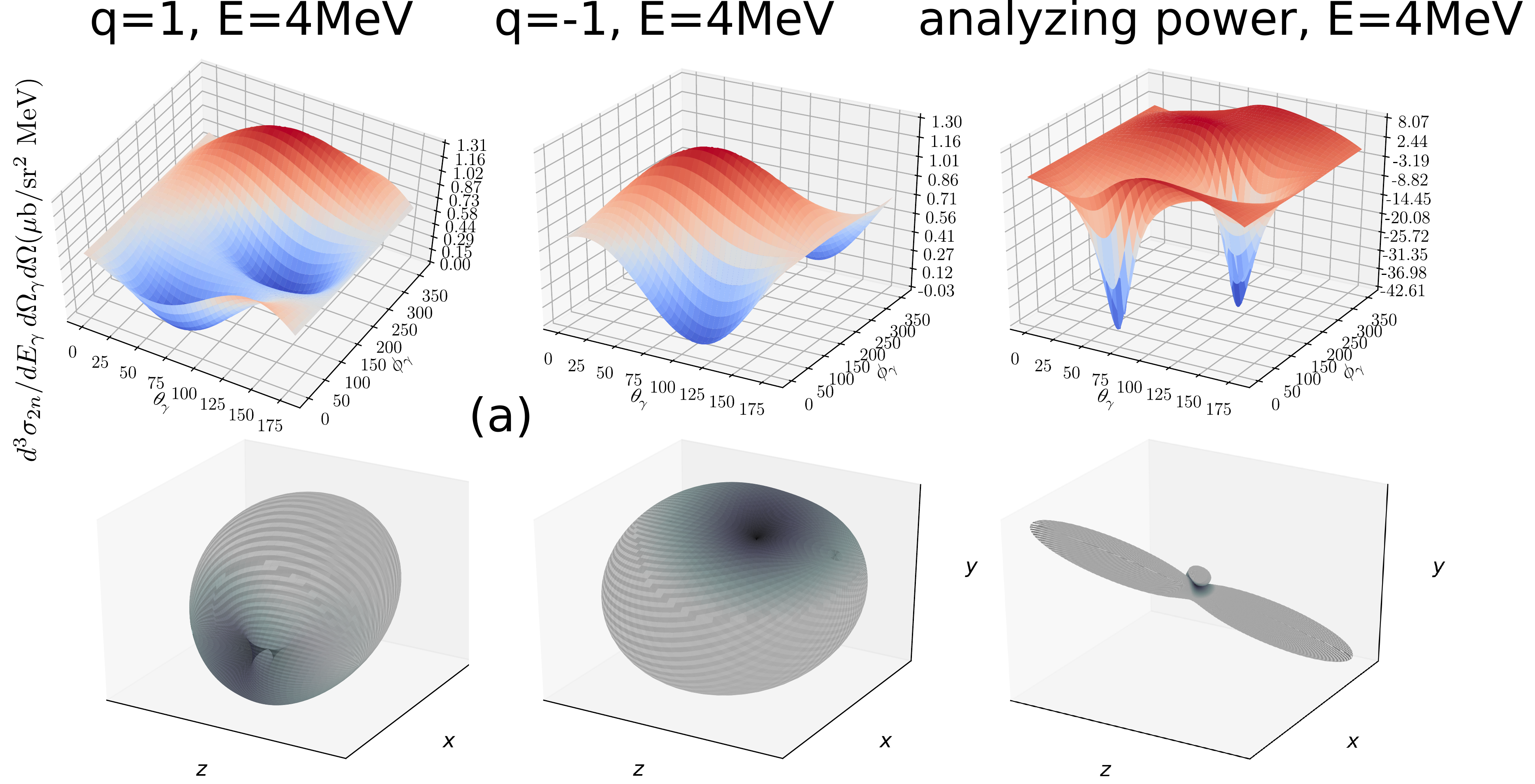}}
	\caption{({\bf{a}}) Polarization observables in both spherical and cartesian coordinates associated with the $\gamma$-emission of Cooper pair tunneling calculated with the same kinematic conditions as in the previous figures, and making use of Eqs. (\ref{eq:2600}) and (\ref{eq:3600}).}\label{fig:pola}
\end{figure}

\begin{figure}
	\centerline{\includegraphics*[width=8cm,angle=0]{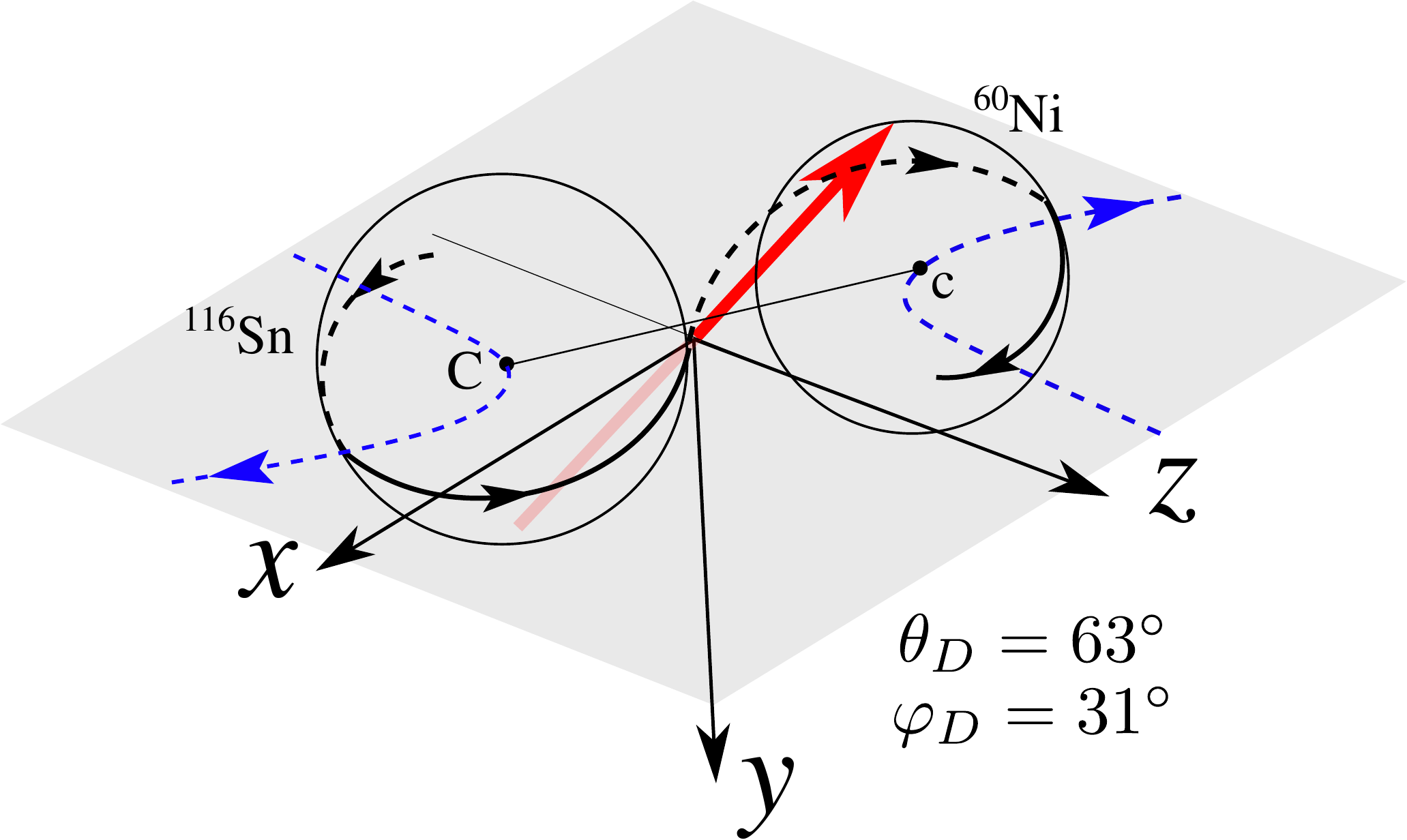}}
	\caption{Representation of the dipole $\mathbf D$ (red arrowed) defined in Eq. (\ref{eq:60}) for $E_\gamma=4$ MeV, and $\theta_{cm}=140^\circ$. The modulus $|\mathbf D|=e_{eff}\times10.56$ fm has been calculated with Eq. (\ref{notes_eq:2}), while the angles defining its orientation where obtained making use of the expressions given in Eq. (\ref{notes_eq:307}). The dashed (blue) curves are schematic representations of the Sn (lower left corner) and Ni (upper right corner) trajectories in the c.m. frame. Black arrowed lines correspond to the smooth trajectory of each of the two Cooper pair transfer partners (see Fig. 8, p. 326 of \cite{Broglia:04a}).}\label{notes_fig3}
\end{figure}

\begin{figure}[h!]
\centerline{\includegraphics*[width=6cm,angle=0]{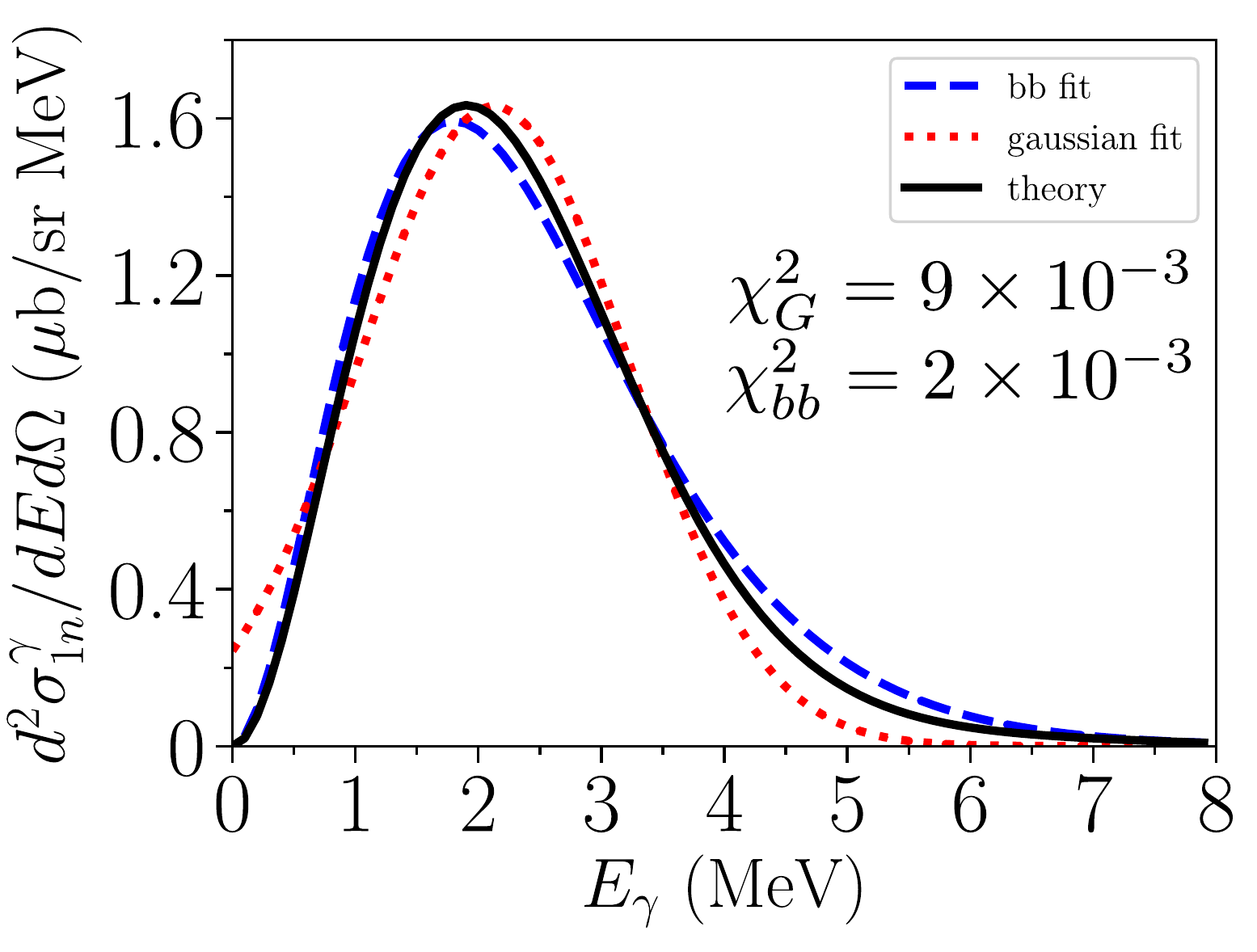}}
	\caption{Gamma-strength function corresponding to the incoherent contributions of the  single-quasiparticle transitions to  $^{61}$Ni and $^{115}$Sn states with  $E_{qp}\lesssim$2.5 MeV. The Gaussian and blackbody ($bb$) line shape fits leads to  the ratio $\chi^2_{bb}/\chi^2_{G}=2\times10^{-3}/(9\times10^{-3})=0.22$, the associated blackbody temperature   being $T=0.64$ MeV.}
  \label{fig1pt}
\end{figure}

\begin{figure}[h]
	\centerline{\includegraphics*[width=8cm,angle=0]{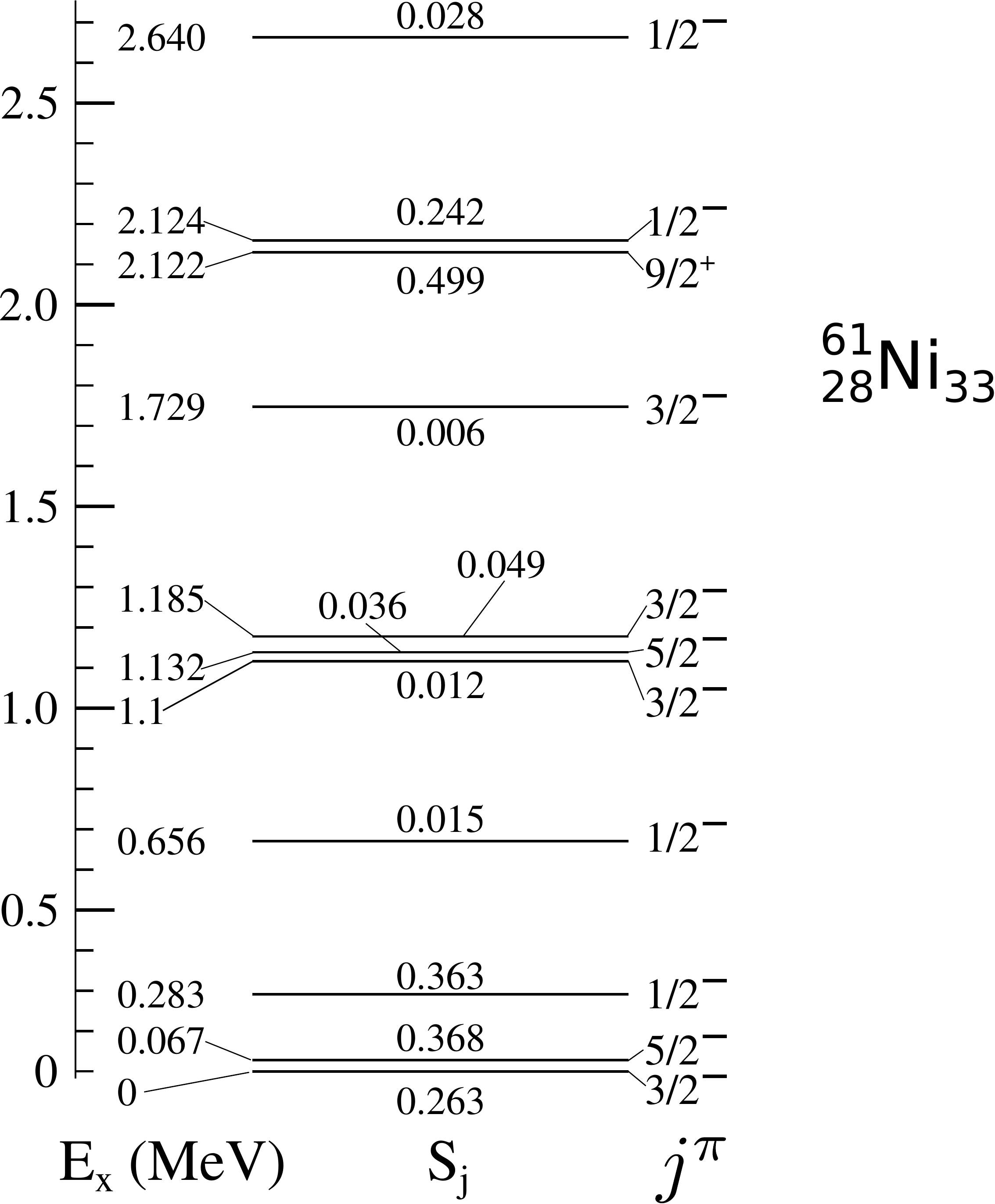}}
	\caption{Single-particle levels of $^{61}$Ni included in the analysis of the reaction $^{116}$Sn+$^{60}$Ni$\to^{115}$Sn+$^{61}$Ni \cite{Montanari:14}. The excitation energy, spectroscopic factor and spin and parity are given for each level \cite{Lee:09}.}
    \label{fig:5x}  
  \end{figure}


\begin{figure}[ht]
\begin{center}
\includegraphics[width=10.cm]{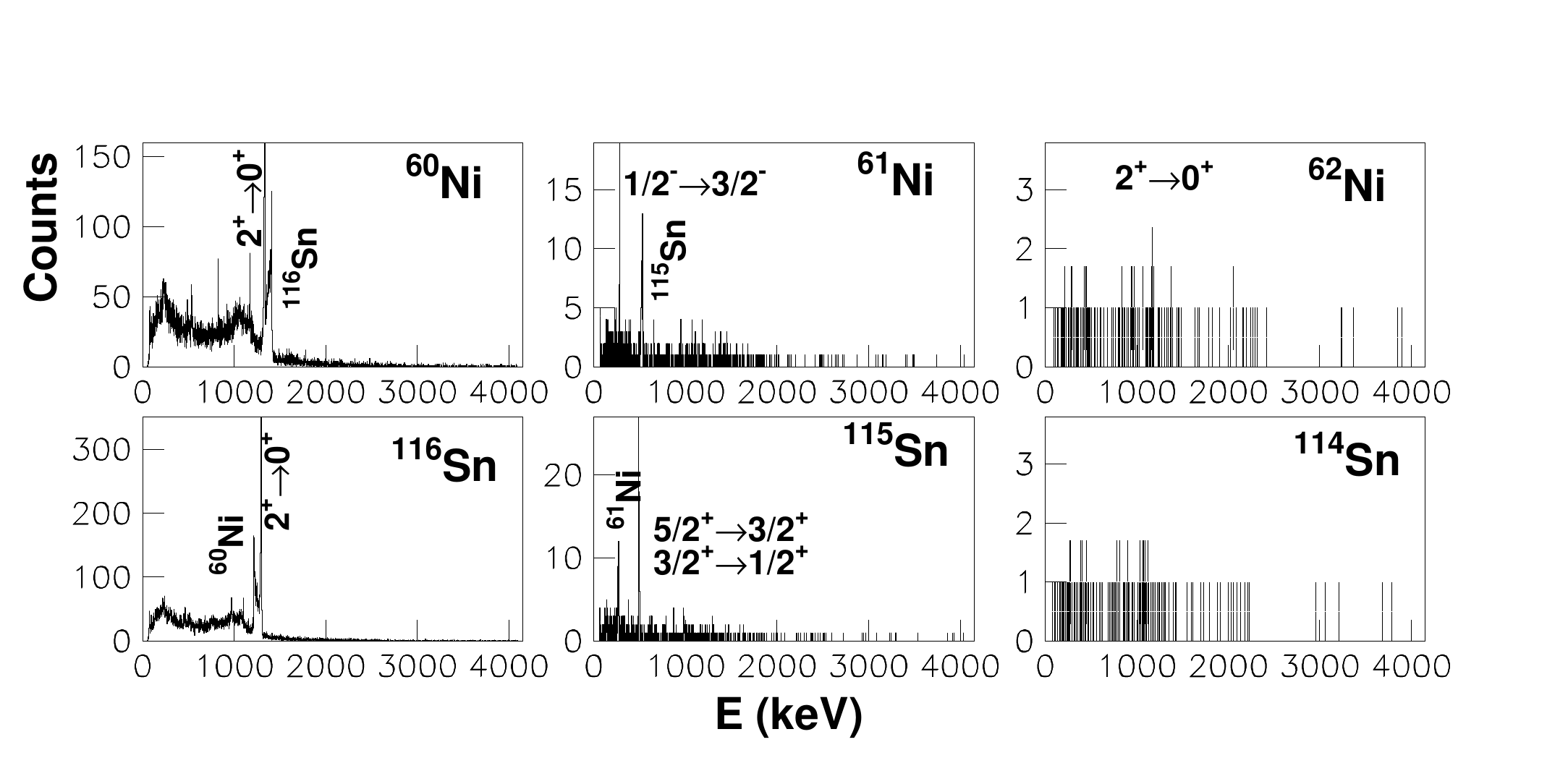}
\end{center}
\caption{
 Top panels:  Doppler corrected $\gamma$ 
 spectra for $^{60}$Ni, $^{61}$Ni and $^{62}$Ni  
 detected in PRISMA in coincidence with the AGATA Demonstrator, 
 displayed in the full energy range up to 4 MeV.  
 Bottom panels: spectra 
 Doppler corrected for the heavy binary 
 partners $^{116}$Sn, $^{115}$Sn and $^{114}$Sn (the bin is 1 keV wide). 
 The strongest transitions are labelled with the    
 spin and parity of initial and final states.   
 The broader peaks correspond to the wrongly Doppler corrected reaction 
 partner. 
 Details of this measurement, performed in direct kinematics, 
 are in Ref. \cite{Montanari:16}.}
\label{fig:g2}
\end{figure}

 \begin{figure}[ht]
\begin{center}
\includegraphics[width=9cm]{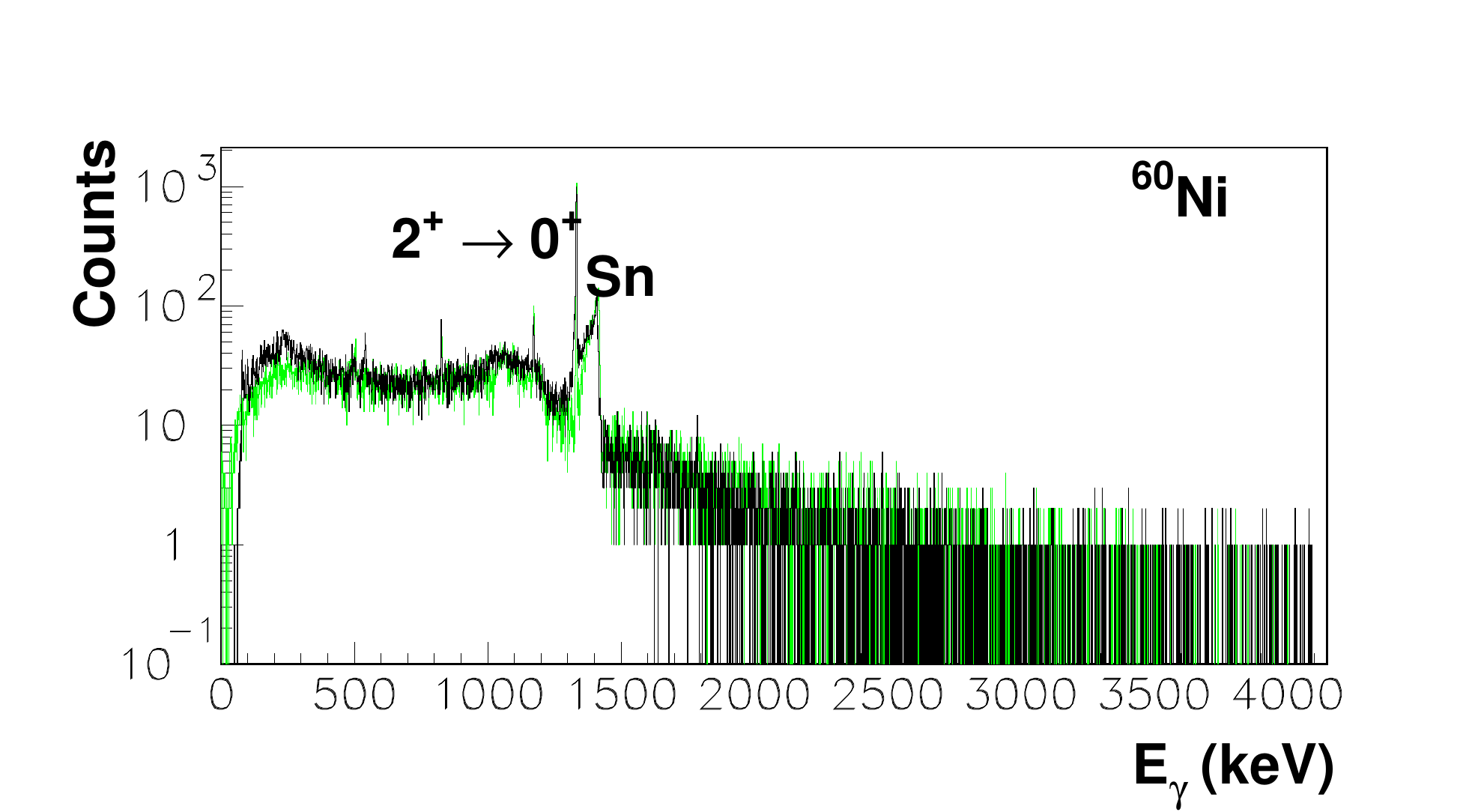}
\end{center}
\caption{
 Experimental (black histogram) and simulated 
 (green histogram) $\gamma$-ray spectrum for $^{60}$Ni.
 The simulation has been done 
 for the Agata Demonstrator configuration, assuming four triple clusters.
 The simulated spectrum was normalized to the measured one in the region 
 of the discrete transitions.
}
\label{fig:sim1}
\end{figure}

  \begin{figure}[ht]
\begin{center}
\includegraphics[width=9cm]{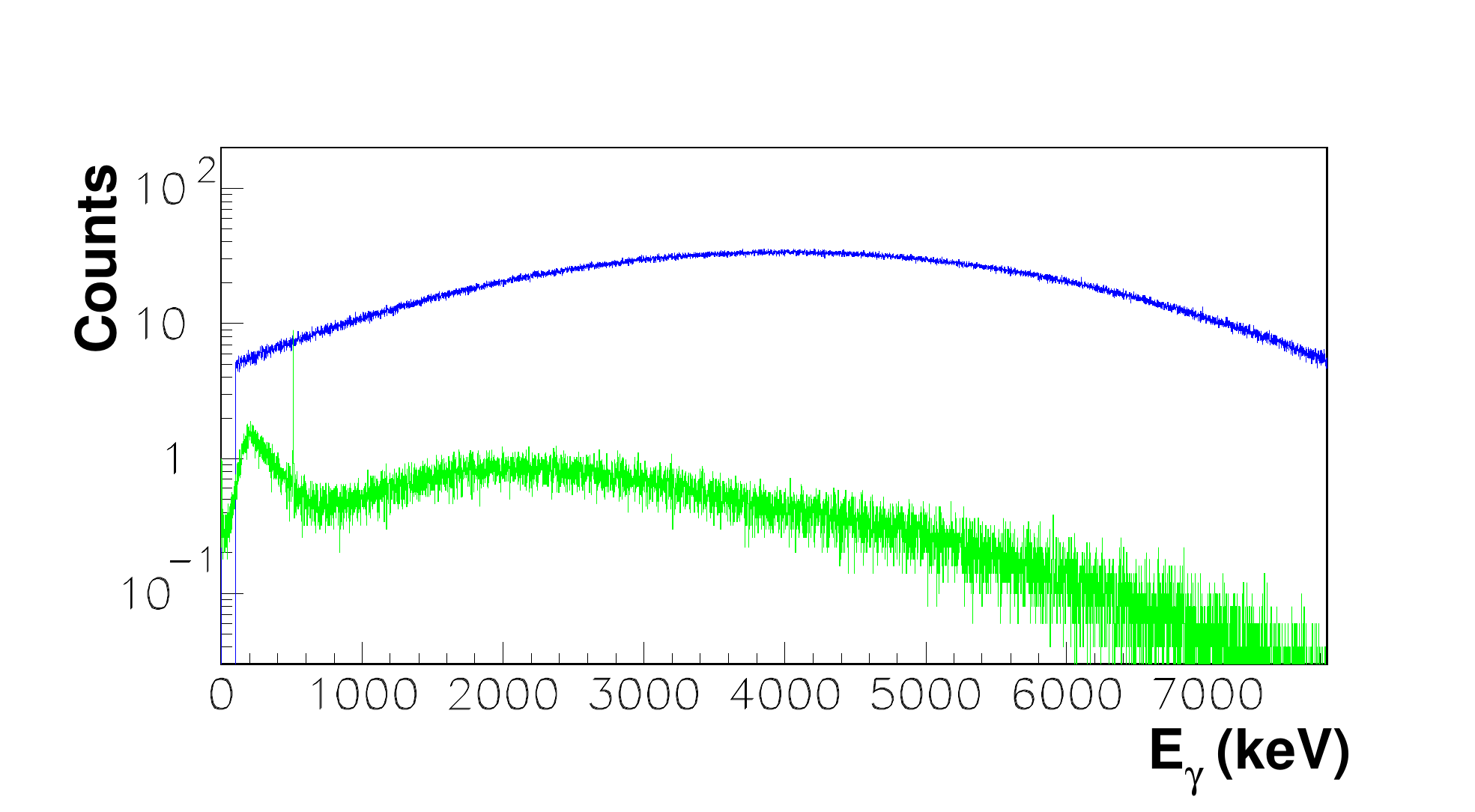}
\end{center}
\caption{Simulated $\gamma$ spectra of $^{62}$Ni.  
 The blue histogram is the gamma-strength function, input distribution,  while 
 the green histogram is the final distribution,  
 reconstructed by using the AGATA tracking algorithm. 
 It is of note that, for didactical reasons,  total counts of the output (and input) distributions are a factor 
 $\approx$ 1000 larger than the one of Fig. \ref{fig:tracked2} 
 (magenta histogram). 
}
\label{fig:tracked}
\end{figure}

 \begin{figure}[ht]
\begin{center}
\includegraphics[width=9cm]{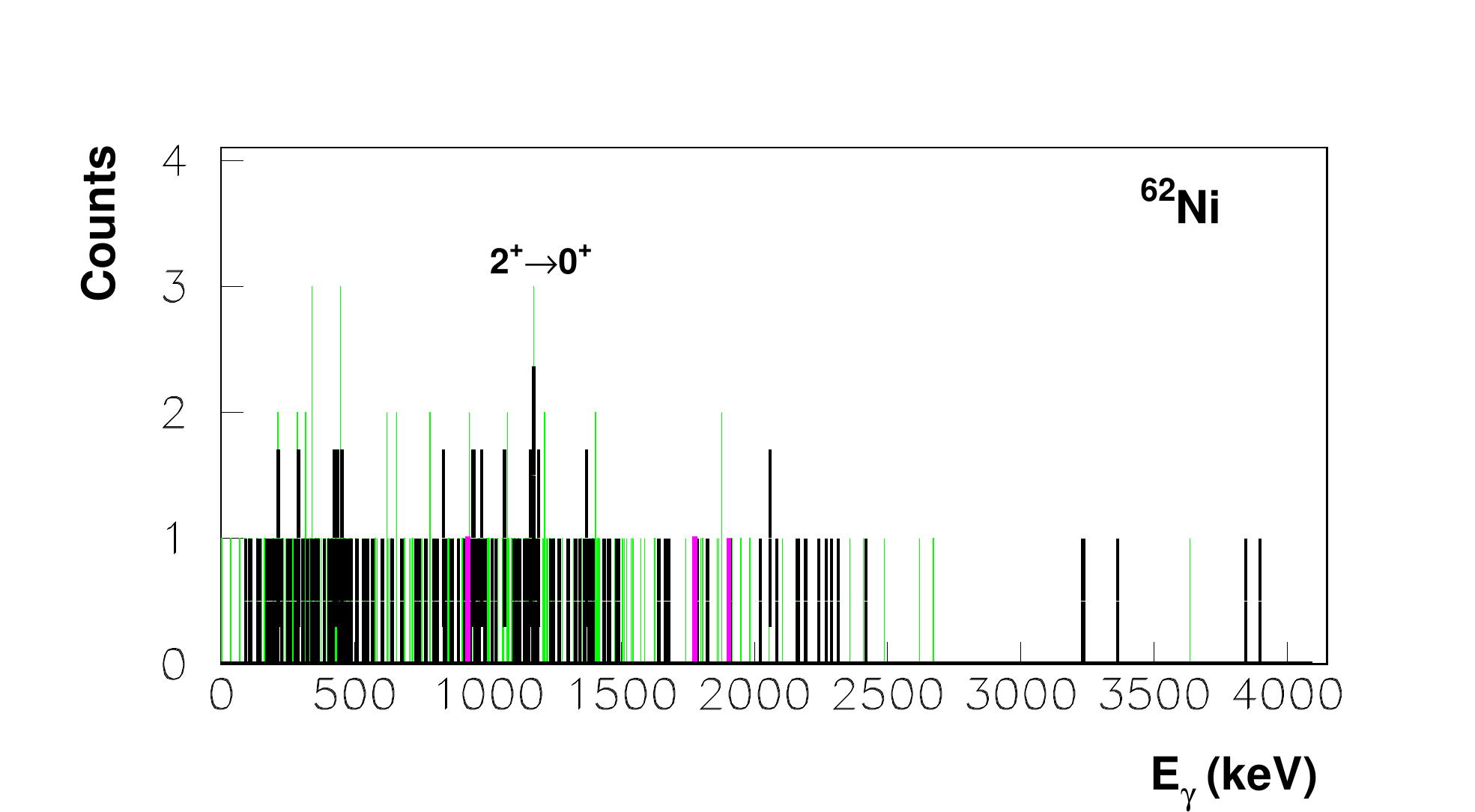}
\end{center}
\caption{Spectrum for $^{62}$Ni. Data (black) are compared with  
 the simulated distribution as discussed in connection with 
 Fig. \ref{fig:sim1} (green). 
 The simulation of the predicted strength function (magenta)  
 incorporates the predicted gamma-strength function. 
}
\label{fig:tracked2}
\end{figure}


\begin{figure}[h]
	\centerline{\includegraphics*[width=6cm,angle=0]{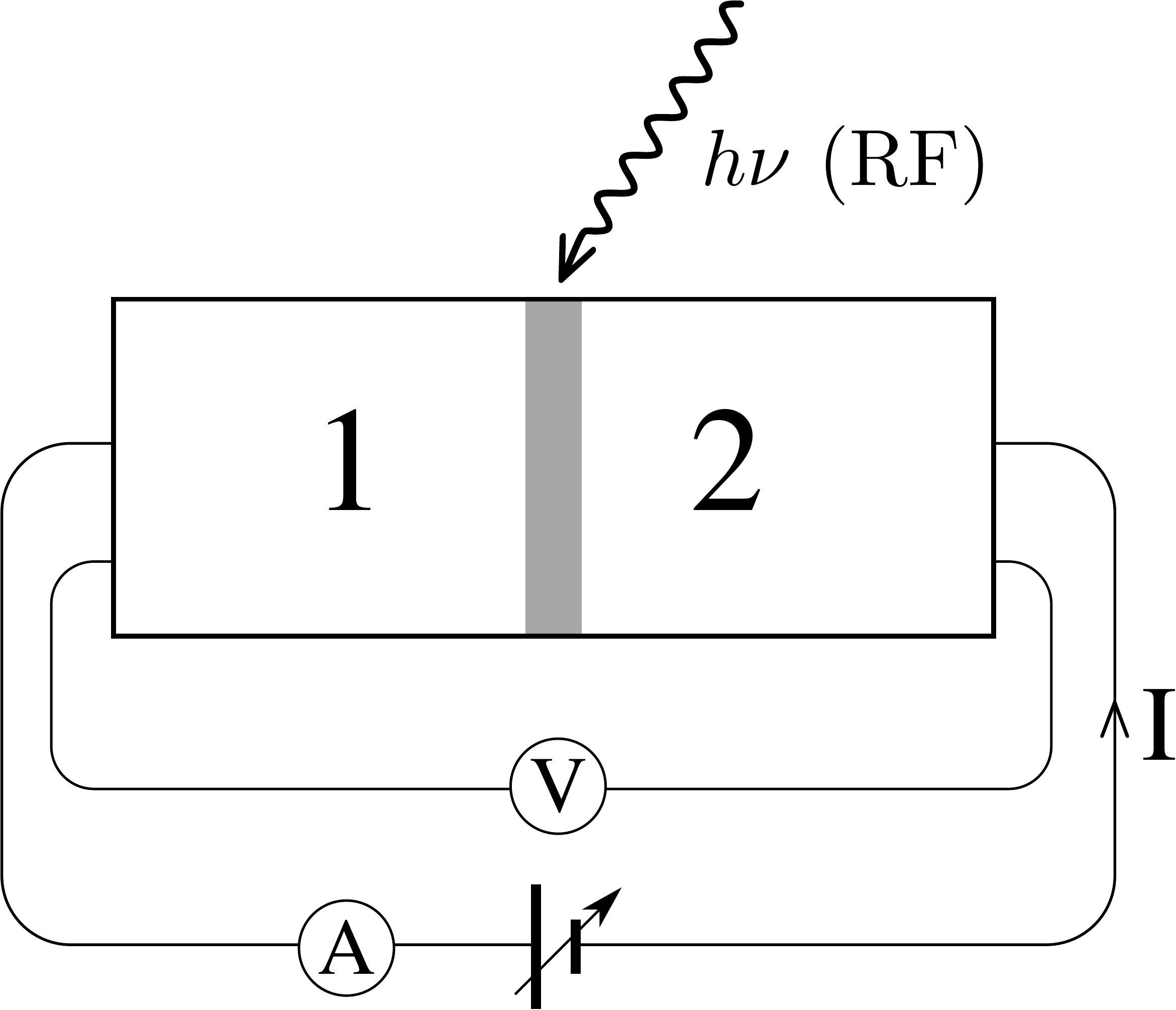}}
	\caption{The selected dc value of $I$ fixes the value of the dc bias $V_0$ of the junction, while $V_1$ is determined by subjecting the junction to  photons of microwave frequency $\nu$ and energy $h\nu$ (see Apps. \ref{AppB} and \ref{AppE}).}
    \label{fig:B1}  
\end{figure} 
  \begin{figure}[h]
	\centerline{\includegraphics*[width=6cm,angle=0]{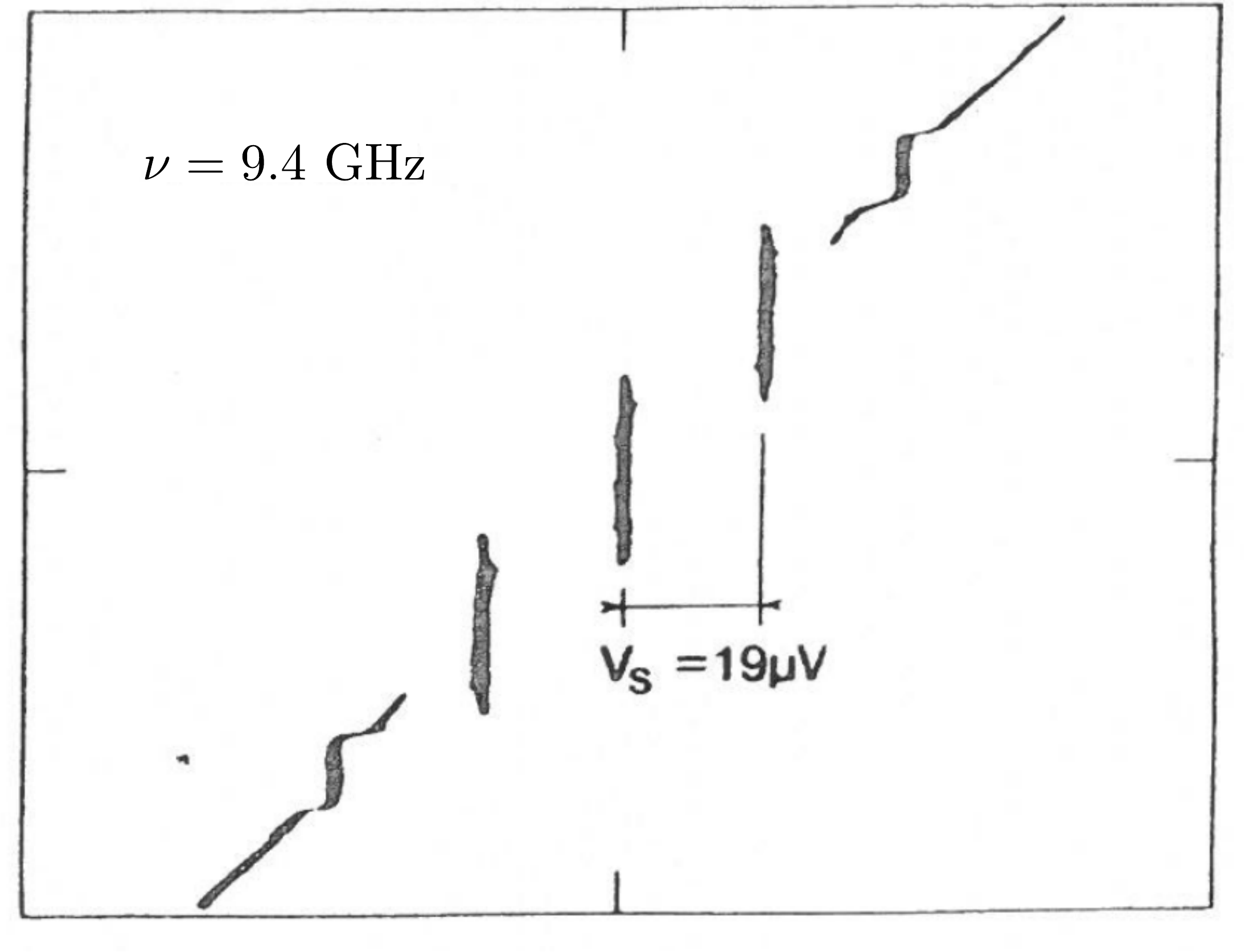}}
	\caption{ Oscilloscope presentation of current-versus-voltage characteristics of a tunnel junction at 4.2 K formed by an Al tip on a (La$_{0.925}$Sr$_{0.075}$)$_2$CuO$_4$ sample (\cite{Esteve:87}). Steps induced by incident microwave radiation at 9.4 GHz, implying $V_0=V_S=kK_J\nu=k\times 2.07\times 10^{-12}$ mV s$\times 9.4\times 10^9$ s$^{-1}=19.4\;\mu$V$\times k$ ($k=0,\pm 1,\pm 2, \dots$).}
    \label{fig:B2}  
\end{figure}

\begin{figure}[h]
	\centerline{\includegraphics*[width=6cm,angle=0]{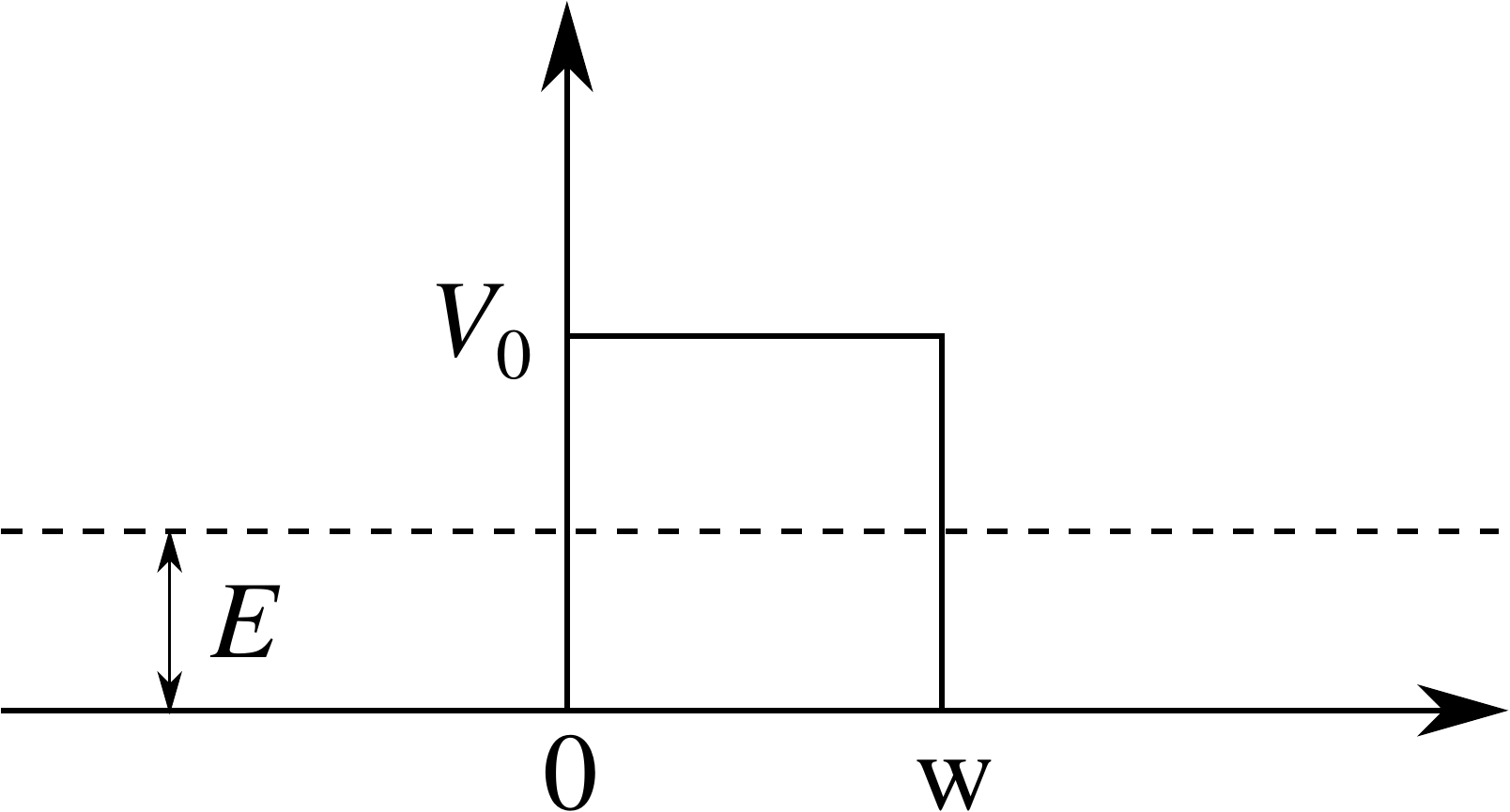}}
	\caption{Square potential barrier.}
    \label{fig:barrier}  
\end{figure}

\begin{figure}[h]
	\centerline{\includegraphics*[width=4cm,angle=0]{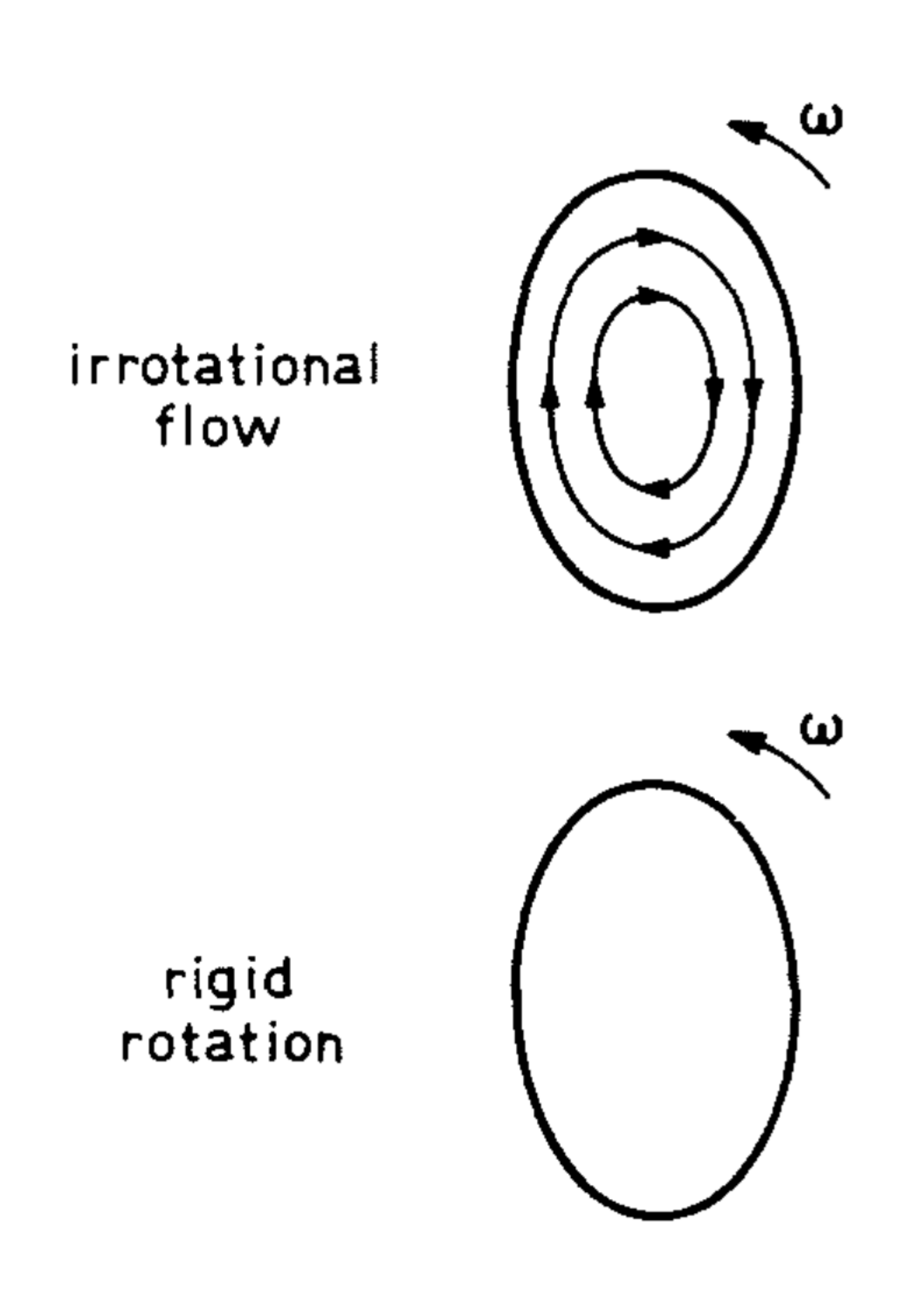}}
	\caption{Flow patterns for rotation of ellipsoidal bodies as seen from the rotating (body-fixed) frame of reference.}
    \label{fig:C1}  
\end{figure} 

\begin{figure}[h]
  \centerline{\includegraphics*[width=9cm,angle=0]{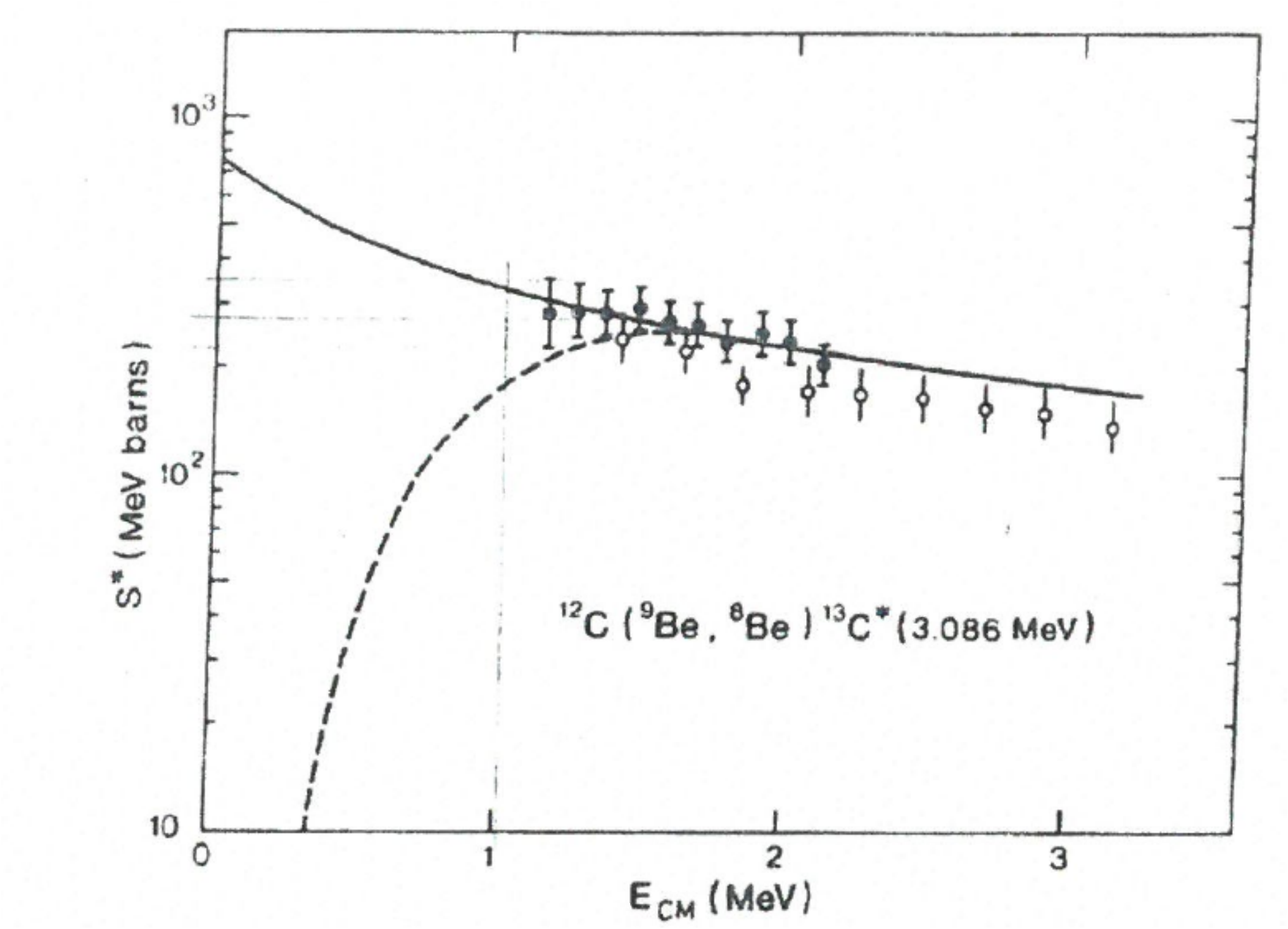}}
	\caption[.]{ The neutron stripping of $^9$Be on $^{12}$C at low bombarding energies. The astrophysical factor
      $
        S^*=E(\alpha)e^{i\eta_\alpha\arctan \kappa/k_\alpha}\sigma,
      $
      where $\sigma$ is the total transfer cross section is plotted as a function of the center-of-mass energy $E_{CM}=E(\alpha)$. The experimental data are taken from Caltech (open circles) and Melbourne (filled circles). The full drawn curve for $E_{CM}<1.5$ MeV indicates the theoretical low-energy limit, while the full drawn curve for $E>1.5$ MeV shows the semiclassical result as calculated with a spectroscopic factor $S(^9\text{Be}(1p_{1/2}))\cdot S(^{13}\text{C}(1s_{1/2}))=0.45 $. The dashed curve indicates the prolongation of this result into the region below 1.5 MeV where it is not expected to apply. The classical distance of closest approach in a head-on collision is 35 fm at 1 MeV (for details see \cite{Broglia:04a}).
    }
    \label{fig:D1}  
\end{figure}

\begin{figure}[h]
	\centerline{\includegraphics*[width=10cm,angle=0]{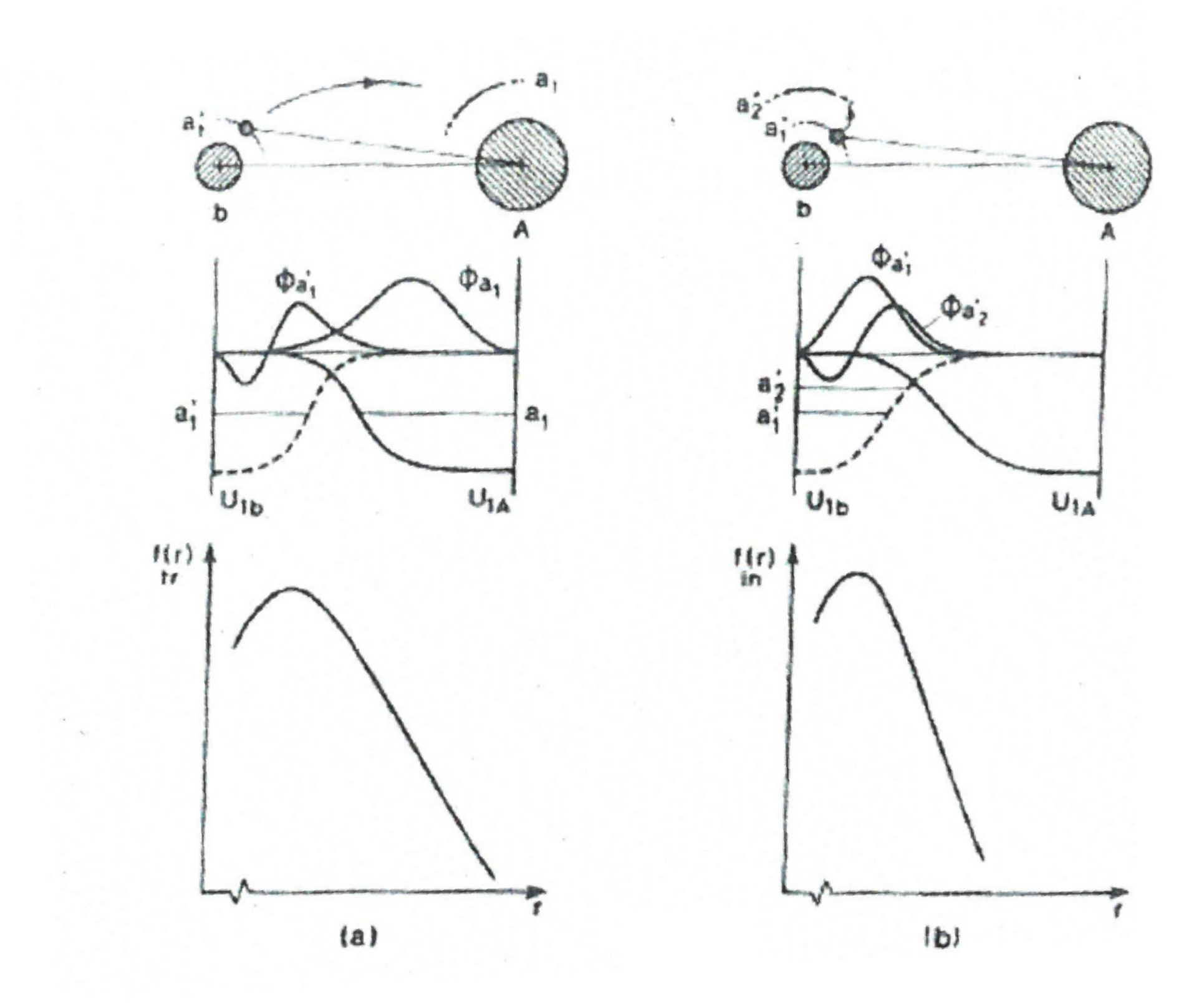}}
	\caption{Schematic representation  of the radial dependence of the one-particle transfer and inelastic form factor. In (a) a nucleon moving in the orbital with quantum numbers $a'_1$ in the projectile $a$ is transferred under the action of the shell model potential $U_{1A}$ to the target nucleus $A$ into an orbital $a_1$. The dependence of the form factor on the distance between the two nuclei is determined by the overlap of the product of the single-particle wavefunctions $\phi_{a'_1}$ and $\phi_{a_1}$ with the potential $U_{1A}$. A schematic representation of this dependence is given at the bottom of (a). In (b) a nucleon in the projectile $a$ is excited under the influence of the target field $U_{1A}$ from the single-particle orbital with quantum numbers $a'_1$ to the orbital with quantum numbers $a'_2$. The dependence of the form factor on the distance between the cores is here determined by the overlap of the product of the wavefunctions $\phi_{a'_1}$ and $\phi_{a'_2}$ with the potential $U_{1A}$. A representation of this dependence is shown at the bottom of (b) \cite{Broglia:04a}.}
    \label{fig:D2}  
\end{figure} 
\begin{figure}[h]
	\centerline{\includegraphics*[width=7cm,angle=0]{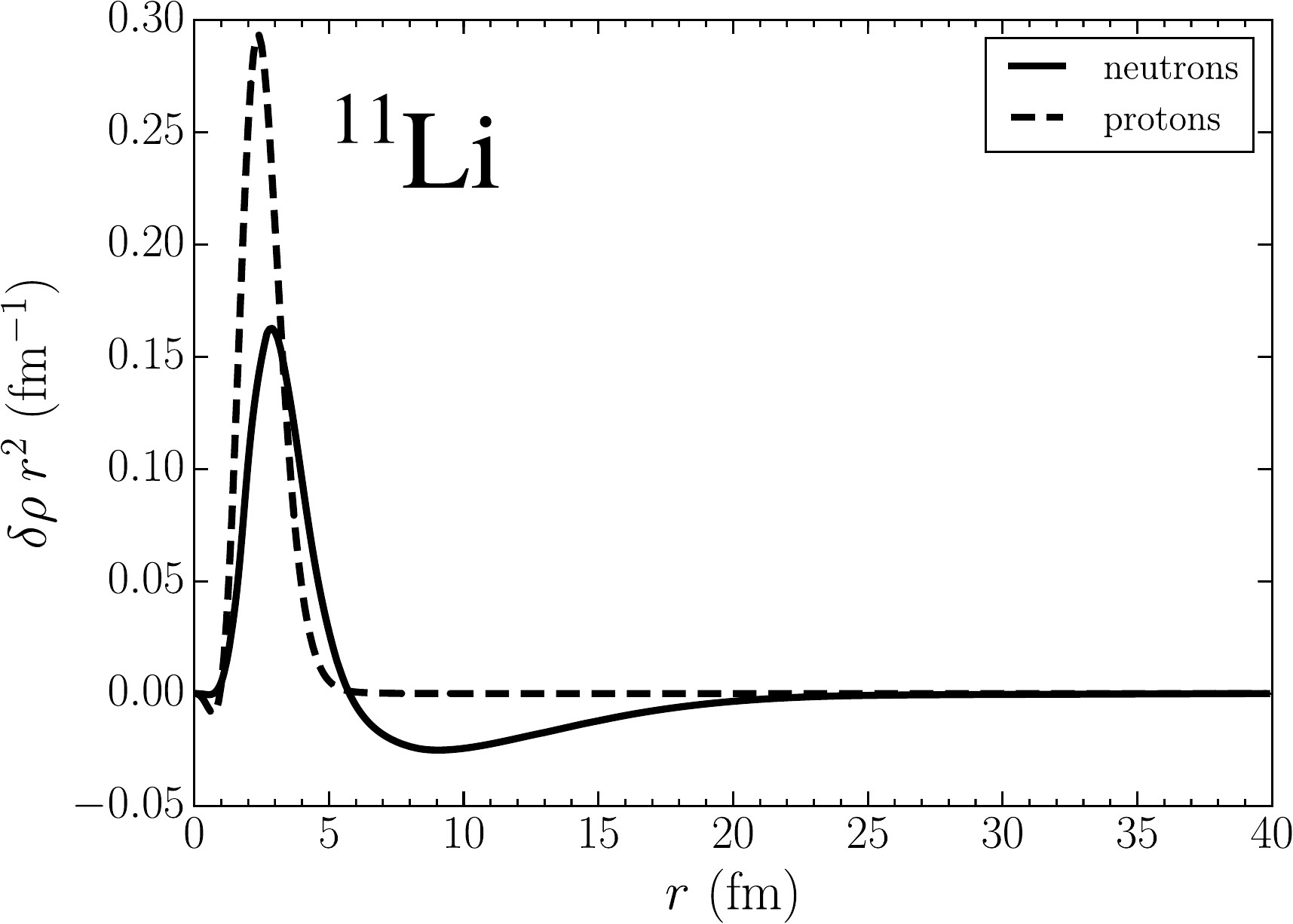}}
	\caption{Transition density of the soft $E1$-mode (pygmy dipole resonance) of $^{11}$Li (for details see \cite{Broglia:19}).}
    \label{fig:D3}  
\end{figure} 
\begin{figure}[h]
	\centerline{\includegraphics*[width=7cm,angle=0]{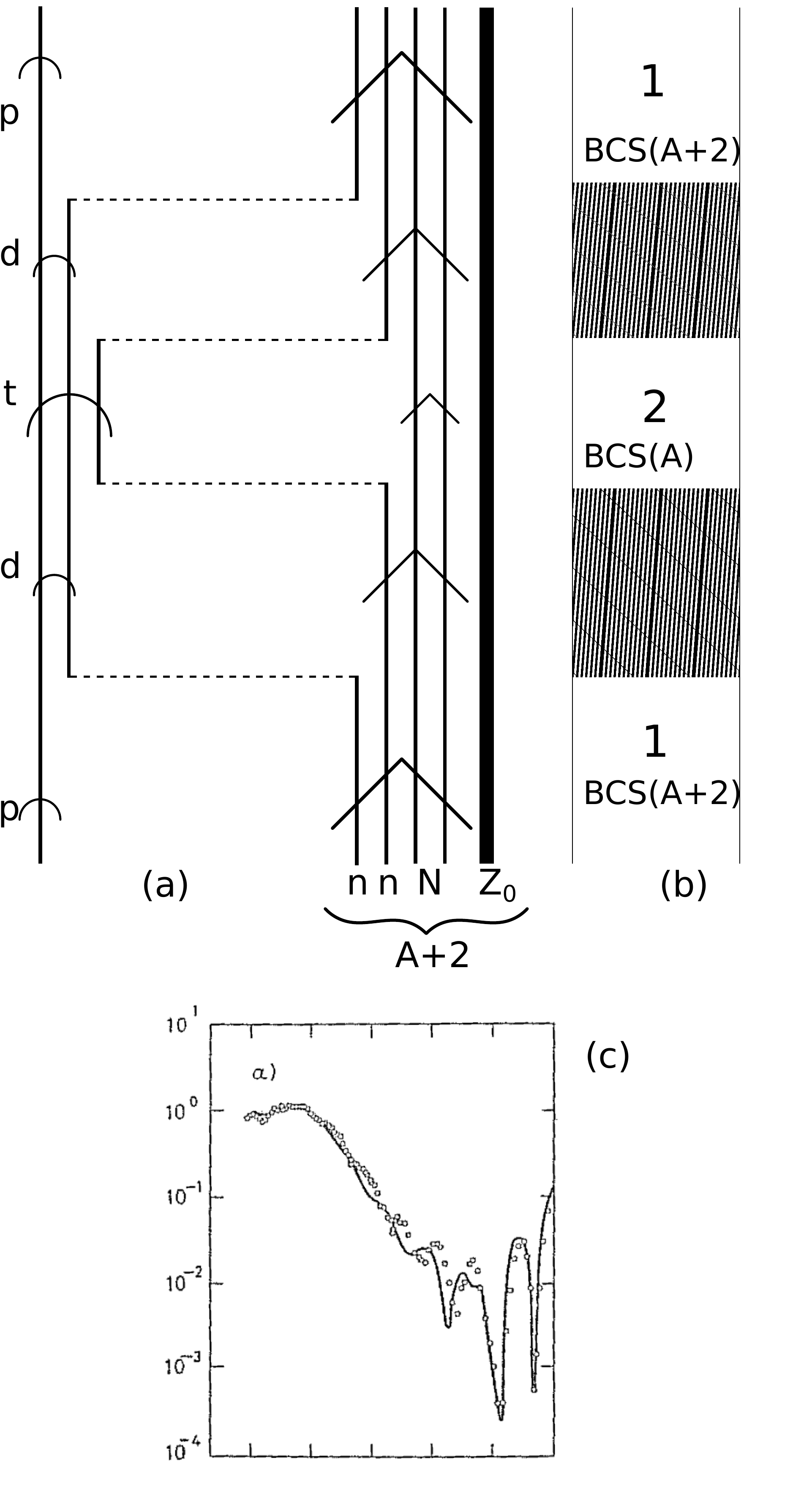}}
	\caption{Gedanken experiment concerning the possibility of observing weak coupling coherence phenomena between the states $\ket{BCS (A+2)}$ and $\ket{BCS (A)}$ in an elastic reaction involving superfluid nuclei (\textbf{a}), e.g. $p+^{120}$Sn$\to p+^{120}$Sn, the system  $^{119}$Sn+$d$ acting as a dynamical barrier (hatched areas arguably play role of that of dioxide layers in Josephson junctions) between the two even $N$ superfluid systems arising from the successive transfer of two nucleons (\textbf{b}) and eventually allowing for a time dependent gauge phase difference between the $(A+2)$ and $A$ superfluid systems, thus leading, in the case in which $Q$-value effects are appropriate, to an oscillating enhancement of the elastic cross section at large angles as observed, for quite different reasons, in the case of the elastic angular distribution of the reaction $^{16}$O+$^{28}$Si (\textbf{c}) (cf. \cite{Pollarolo:84}).} 
    \label{fig:D4}  
  \end{figure}


\clearpage






\begin{thebibliography}{10}

\bibitem{Bardeen:57a}
J.~Bardeen, L.~N. Cooper, and J.~R. Schrieffer.
\newblock Microscopic theory of superconductivity.
\newblock {\em Physical Review}, 106:162, 1957.

\bibitem{Bardeen:57b}
J.~Bardeen, L.~N. Cooper, and J.~R. Schrieffer.
\newblock Theory of superconductivity.
\newblock {\em Physical Review}, 108:1175, 1957.

\bibitem{Bohr:58}
A.~Bohr, B.~R. Mottelson, and D.~Pines.
\newblock Possible analogy between the excitation spectra of nuclei and those
  of the superconducting metallic state.
\newblock {\em Physical Review}, 110:936, 1958.

\bibitem{Mahaux:85}
C.~Mahaux, P.~F. Bortignon, R.~A. Broglia, and C.~H. Dasso.
\newblock Dynamics of the shell model.
\newblock {\em Physics Reports}, 120:1--274, 1985.

\bibitem{Josephson:62}
B.~D. Josephson.
\newblock Possible new effects in superconductive tunnelling.
\newblock {\em Phys. Lett.}, 1:251, 1962.

\bibitem{Anderson:64}
P.~W. Anderson.
\newblock Superconductivity.
\newblock {\em Science}, 144:373, 1964.

\bibitem{Dietrich:70}
K.~Dietrich.
\newblock {On a nuclear Josephson effect in heavy ion scattering}.
\newblock {\em Physics Letters B}, 32(6):428, 1970.

\bibitem{Dietrich:71b}
K.~Dietrich.
\newblock {Semiclassical theory of a nuclear Josephson effect in reactions
  between heavy ions}.
\newblock {\em Annals of Physics}, 66:480, 1971.

\bibitem{Hara:71}
K.~Hara.
\newblock {On the Josephson-current in heavy-ion reactions}.
\newblock {\em Physics Letters B}, 35:198, 1971.

\bibitem{Kleber:71}
M.~Kleber and H.~Schmidt.
\newblock Josephson effect in nuclear reactions.
\newblock {\em Zeitschrift f{\"u}r Physik}, 245:68, 1971.

\bibitem{Weiss:79}
H.~Weiss.
\newblock {Semiclassical description of two-nucleon transfer between superfluid
  nuclei}.
\newblock {\em Phys. Rev. C}, 19:834, 1979.

\bibitem{vonOertzen:01}
W.~von Oertzen and A.~Vitturi.
\newblock Pairing correlations of nucleons and multi--nucleon transfer between
  heavy nuclei.
\newblock {\em Reports on Progress in Physics}, 64:1247, 2001.

\bibitem{Potel:21}
G.~Potel, F.~Barranco, E.~Vigezzi, and R.~A. Broglia.
\newblock {Quantum entanglement in nuclear Cooper-pair tunneling with
  $\ensuremath{\gamma}$ rays}.
\newblock {\em Phys. Rev. C}, 103:L021601, 2021.

\bibitem{Montanari:14}
D.~Montanari, L.~Corradi, S.~Szilner, G.~Pollarolo, E.~Fioretto, G.~Montagnoli,
  F.~Scarlassara, A.~M. Stefanini, S.~Courtin, A.~Goasduff, F.~Haas,
  D.~Jelavi\'{c}~Malenica, C.~Michelagnoli, T.~Mijatovi\ifmmode~\acute{c}\else
  \'{c}\fi{}, N.~Soi\ifmmode~\acute{c}\else \'{c}\fi{}, C.~A. Ur, and
  M.~Varga~Pajtler.
\newblock {Neutron Pair Transfer in $^{60}\mathrm{Ni}+^{116}\mathrm{Sn}$ Far
  below the Coulomb Barrier}.
\newblock {\em Phys. Rev. Lett.}, 113:052501, 2014.

\bibitem{Montanari:16}
D.~Montanari, L.~Corradi, S.~Szilner, G.~Pollarolo, A.~Goasduff,
  T.~Mijatovi\ifmmode~\acute{c}\else \'{c}\fi{}, D.~Bazzacco, B.~Birkenbach,
  A.~Bracco, L.~Charles, S.~Courtin, P.~D\'esesquelles, E.~Fioretto, A.~Gadea,
  A.~G\"orgen, A.~Gottardo, J.~Grebosz, F.~Haas, H.~Hess, D.~Jelavi\ifmmode
  \acute{c}\else~\'{c}\fi{} Malenica, A.~Jungclaus, M.~Karolak, S.~Leoni,
  A.~Maj, R.~Menegazzo, D.~Mengoni, C.~Michelagnoli, G.~Montagnoli, D.~R.
  Napoli, A.~Pullia, F.~Recchia, P.~Reiter, D.~Rosso, M.~D. Salsac,
  F.~Scarlassara, P.-A. S\"oderstr\"om, N.~Soi\ifmmode~\acute{c}\else
  \'{c}\fi{}, A.~M. Stefanini, O.~Stezowski, Ch. Theisen, C.~A. Ur, J.~J.
  Valiente-Dob\'on, and M.~Varga~Pajtler.
\newblock Pair neutron transfer in
  $^{60}\text{Ni}+\phantom{\rule{0.16em}{0ex}}^{116}\text{Sn}$ probed via
  $\ensuremath{\gamma}$-particle coincidences.
\newblock {\em Phys. Rev. C}, 93:054623, 2016.

\bibitem{Magierski:21}
P.~Magierski.
\newblock {The Tiniest Superfluid Circuit in Nature}.
\newblock {\em Physics}, 14:27, 2021.

\bibitem{Cooper:56}
L.~N. Cooper.
\newblock {Bound Electron Pairs in a Degenerate Fermi Gas}.
\newblock {\em Phys. Rev.}, 104:1189, 1956.

\bibitem{Schrieffer:64}
J.~R. Schrieffer.
\newblock {\em Superconductivity}.
\newblock Benjamin, New York, 1964.

\bibitem{Penrose:51}
O.~Penrose.
\newblock {Bose-Einstein Condensation and Liquid Helium}.
\newblock {\em Philosophical Magazine}, 42:1373, 1951.

\bibitem{Penrose:56}
O.~Penrose and L.~Onsager.
\newblock {Bose-Einstein Condensation and Liquid Helium}.
\newblock {\em Phys. Rev.}, 104:576, 1956.

\bibitem{Yang:62}
C.~N. Yang.
\newblock {Concept of Off-Diagonal Long-Range Order and the Quantum Phases of
  Liquid He and of Superconductors}.
\newblock {\em Rev. Mod. Phys.}, 34:694, 1962.

\bibitem{Anderson:96}
P.~W. Anderson.
\newblock {Off-diagonal long-range order and flux quantization}.
\newblock In H.~Holden and S.~Kjellstrup~Ratkje, editors, {\em {The collected
  works of Lars Onsager}}, page 729. World Scientific, Singapore, 1996.

\bibitem{Pippard:53}
A.~B. Pippard.
\newblock An experimental and theoretical study of the relation between
  magnetic field and current in a superconductor.
\newblock {\em Proceedings of the Royal Society A}, 216:547, 1953.

\bibitem{Mottelson:02}
B.~Mottelson.
\newblock Elementary features of nuclear structure.
\newblock In H.~Niefenecker, J.~P. Blaizot, G.~F. Bertsch, W.~Weise, and
  F.~David, editors, {\em {Trends in Nuclear Physics, 100 years later, Les
  Houches, Session LXVI}}, page~25. Elsevier, Amsterdam, 1998.

\bibitem{Kubota:20}
Y.~Kubota, A.~Corsi, G.~Authelet, H.~Baba, C.~Caesar, D.~Calvet, A.~Delbart,
  M.~Dozono, J.~Feng, F.~Flavigny, J.-M. Gheller, J.~Gibelin, A.~Giganon,
  A.~Gillibert, K.~Hasegawa, T.~Isobe, Y.~Kanaya, S.~Kawakami, D.~Kim,
  Y.~Kikuchi, Y.~Kiyokawa, M.~Kobayashi, N.~Kobayashi, T.~Kobayashi, Y.~Kondo,
  Z.~Korkulu, S.~Koyama, V.~Lapoux, Y.~Maeda, F.~M. Marqu\'es, T.~Motobayashi,
  T.~Miyazaki, T.~Nakamura, N.~Nakatsuka, Y.~Nishio, A.~Obertelli, K.~Ogata,
  A.~Ohkura, N.~A. Orr, S.~Ota, H.~Otsu, T.~Ozaki, V.~Panin, S.~Paschalis,
  E.~C. Pollacco, S.~Reichert, J.-Y. Rouss\'e, A.~T. Saito, S.~Sakaguchi,
  M.~Sako, C.~Santamaria, M.~Sasano, H.~Sato, M.~Shikata, Y.~Shimizu,
  Y.~Shindo, L.~Stuhl, T.~Sumikama, Y.~L. Sun, M.~Tabata, Y.~Togano,
  J.~Tsubota, Z.~H. Yang, J.~Yasuda, K.~Yoneda, J.~Zenihiro, and T.~Uesaka.
\newblock Surface localization of the dineutron in $^{11}\mathrm{Li}$.
\newblock {\em Phys. Rev. Lett.}, 125:252501, 2020.

\bibitem{Shimizu:89}
Y.~R. Shimizu, J.~D. Garrett, R.~A. Broglia, M.~Gallardo, and E.~Vigezzi.
\newblock Pairing fluctuations in rapidly rotating nuclei.
\newblock {\em Reviews of Modern Physics}, 61:131, 1989.

\bibitem{Meissner:33}
W.~Meissner and R.~Ochsenfeld.
\newblock {Ein neuer Effekt bei Eintritt der Supraleit\"ahigkeit}.
\newblock {\em Naturw.}, 21:787, 1933.

\bibitem{Tinkham:96}
M.~Tinkham.
\newblock {\em Introduction to Superconductivity}.
\newblock Mc Graw-Hill, New York, 1980.

\bibitem{Waldram:96}
J.~R. Waldram.
\newblock {\em Superconductivity of metals and cuprates}.
\newblock Institute of Physics, Bristol, 1996.

\bibitem{Vollhardt:90}
D.~Vollhardt and P.~W{\"{o}}lfle.
\newblock {\em The superfluid phases of Helium 3}.
\newblock Taylor and Francis, London, 1990.

\bibitem{Cohen:62}
M.~H. Cohen, L.~M. Falicov, and J.~C. Phillips.
\newblock Superconductive tunneling.
\newblock {\em Phys. Rev. Lett.}, 8:316, 1962.

\bibitem{Anderson:64b}
P.~W. Anderson.
\newblock Special effects in superconductivity.
\newblock In E.~R. Caianiello, editor, {\em The Many-Body Problem, Vol.2}, page
  113. Academic Press, New York, 1964.

\bibitem{Pippard:12}
A.~B. Pippard.
\newblock {The historical context of Josephson discovery}.
\newblock In H.~Rogalla and P.~H. Kes, editors, {\em 100 years of
  superconductivity}, page~30. CRC Press, Taylor and Francis, FL, 2012.

\bibitem{Anderson:63}
P.~W. Anderson and J.~M. Rowell.
\newblock {Probable observation of the Josephson superconducting tunneling
  effect}.
\newblock {\em Physical Review Letters}, 10:230, 1963.

\bibitem{Rogalla:12}
H.~Rogalla and P.~H. Kes, editors.
\newblock {\em 100 years of superconductivity}.
\newblock CRC Press, Taylor and Francis, FL, 2012.

\bibitem{Bohr:88}
A.~Bohr and O.~Ulfbeck.
\newblock Quantal structure of superconductivity gauge angle.
\newblock In {\em {First Tops\o{}e summer School on Superconductivity and
  Workshop on Superconductors}}, Roskilde, Denmark Riso/M/2756, 1988.

\bibitem{Barranco:99}
F.~Barranco, R.~A. Broglia, G.~Gori, E.~Vigezzi, P.~F. Bortignon, and
  J.~Terasaki.
\newblock Surface vibrations and the pairing interaction in nuclei.
\newblock {\em Phys. Rev. Lett.}, 83:2147, 1999.

\bibitem{Terasaki:02b}
Jun Terasaki, F.~Barranco, E.~Vigezzi, R.~A. Broglia, and P.~F. Bortignon.
\newblock Effect of particle-phonon coupling on pairing correlations in finite
  systems -- the atomic nucleus --.
\newblock {\em Progress of Theoretical Physics}, 108:495, 2002.

\bibitem{Saperstein:12}
E.~E. Saperstein and M.~Baldo.
\newblock {Microscopic Origin of Pairing}.
\newblock In R.~A. Broglia and V.~Zelevinsky, editors, {\em 50 Years of Nuclear
  BCS}, page 263. World Scientific, Singapore, 2013.

\bibitem{Avdenkov:12}
A.~Avdeenkov and S.~Kamerdzhiev.
\newblock {Phonon Coupling and the Single-Particle Characteristics of Sn
  Isotopes}.
\newblock In R.~A. Broglia and V.~Zelevinsky, editors, {\em 50 Years of Nuclear
  BCS}, page 274. World Scientific, Singapore, 2013.

\bibitem{Lombardo:12}
U.~Lombardo, H.~J. Schulze, and W.~Zuo.
\newblock {Induced Pairing Interaction in Neutron Star Matter}.
\newblock In R.~A. Broglia and V.~Zelevinsky, editors, {\em 50 Years of Nuclear
  BCS}, page 338. World Scientific, Singapore, 2013.

\bibitem{Heisenberg:25}
W.~Heisenberg.
\newblock {\"{U}ber} quantentheoretische umdeutung kinematischer und
  mechanischer beziehungen.
\newblock {\em {Z. Phys.}}, 33:879, 1925.

\bibitem{Born:25a}
M.~Born and P.~Jordan.
\newblock Zur quantenmechanik.
\newblock {\em {Zeitschr. Phys.}}, 34:858, 1925.

\bibitem{Born:25b}
M.~Born, P.~Jordan, and W.~Heisenberg.
\newblock {Zur Quantenmechanik II}.
\newblock {\em {Zeitschr. Phys.}}, 35:557, 1926.

\bibitem{Broglie:25}
L.~de~Broglie.
\newblock Recherches sur la {th\'eorie} des quanta.
\newblock {\em Ann. Phys.}, 10:22, 1925.

\bibitem{Kramer:20}
D.~Kramer.
\newblock World’s physics instruments turn their focus to covid-19.
\newblock {\em Physics Today}, 73(5):22, 2020.

\bibitem{Bohr:76}
A.~Bohr.
\newblock {\em Rotational Motion in Nuclei, in Les Prix Nobel en 1975}.
\newblock Imprimerie Royale Norstedts Tryckeri, Stockholm, 1976.
\newblock p. 59.

\bibitem{Bayman:82}
B.~F. Bayman and J.~Chen.
\newblock One-step and two-step contributions to two-nucleon transfer
  reactions.
\newblock {\em Phys. Rev. C}, 26:1509, 1982.

\bibitem{Potel:13}
G.~Potel, A.~Idini, F.~Barranco, E.~Vigezzi, and R.~A. Broglia.
\newblock Cooper pair transfer in nuclei.
\newblock {\em {Rep. Prog. Phys.}}, 76:106301, 2013.

\bibitem{Gotz:75}
U.~{G\"{o}tz}, M.~Ichimura, R.~A. Broglia, and A.~Winther.
\newblock {Reaction mechanism of two-nucleon transfer between heavy ions}.
\newblock {\em Phys. Rep.}, 16:115, 1975.

\bibitem{Thompson:88}
I.~J. Thompson.
\newblock Coupled reaction channels calculations in nuclear physics.
\newblock {\em Comput. Phys. Rep.}, 7:167, 1988.

\bibitem{Broglia:04a}
R.~A. Broglia and A.~Winther.
\newblock {\em Heavy Ion Reactions}.
\newblock Westview Press, Boulder, CO., 2004.

\bibitem{Brink:68}
D.~M. Brink and G.~R. Satchler.
\newblock {\em Angular Momentum}.
\newblock Clarendon Press, 1968.

\bibitem{Rogovin:76}
D.~Rogovin and M.~Scully.
\newblock Superconductivity and macroscopic quantum phenomena.
\newblock {\em Physics Reports}, 25:175, 1976.

\bibitem{Broglia:87}
R.~A. Broglia, T.~{D\o{}ssing}, B.~Lauritzen, and B.~R. Mottelson.
\newblock Nuclear rotational damping: Finite-system analogue to motional
  narrowing in nuclear magnetic resonances.
\newblock {\em Phys. Rev. Lett.}, 58:326, 1987.

\bibitem{Brink:05}
D.~M. Brink and R.~A. Broglia.
\newblock {\em Nuclear Superfluidity}.
\newblock Cambridge University Press, Cambridge, 2005.

\bibitem{Satchler:80}
G.R. Satchler.
\newblock {\em Introduction to Nuclear Reactions}.
\newblock Mc Millan, New York, 1980.

\bibitem{Dedrick:60}
K.~G. Dedrick.
\newblock Jacobians and relativistic kinematics.
\newblock {\em Stanford University, Report No. M-226}, 1960.

\bibitem{Broglia:22}
R.~A. Broglia, F.~Barranco, G.~Potel, and E.~Vigezzi.
\newblock Transient joule- and (ac) josephson-like photon emission in one- and
  two- nucleon tunneling processes between superfluid nuclei: blackbody and
  coherent spectral functions.
\newblock {\em arxiv.2202.13193 [nucl-th]}, 2022.

\bibitem{Szilner:07}
S.~Szilner, C.~A. Ur, L.~Corradi, N.~M\ifmmode~\u{a}\else \u{a}\fi{}rginean,
  G.~Pollarolo, A.~M. Stefanini, S.~Beghini, B.~R. Behera, E.~Fioretto,
  A.~Gadea, B.~Guiot, A.~Latina, P.~Mason, G.~Montagnoli, F.~Scarlassara,
  M.~Trotta, G.~de Angelis, F.~Della Vedova, E.~Farnea, F.~Haas, S.~Lenzi,
  S.~Lunardi, R.~M\ifmmode~\u{a}\else \u{a}\fi{}rginean, R.~Menegazzo, D.~R.
  Napoli, M.~Nespolo, I.~V. Pokrovsky, F.~Recchia, M.~Romoli, M.-D. Salsac,
  N.~Soi\ifmmode~\acute{c}\else \'{c}\fi{}, and J.~J. Valiente-Dob\'on.
\newblock Multinucleon transfer reactions in closed-shell nuclei.
\newblock {\em Phys. Rev. C}, 76:024604, 2007.

\bibitem{Corradi:09}
L~Corradi, G~Pollarolo, and S~Szilner.
\newblock Multinucleon transfer processes in heavy-ion reactions.
\newblock {\em Journal of Physics G: Nuclear and Particle Physics},
  36(11):113101, 2009.

\bibitem{Gadea:11}
A.~Gadea, E.~Farnea, J.J. Valiente-Dobón, B.~Million, D.~Mengoni, D.~Bazzacco,
  F.~Recchia, A.~Dewald, Th. Pissulla, W.~Rother, G.~{de Angelis}, A.~Austin,
  S.~Aydin, S.~Badoer, M.~Bellato, G.~Benzoni, L.~Berti, R.~Beunard,
  B.~Birkenbach, E.~Bissiato, N.~Blasi, C.~Boiano, D.~Bortolato, A.~Bracco,
  S.~Brambilla, B.~Bruyneel, E.~Calore, F.~Camera, A.~Capsoni, J.~Chavas,
  P.~Cocconi, S.~Coelli, A.~Colombo, D.~Conventi, L.~Costa, L.~Corradi,
  A.~Corsi, A.~Cortesi, F.C.L. Crespi, N.~Dosme, J.~Eberth, S.~Fantinel,
  C.~Fanin, E.~Fioretto, Ch. Fransen, A.~Giaz, A.~Gottardo, X.~Grave,
  J.~Grebosz, R.~Griffiths, E.~Grodner, M.~Gulmini, T.~Habermann, C.~He,
  H.~Hess, R.~Isocrate, J.~Jolie, P.~Jones, A.~Latina, E.~Legay, S.~Lenzi,
  S.~Leoni, F.~Lelli, D.~Lersch, S.~Lunardi, G.~Maron, R.~Menegazzo,
  C.~Michelagnoli, P.~Molini, G.~Montagnoli, D.~Montanari, O.~Möller, D.R.
  Napoli, M.~Nicoletto, R.~Nicolini, M.~Ozille, G.~Pascovici, R.~Peghin,
  M.~Pignanelli, V.~Pucknell, A.~Pullia, L.~Ramina, G.~Rampazzo, M.~Rebeschini,
  P.~Reiter, S.~Riboldi, M.~Rigato, C.~{Rossi Alvarez}, D.~Rosso, G.~Salvato,
  J.~Strachan, E.~Sahin, F.~Scarlassara, J.~Simpson, A.M. Stefanini,
  O.~Stezowski, F.~Tomasi, N.~Toniolo, A.~Triossi, M.~Turcato, C.A. Ur,
  V.~Vandone, R.~Venturelli, F.~Veronese, C.~Veyssiere, E.~Viscione,
  O.~Wieland, A.~Wiens, F.~Zocca, and A.~Zucchiatti.
\newblock {Conceptual design and infrastructure for the installation of the
  first AGATA sub-array at LNL}.
\newblock {\em Nuclear Instruments and Methods in Physics Research Section A:
  Accelerators, Spectrometers, Detectors and Associated Equipment}, 654:88,
  2011.

\bibitem{Bohr:64}
A.~Bohr.
\newblock Elementary modes of excitation and their coupling.
\newblock In {\em {Comptes Rendus du Congr\`{e}s International de Physique
  Nucl\'{e}aire}}, volume~1, page 487. Centre National de la Recherche
  Scientifique, 1964.

\bibitem{Bohr:75}
A.~Bohr and B.~R. Mottelson.
\newblock {\em Nuclear Structure, Vol.II}.
\newblock Benjamin, New York, 1975.

\bibitem{Hogassen:61}
J.~H\"ogaasen-Feldman.
\newblock A study of some approximations of the pairing force.
\newblock {\em Nuclear Physics}, 28:258, 1961.

\bibitem{Bes:66}
D.~R. B{\`{e}}s and R.~A. Broglia.
\newblock Pairing vibrations.
\newblock {\em Nucl. Phys.}, 80:289, 1966.

\bibitem{Papenbrock:22}
T.~Papenbrock.
\newblock Effective field theory of pairing rotations.
\newblock {\em arXiv:2202.13146v1 [nucl-th]}, 2022.

\bibitem{Nilsson:55}
S.~G. Nilsson.
\newblock Binding states of individual nucleons in strongly deformed nuclei.
\newblock {\em Mat. Fys. Medd. Dan. Vid. Selsk.}, 29, 1955.

\bibitem{Feynman:63}
R.~P. Feynman.
\newblock {\em {Lectures on Physics}}, volume~3.
\newblock Addison-Wesley, Reading, Mass., 1963.

\bibitem{Potel:13b}
G.~Potel, A.~Idini, F.~Barranco, E.~Vigezzi, and R.~A. Broglia.
\newblock {Quantitative study of coherent pairing modes with two--neutron
  transfer: Sn isotopes}.
\newblock {\em {Phys. Rev. C}}, 87:054321, 2013.

\bibitem{Cavallaro:17}
M.~Cavallaro, M.~De~Napoli, F.~Cappuzzello, S.~E.~A. Orrigo, C.~Agodi,
  M.~Bond\'{\i}, D.~Carbone, A.~Cunsolo, B.~Davids, T.~Davinson, A.~Foti,
  N.~Galinski, R.~Kanungo, H.~Lenske, C.~Ruiz, and A.~Sanetullaev.
\newblock Investigation of the $^{10}\mathrm{Li}$ shell inversion by neutron
  continuum transfer reaction.
\newblock {\em Phys. Rev. Lett.}, 118:012701, 2017.

\bibitem{Tanihata:08}
I.~Tanihata, M.~Alcorta, D.~Bandyopadhyay, R.~Bieri, L.~Buchmann, B.~Davids,
  N.~Galinski, D.~Howell, W.~Mills, S.~Mythili, R.~Openshaw, E.~Padilla-Rodal,
  G.~Ruprecht, G.~Sheffer, A.~C. Shotter, M.~Trinczek, P.~Walden, H.~Savajols,
  T.~Roger, M.~Caamano, W.~Mittig, P.~Roussel-Chomaz, R.~Kanungo, A.~Gallant,
  M.~Notani, G.~Savard, and I.~J. Thompson.
\newblock Measurement of the two-halo neutron transfer reaction
  {$^1$H($^{11}$Li,$^{9}$Li)$^3$H} at {3A} {MeV}.
\newblock {\em Phys. Rev. Lett.}, 100:192502, 2008.

\bibitem{Fortune:94}
H.~T. Fortune, G.-B. Liu, and D.~E. Alburger.
\newblock {(\textit{sd}${)}^{2}$ states in $^{12}\mathrm{Be}$}.
\newblock {\em Phys. Rev. C}, 50:1355, 1994.

\bibitem{Schmitt:13}
K.~T. Schmitt, K.~L. Jones, S.~Ahn, D.~W. Bardayan, A.~Bey, J.~C. Blackmon,
  S.~M. Brown, K.~Y. Chae, K.~A. Chipps, J.~A. Cizewski, K.~I. Hahn, J.~J.
  Kolata, R.~L. Kozub, J.~F. Liang, C.~Matei, M.~Matos, D.~Matyas, B.~Moazen,
  C.~D. Nesaraja, F.~M. Nunes, P.~D. O'Malley, S.~D. Pain, W.~A. Peters, S.~T.
  Pittman, A.~Roberts, D.~Shapira, J.~F. Shriner, M.~S. Smith, I.~Spassova,
  D.~W. Stracener, N.~J. Upadhyay, A.~N. Villano, and G.~L. Wilson.
\newblock {Reactions of a ${}^{10}$Be beam on proton and deuteron targets}.
\newblock {\em Phys. Rev. C}, 88:064612, 2013.

\bibitem{Bassani:65}
G.~Bassani, N.~M. Hintz, C.~D. Kavaloski, J.~R. Maxwell, and G.~M. Reynolds.
\newblock {$(p,t)$ Ground-State $L=0$ Transitions in the Even Isotopes of Sn
  and Cd at 40 MeV, $N=$62 to 74}.
\newblock {\em Phys. Rev.}, 139:B830, 1965.

\bibitem{Bechara:75}
M.~J. Bechara and O.~Dietzsch.
\newblock States in $^{121}\mathrm{Sn}$ from the $^{120}\mathrm{Sn}(d,
  p)^{121}\mathrm{Sn}$ reaction at 17 {MeV}.
\newblock {\em Phys. Rev. C}, 12:90, 1975.

\bibitem{Guazzoni:04}
P.~Guazzoni, L.~Zetta, A.~Covello, A.~Gargano, G.~Graw, R.~Hertenberger, H.-F.
  Wirth, and M.~Jaskola.
\newblock {High-resolution study of the $^{116}$Sn $( p,t )$ reaction and shell
  model structure of $^{114}$Sn}.
\newblock {\em Phys. Rev. C}, 69:024619, 2004.

\bibitem{Broglia:19}
R.~A. Broglia, F.~Barranco, A.~Idini, G.~Potel, and E.~Vigezzi.
\newblock Pygmy resonances: what's in a name?
\newblock {\em Phys. Scr.}, 94:114002, 2019.

\bibitem{Bertsch:05}
G.~F. Bertsch and R.~A. Broglia.
\newblock {\em Oscillations in Finite Quantum Systems}.
\newblock Cambridge University Press, Cambridge, 2005.

\bibitem{Poole:95}
C.~P. Poole, H.~A. Farach, and R.~J. Creswick.
\newblock {\em Superconductivity}.
\newblock Academic Press, New York, 1995.

\bibitem{Giaver:73}
I.~Giaever.
\newblock Electron tunneling and superconductivity.
\newblock In {\em Le Prix Nobel en 1973}, page~84. Norstedt, P.A. and
  S{\"{o}}ner, 1973.

\bibitem{Lee:09}
J.~Lee, M.~B. Tsang, W.~G. Lynch, M.~Horoi, and S.~C. Su.
\newblock {Neutron spectroscopic factors of Ni isotopes from transfer
  reactions}.
\newblock {\em Phys. Rev. C}, 79:054611, 2009.

\bibitem{Esteve:87}
D~Est{\`{e}}ve, J.~M Martinis, C~Urbina, M.~H Devoret, G~Collin, P~Monod,
  M~Ribault, and A~Revcolevschi.
\newblock {Observation of the a.c. Josephson Effect Inside Copper-Oxide-Based
  Superconductors}.
\newblock {\em Europhysics Letters ({EPL})}, 3:1237, 1987.

\bibitem{Pollarolo:84}
G.~Pollarolo and R.~A. Broglia.
\newblock {Microscopic Description of the Backward Rise of the Elastic Angular
  Distribution $^{16}$O+$^{28}$Si}.
\newblock {\em Nuovo Cimento}, 81:278, 1984.

\end{thebibliography}
\end{document}